\documentclass[a4paper,twocolumn,11pt,accepted=2023-05-23,amsmath]{quantumarticle}
\pdfoutput=1

\usepackage{amsfonts}
\usepackage[utf8]{inputenc}
\usepackage[english]{babel}
\usepackage[T1]{fontenc}
\usepackage{amsmath}
\usepackage[numbers,sort&compress]{natbib}

\usepackage{graphicx}
\usepackage[T1]{fontenc}

\usepackage{amsmath, amsthm}
\usepackage{mathrsfs}
\usepackage{hyperref}
\usepackage{url}
\usepackage{braket}

\usepackage[dvipsnames]{xcolor}
\definecolor{C1}{RGB}{52, 89, 149}
\definecolor{C2}{RGB}{251, 77, 61}
\definecolor{C3}{RGB}{3, 206, 164}
\definecolor{C4}{RGB}{202, 21, 81}
\hypersetup{colorlinks=true, linkcolor=C2, citecolor=C2, urlcolor=C2}
\usepackage{dsfont}
\usepackage{epstopdf}

\newcommand*{\id}{\mathds{1}}
\newcommand*{\f}{\frac}

\newcommand*{\mc}{\mathcal}
\newcommand*{\dg}{\dagger}
\newcommand*{\ex}{\mathrm{e}}

\newcommand*{\nn}{\nonumber}
\DeclareMathOperator{\tr}{tr}

\newtheorem{theorem}{Theorem}
\newtheorem{result}[theorem]{Result}

\definecolor{mathematicaorange}{rgb}{0.880722, 0.611041, 0.142051} 

\definecolor{mathematicablue}{rgb}{0.368417, 0.506779, 0.709798} 

\definecolor{mathematicayellow}{rgb}{0.880722, 0.611041, 0.142051} 

\definecolor{mathematicared}{rgb}{0.922526, 0.385626, 0.209179} 

\definecolor{mathematicagreen}{rgb}{0.560181, 0.691569, 0.194885} 

\begin{document}
\title[]{Relaxation of Multitime Statistics in Quantum Systems} 

\author{Neil Dowling}
\email[]{neil.dowling@monash.edu}
\address{School of Physics \& Astronomy, Monash University, Victoria 3800, Australia}

\author{Pedro Figueroa--Romero}
\address{Hon Hai Quantum Computing Research Center, Taipei, Taiwan}

\author{Felix A. Pollock}
\address{School of Physics \& Astronomy, Monash University, Victoria 3800, Australia}

\author{Philipp Strasberg}
\address{F\'isica Te\`orica: Informaci\'o i Fen\`omens Qu\`antics, Departament de F\'isica, Universitat Aut\`onoma de Barcelona, 08193 Bellaterra (Barcelona), Spain}

\author{Kavan Modi}
\email[]{kavan.modi@monash.edu}
\address{School of Physics \& Astronomy, Monash University, Victoria 3800, Australia}

\begin{abstract}
Equilibrium statistical mechanics provides powerful tools to understand physics at the macroscale. Yet, the question remains how this can be justified based on a microscopic quantum description. Here, we extend the ideas of pure state quantum statistical mechanics, which focus on single time statistics, to show the equilibration of isolated quantum processes. Namely, we show that most multitime observables for sufficiently large times cannot distinguish a nonequilibrium process from an equilibrium one, unless the system is probed for an extremely large number of times or the observable is particularly fine-grained. A corollary of our results is that the degree of non-Markovianity and other multitime characteristics of a nonequilibrium process also equilibrate.
\end{abstract}


\maketitle

\section{Introduction}
Conventional open quantum system theory~\cite{Rivas2012,Rotter2015} emphasizes understanding single-time expectation values $\braket{X(t)}_\rho := \tr[X(t) \rho]$, where $X(t)$ is a linear operator in the Heisenberg picture, and $\rho$ is a density operator describing some initial state. However, such single-time expectation values only reveal a small fraction of the available information about a quantum process compared to multitime correlation functions of the form 
\begin{equation}
    \label{eq:nPoint}
    \braket{X_k(t_k)\! \cdots \! X_1(t_1) }_\rho = \tr\left[X_k(t_k)\! \cdots\! X_1(t_1) \rho\right]
\end{equation}
which appear, for example, in linear response theory~\cite{pottier2010nonequilibrium, Kubo1966, Weiss2012}, the formalism of non-equilibrium Green's functions~\cite{stefanuccivanleeuwen2013}, the quantum regression theorem~\cite{Lax1967}, and also play an integral role in our modern understanding of quantum stochastic processes and (non-)Markovianity, as we explain later in greater detail~\cite{processtensor, processtensor2, Wiseman2018}. 

    \begin{figure}[t]
\centering
\includegraphics[width=0.48\textwidth]{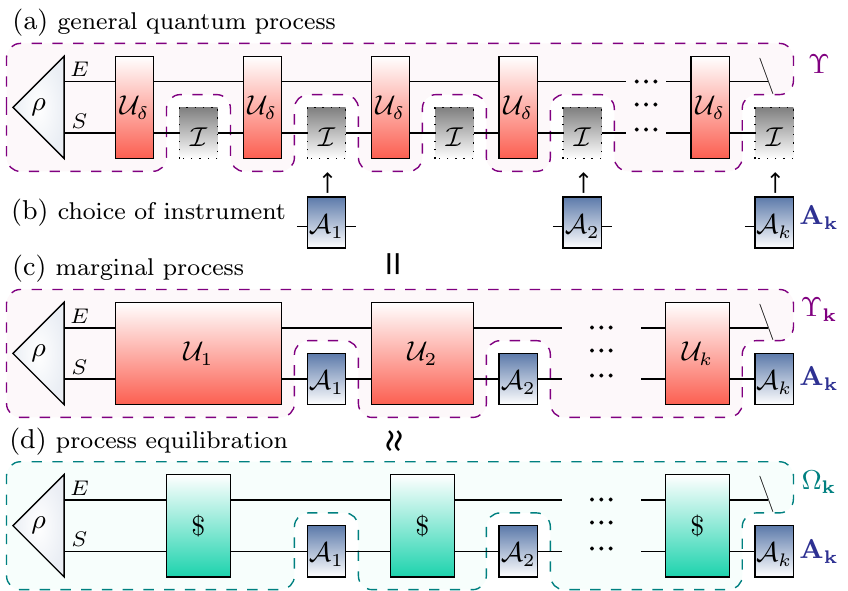}
\caption{ (a) A general quantum process $\Upsilon$ {composed of infinitesimal unitary evolution operators  $\mc{U}_\delta$ and some initial state $\rho$}. (b) Multitime instrument $\textbf{A}_\textbf{k}$, corresponding to any sequence of quantum operations (including measurements), is applied at times $\textbf{k}$ on the accessible system Hilbert space labeled $S$, with implicit identity operators $\mc{I}$ in-between. (c) The resultant expectation value is sampled from the marginal process $\Upsilon_\textbf{k}$ (purple dotted comb). (d) This is indistinguishable from an equilibrium process $\Omega$ (green dotted comb), composed of dephasing $\$$ rather than unitary evolution, for most $\textbf{A}_\textbf{k}$.} 
\label{fig:ProcessTensor}
\end{figure}

Due to the greater informational content of multitime correlation functions, they may reveal richer non-equilibrium features. We can naturally ask then, if such objects tend towards an equilibrium quantity in generic many-body situations? We can phrase this question in a rigorous manner by reformulating Eq.~\eqref{eq:nPoint} in terms of an expectation value of a single object,
\begin{equation}
\begin{split} \label{eq:nPointGamma}
    \braket{X_k(t_k)\cdots X_1(t_1) }_\rho &= \tr[\mathsf{O}_{\textbf{k}} \Upsilon_{\textbf{k}}] =: \braket{\mathsf{O}_{\textbf{k}}}_{\Upsilon},
\end{split}
\end{equation}
where $\Upsilon_{\textbf{k}}$ and $\mathsf{O}_{\textbf{k}}$ are tensor representations of a quantum process and a multitime quantum observable, respectively, over times $\textbf{k}:=\{t_1, \dots, t_k\}$. That is, $\mathsf{O}_{\textbf{k}}$ encodes the time-ordered sequence of operators $\{X_k, X_{k-1}, \dots, X_1\}$, now in the Schr\"odinger picture, applied at times $\textbf{k}$. Here, $\Upsilon$ encompasses all time evolution operators and an initial state $\rho$, with the tensor $\Upsilon_\textbf{k}$ being the `marginal process' encompassing only times $\textbf{k}$~\cite{Milz2020kolmogorovextension, milz2020quantum}, as depicted in Fig.~\ref{fig:ProcessTensor} (a)-(c), and detailed below.

Recasting the above question, we are asking whether we could replace the process $\Upsilon$, which contains time dependencies, with a corresponding time-independent \emph{equilibrated process} $\Omega$, such that
\begin{equation}\label{eq:equilibration}
    \braket{\mathsf{O}_{\textbf{k}}}_{\Upsilon} \approx \braket{\mathsf{O}_{\textbf{k}}}_{\Omega}\quad \mbox{for most } \textbf{k}.
\end{equation}
If so, how can we define $\Omega$? for which observables is this valid? and how universal is this equilibration?

In this work, we show that Eq.~\eqref{eq:equilibration} indeed holds for generic many-body quantum processes. Further, in a follow up work we show that Eq.~\eqref{eq:equilibration} holds over finite time intervals~\cite{finitetime}. In both cases, the error in the approximation is shown to be smaller than the inverse \emph{effective dimension} (participation ratio) of a quantum state $\sigma$ with respect to a Hamiltonian $H=\sum_{n} E_{n} P_{n} $, defined as 
\begin{equation} \label{eq:deff}
    d_{\mathrm{eff}}[\sigma] \! := \! \frac{1}{\tr[\$(\sigma)^2]}, \, \,  \mbox{with} \, \, \, \, \$(\cdot) \! := \! \sum_{n} P_{n} (\cdot) P_{n}
\end{equation} 
being the dephasing map with respect to $H$. The effective dimension naturally appears in many contexts~\cite{Linden2008,Neuenhahn2012,Zanardi2010,Bochieri1957,Gogolin}. For instance, the recurrence time for an isolated quantum system typically scales exponentially with $d_{\rm eff}$~\cite{venuti2015recurrence}. Intuitively then, it acts as a quantifier for the validity of a statistical description of this system. 
A system with a small effective dimension essentially behaves as a small quantum system, and so is not expected to display thermodynamic behavior. On the other hand, in realistic many-body situations the effective dimension scales exponentially with system size~\cite{Reimann_2008, Linden2008}.

Our results extend and complement related work, such as Ref.~\cite{Alhambra2020-fj} where they prove a thermal two-point correlation function equilibration bound, which can be seen as a special case of our general results Eq.~\eqref{eq:equilibration}. Formulating our results in terms of instruments and process tensors has the advantage of allowing us to determine equilibration of operationally meaningful multitime properties, such as the degree of non-Markovianity (see Section~\ref{sec:markov}), and beyond~\cite{finitetime}. In both instances, however, it is the large effective dimension that dictates equilibration and the related phenomenon of Markovianisation~\cite{Pedro2021}. We comment further on the relation between Ref.~\cite{Alhambra2020-fj} and the present work in Section~\ref{sec:markov}, and in Appendix~\ref{ap:MainResultProof}.

Eq.~\eqref{eq:equilibration} is addressing the foundational problem at the heart of quantum statistical physics of how equilibrium conditions emerge from underlying unitary dynamics. Here, we show that to witness nonequilibrium signatures of a process describing many-body dynamics, one requires an astronomically large number of measurements. This further substantiates the widely held belief that quantum mechanics alone should be enough to derive statistical mechanics and the emergence of thermodynamics, without any additional assumptions~\cite{gemmer2009, Gogolin,Rigol2016, Mori_2018}.

We begin by casting generic quantum dynamics as a \emph{process tensor} $\Upsilon$~\cite{processtensor,processtensor2, Costa2016,milz2020quantum}, independent of the observables, using the quantum combs formalism~\cite{Chiribella2009physrevA}. This innovation then allows us to unambiguously define a corresponding equilibrium process $\Omega$, which we show to be nearly indistinguishable from $\Upsilon$. We will show that Eq.~\eqref{eq:equilibration} holds for multitime correlation functions under general conditions, extending the usual paradigm of equilibration which largely considers only single time observables~\cite{Tasaki_1998, Reimann_2008, Linden2008, ShortSystemsAndSub} (\cite{Alhambra2020-fj} is an exception to this).
However, we go further by providing a platform for comparing nonequilibrium \emph{processes} with equilibrium ones. For instance, we show that the non-Markovianity of a process is restricted by the process equilibration bounds, while in a different work we show that this extends to a range of other general multitime features, such as classicality of measurement statistics~\cite{finitetime}. While equilibration is a necessary condition for thermalization, other assumptions are needed for this equilibrium state to be of Gibbs form~\cite{Gogolin, Ueda2020}. Our results on process equilibration and the restriction of non-Markovianity explore a different facet of the question of thermalization of isolated systems, indicating a stability of equilibrium against perturbations, and a relaxation of quantum memory.

\section{Tensor Representation of Quantum Processes}
Consider an isolated finite-dimensional quantum system as in Fig.~\ref{fig:ProcessTensor} (c). Let the initial state of the system, $\rho$, evolve to time $t_1$ due to unitary evolution generated by Hamiltonian $H$. We write the unitary dynamics from $t_{\ell-1}$ to $t_\ell$ in its superoperator form as $\mc{U}_\ell(\cdot):= \ex^{-iH \Delta{t}_\ell}(\cdot)\,\ex^{iH \Delta{t}_\ell}$, where $\Delta{t}_\ell = t_\ell - t_{\ell-1}$. In accordance to Eq.~\eqref{eq:nPoint}, we interrogate the system at times ${\textbf{k}} := \{t_1,\dots,t_k\}$ with operations $\{\mc{A}_{1},\dots,\mc{A}_{k}\}$. The $\mc{A}_i$ are called instruments~\cite{Davies1970} and are completely positive (CP) maps, representing any physical operations in a laboratory. In principle, these act on an experimentally accessible space, which we refer to as the system ($S$) and we describe the rest as the environment ($E$), such that $\mc{A} \equiv \mc{A}_S \otimes \mc{I}_E$. If the $\mc{A}_i$ are further specified trace preserving (TP), they may represent a complete set of outcomes of the measurement (or simply a quantum channel), and otherwise generally correspond to a subset of outcomes. This is the most general way to describe measurement and observables in quantum theory.

We can then cast the full process as
\begin{align}
\label{eq:multitime expectation}
    \tr[\mc{A}_{k}\mc{U}_k\cdots \mc{A}_{1}\mc{U}_1(\rho)]&=\tr[\Upsilon_{\textbf{k}} \textbf{A}_{\textbf{k}}^{\mathrm{T}}] \nonumber \\
    &=: \braket{\textbf{A}_{\textbf{k}}}_{\Upsilon}.
\end{align}
Here, $\Upsilon_{\textbf{k}}$ is called a \emph{process tensor} and $\textbf{A}_{\textbf{k}}$ is a multitime instrument, explicitly defined in Eq.~\eqref{eq:LinkProdProcessTensor}. To see how Eq.~\eqref{eq:multitime expectation} can be decomposed into these objects, see Fig.~\ref{fig:ProcessTensor}. Further explanation of this can be found in Appendix~\ref{ap:link}.

Any observable can be decomposed as a linear combination of a family of instruments, and so any correlation function~\eqref{eq:nPointGamma} can be obtained from corresponding expectation values of instruments:
\begin{equation} \label{eq:GammaDef}
    \mathsf{O}_{\textbf{k}}^{\mathtt{T}} \! = \! \sum_i \! \alpha_i \textbf{A}_{\textbf{k}}^{(i)},
    \ \ \Rightarrow \ \
    \braket{\mathsf{O}_{\textbf{k}}^{\mathtt{T}} }_\Upsilon \!=\! \sum_i \!\alpha_i \langle \textbf{A}_{\textbf{k}}^{(i)} \rangle_{\Upsilon_{\textbf{k}}}
\end{equation}
with $\alpha_i \!\in\! \mathbb{C}$. This is shown explicitly in Appendix~\ref{appendix:nPoint}. The equilibration of Eq.~\eqref{eq:nPoint} follows as a corollary of our main results due to Eq.~\eqref{eq:GammaDef}, while instruments have the conceptual advantage of being meaningful interventions in a laboratory.

Both $\Upsilon_{\textbf{k}}$ and $\textbf{A}_{\textbf{k}}$ are quantum combs~\cite{Chiribella_2008}, and naturally arise in many physical contexts~\cite{Hardy_2007, hardy2012, hardy2016operational, Cotler2018,Werner2005, Caruso2014, Portmann2017, CostaBornRule, CostaOreshkov2012ER, PhysRevE.100.022127, arXiv:1811.03722,PhysRevE.100.022135, PhysRevLett.123.180604, strasberg2019, Milz2020prx}; with further details to be found in Ref.~\cite{finitetime}. Processes and the multitime temporal correlation functions computable from them are highly relevant to modern experiments~\cite{Chernyak2006,Engel2007,Krumm2016,Moreva2017,Miller2017,Ringbauer2018,White2020,white2022manybody}.

Eq.~\eqref{eq:multitime expectation} is the multitime generalization of the Born rule~\cite{Chiribella2009physrevA, CostaOreshkov2012ER, CostaBornRule}, where $\Upsilon_{\textbf{k}}$ plays the role of a state (where it has the properties of an unnormalized density matrix) and $\textbf{A}_{\textbf{k}}$ that of a measurement, accounting for the invasive nature of measurements in quantum mechanics~\cite{Davies1970}. When $k=1$, we obtain the usual quantum mechanical expectation value with $\textbf{A}_\textbf{k} \to \mathtt{A}_1$, which is a measurement operator: an element of a \emph{positive operator valued measure} (POVM). Equipped with these concepts, we are now in position to define an equilibrated process.

\section{Equilibrium Process}
The key advantage of using the process tensor formalism is that all correlations and dynamics are stored in a single object, $\Upsilon_{\textbf{k}}$. This allows us to define an equilibrium process $\Omega_{\textbf{k}}$, corresponding to the nonequilibrium $\Upsilon_{\textbf{k}}$, as the infinite-time average over all time intervals $\textbf{k}$,
\begin{equation}
    \begin{split}
\label{eq:LinkProdProcessTensor1}
        \Omega_{\textbf{k}} := \overline{\Upsilon_{\textbf{k}}}^\infty &= \Big(\prod_{i=1}^k  \lim_{T_i\to\infty}
        \f{1}{T_i}
        \int_0^{T_i}
        d(\Delta t_i) \Big)
        \Upsilon_{\textbf{k}}
    \end{split}
\end{equation}
That is, the equilibrium process is defined by replacing all unitary evolutions with dephasing (defined in Eq.~\eqref{eq:deff}), as depicted in Fig.~\ref{fig:ProcessTensor} (c)-(d). The equilibrium process $\Omega_{\textbf{k}}$ is the reference with which we will compare an arbitrary process to, in order to define process equilibration. A multitime expectation value on such a process then corresponds to
\begin{equation} \label{eq:equilExp}
     \braket{\textbf{A}_{\textbf{k}}}_{\Omega}:=\tr[\mc{A}_{k}\$\cdots \mc{A}_{1}\$(\rho)]=\tr[\Omega_{\textbf{k}}\textbf{A}_{\textbf{k}}^{\mathrm{T}}].
\end{equation}
The key feature here is that the expectation value is an intervention-time independent quantity.

Up until now we have defined $k$-step processes. However, both $\Upsilon$ and $\Omega$ are defined over all times~\cite{Milz2020kolmogorovextension, milz2020quantum}. In fact, it is the instrument that is defined on $k$-steps, such that Eq.~\eqref{eq:multitime expectation} means that we are sampling from a marginal process $\Upsilon_{\textbf{k}} \subset \Upsilon$~\cite{Milz2020kolmogorovextension}. This allows the physical interpretation of $\braket{\textbf{A}_\textbf{k}}_\Omega$ as the average of $\braket{\textbf{A}_\textbf{k}}_\Upsilon$ over all possible interrogation times an experimenter could choose to apply the multitime instrument $\textbf{A}_\textbf{k}$; see Fig.~\ref{fig:ProcessTensor}. This marginalization to finite number of measurement times can be interpreted as a coarse graining in time, which along with coarse graining in space is needed for general process equilibration, as we will see in the results that follow. First, we quantify how well a given $k$-time instrument can tell apart $\Upsilon$ from $\Omega$.

\section{Equilibration of Multitime Observables}
Dynamical equilibration of quantum processes says that for almost all $\textbf{k}$, a nonequilibrium process $\Upsilon$ cannot be distinguished from the corresponding equilibrium process $\Omega$. This is stronger than usual notions of equilibrium, as it encompasses multitime observables, and allows for almost arbitrary interventions. For all following results, we assume that the full $SE$ is isolated, initially in the state $\rho$, and so evolves unitarily. Our only assumption on the dynamics is that the time independent Hamiltonian that dictates the $SE$ evolution obeys the non-resonance condition: $E_m-E_n = E_{m^\prime}-E_{n^\prime} \neq 0$ if and only if $m=m^{\prime}$ and $n=n^\prime$. This condition is not so restrictive in many-body systems without contrived symmetries, and can in fact be loosened; see Ref.~\cite{finitetime}. This means that our results are valid for arbitrary non-equilibrium states $\rho$, with entirely minimal restrictions on the dynamics.

Our main results follow from the following bound in terms of the expectation value of any $k$-time instrument $\textbf{A}_{\textbf{k}}$ applied on the system (of dimension $d_S$) of a process $\Upsilon$ and its corresponding equilibrium $\Omega$, 
\begin{gather}
    \label{eq:result1} \overline{|\langle\textbf{A}_{\textbf{k}}\rangle_\Upsilon-\langle \textbf{A}_{\textbf{k}} \rangle_\Omega|^2}^\infty  \leq   \frac{(2^{k}-1)d_S^{2k}} {d_{\mathrm{eff}}[\rho]}\, .
\end{gather}    
A somewhat tighter bound, which depends on intermediate effective dimensions within the process and operator norms of the multitime instrument $\textbf{A}_{\textbf{k}}$, can be found in Appendix~\ref{ap:MainResultProof}, together with a proof of Eq.~\eqref{eq:result1}. While all of the results in this paper may be stated in terms of this tighter bound, we have presented Eq.~\eqref{eq:result1} due to its clear physical interpretation. One can immediately compare the size of the accessible system space to (an estimate of) the effective dimension, to determine if your measurement statistics will equilibrate. Note that if the process ends after only one intervention, the tighter bound reduces to the single-time equilibration result of Ref.~\cite{ShortSystemsAndSub}.

As we have discussed, $d_{\mathrm{eff}}$ typically scales exponentially with the number of particles $N$, and so for macroscopic systems and coarse enough observables the right hand side of this bound is vanishingly small unless one probes the system for a very large number of times, $k \sim N/(2\log_2 d_S)$. This is clearly not possible for realistic many-body situations, where $N \sim \mc{O}(10^{23})$, and where experimental limitations restrict the degrees of freedom that an instrument may access (corresponding to the size of $d_S$). Even constructing an experiment in the lab with $k=100$ time measurements, and determining all correlations between them, is practically unfeasible.

Considering that $\langle\textbf{A}_{\textbf{k}} \rangle_\Upsilon$ is a random variable on $\{(\Delta t_1,\Delta t_2,\dots,\Delta t_k):\Delta t_i\geq0\}$, $\langle\textbf{A}_{\textbf{k}}\rangle_\Omega$ is exactly the mean and Eq.~\eqref{eq:result1} gives an upper bound on the variance. We can therefore use Chebyshev's inequality to bound the probability $\mathbb{P}$ of deviation from equilibrium multitime observables, to arrive at our first main physical result on the equilibration of multitime observables.

\begin{result} \label{result:nCheby}
For any $k$-point correlation function, $\braket{\mathsf{O}_{\textbf{k}}}_{\Upsilon-\Omega}:=\braket{\mathsf{O}_{\textbf{k}}}_{\Upsilon}-\braket{\mathsf{O}_{\textbf{k}}}_{\Omega}$, computed over a process $\Upsilon$ and equilibrated $\Omega$,
\begin{equation} \label{eq:nPointCheby}
    \mathbb{P}\! \Bigg\{\!\left|\braket{\mathsf{O}_{\textbf{k}}}_{\Upsilon-\Omega}\right|\!\geq\! \frac{d_S^k \!\sqrt{2^{k}\!-\!1} \sum_i \! | \alpha_i|}{d_{\mathrm{eff}}^{1/3}[\rho]} \!\Bigg\} \! \leq \! \frac{1}{d_{\mathrm{eff}}^{{1}/{3}}[\rho]}\!
\end{equation}
where $\alpha_i$ are defined in the decomposition Eq.~\eqref{eq:GammaDef}.
\end{result}
A proof of this follows from Eq.~\eqref{eq:result1}, and can be found in Appendix~\ref{ap:chebyProof}. This result states that, for most $k-$time observables $\mathsf{O}_{\textbf{k}}$, $\Upsilon$ and $\Omega$ look the same when the effective dimension is large. This can be interpreted as the statement that for typical many-body systems (with a large effective dimension), the overlap between measurement operators and the system state $\rho$ is unbiased for most times, even with respect to temporally non-local apparatuses. Of course, this is a statement for most times, and does not exclude transient non-equilibrium behavior. For example, soon after a quantum quench, non-equilibrium behavior may of course be apparent; see also Ref.~\cite{Gemmer2020}.

Since no single multi-time observable can differentiate between a nonequilibrium and an equilibrium process on average, what about a family of optimal observables? That is, given an $\Upsilon$, how finely grained does an instrument $\textbf{A}_{\textbf{k}}$ need to be for one to distinguish it from $\Omega$? We will now address this with an alternative result derived from Eq.~\eqref{eq:result1}.

\section{Process Equilibration Through Coarse Observables}
In contrast to Result~\ref{result:nCheby}, rather than fixing an observable, we now consider a set of allowed instruments with which a process could be probed, and ask how well can we differentiate between $\Upsilon$ and $\Omega$ in the best case scenario. Mathematically, these possible measurements are represented by a set $\mc{M}_{k}$ of multitime POVMs on at most $k$ times that can carry memory. We define the \emph{operational diamond norm distance} of two processes as 
\begin{equation}
    D_{\mc{M}_{k}}(\Upsilon,\Omega)  := \tfrac{1}{2} \max_{\textbf{A}_{\textbf{k}} \in \mc{M}_{k}} \sum_{\vec{x}} \left|\braket{\textbf{A}_{\vec{x}}}_{ \Upsilon - \Omega} \right|, \label{eq:diamond}
\end{equation}
where $\vec{x}$ denotes an outcome of the instrument $\textbf{A}_{\textbf{k}}$. In the limit of $\mc{M}_k$ encompassing all possible measurements, this reduces to the generalized diamond norm distance for quantum combs~\cite{Chiribella_2008, taranto2019memory}. This is the natural measure of the distance between processes, and corresponds operationally to the probability of mistaking them when measuring with the most distinguishing instrument available.

From our previous bound, Eq.~\eqref{eq:result1}, one can show that the average multitime operational diamond norm distance between $\Upsilon$ and $\Omega$ satisfies
    \begin{equation}
    \begin{split} \label{eq:Distinguishability}
         \overline{ D_{\mc{M}_{k}}(\Upsilon,\Omega)}^\infty \leq \tfrac{1}{2}\mc{S}_{\mc{M}_{k}} d_S^k \sqrt{\frac{2^{k}-1}{d_{\mathrm{eff}}[\rho]}},
    \end{split}
\end{equation}
where $\mc{S}_{\mc{M}_{k}}$ is the combined total number of outcomes for all measurements in the set $\mc{M}_{k}$, meaning it is proportional to the resolution of an experimental apparatus. This quantity is a multitime analogue of that defined in Ref.~\cite{ShortSystemsAndSub}, with a technical definition given in Eq.~\eqref{eq:S_def}. While $\mc{S}_{\mc{M}_{k}}$ is in general large, a typical macroscopic $d_{\mathrm{eff}}$ will nonetheless out-scale it. Given that the diamond norm is non-negative, we can now use Markov's inequality to give a result on the distinguishability of a given process from an equilibrium one, given a set of possible measurements.
\begin{result} \label{result:2}
For any quantum process $\Upsilon$, with equilibrated process $\Omega$, and a set of multitime measurements $\mc{M}_k$,
    \begin{equation}
    \begin{split} \label{eq:DistinguishabilityMarkov}
         \mathbb{P}\!\left\{
         \!D_{\mc{M}_k}\!(\Upsilon,\Omega) \! \ge \!\frac{\mc{S}_{\mc{M}_k} d_S^k \!\sqrt{2^k-1} \!}{2\, d_{\rm eff}^{{1}/{4}}[\rho] } \!\right\} \!\le\! \frac{1}{d_{\rm eff}^{{1}/{4}}[\rho]}.
    \end{split}
\end{equation}
\end{result}
A proof of Eq.~\eqref{eq:Distinguishability} and this result can be found in Appendix~\ref{ap:distinProof}.
Therefore, even with an optimal set of measurements---that are temporally nonlocal, but with a reasonable number of finite outcomes---a nonequilibrium process looks like an equilibrated process when $d_{\rm eff}$ is large. Only in the limiting case of a highly fine-grained measuring apparatus, could one feasibly distinguish between $\Upsilon$ and $\Omega$.

This latest result readily implies Eq.~\eqref{eq:equilibration}, as long as the constituent instruments of $\mathsf{O}_{\textbf{k}}$ in the decomposition Eq.~\eqref{eq:nPointGamma} are in the span of $\mc{M}_k$, i.e., that they are in principle experimentally realizable. If the total process is on a finite dimensional space, then quantum recurrences occur even for multitime observables. As such, it is also possible to generalize the above results to finite time intervals, but it is nontrivial and will be presented elsewhere~\cite{finitetime}.

\section{Equilibration, Markovianization, \& Thermalization} \label{sec:markov}
Our results on process equilibration, and hence correlation functions, bound the degree of detectable non-Markovianity in $\Upsilon$. 
One can quantify this in an operational sense~\cite{processtensor2, Costa2016, Milz2019completePos} by considering the application of two sequential instruments, $\textbf{A}_{-}$ and $\textbf{A}_{+}$ with respective outcomes $\{ x_- \}$ and $\{ x_+ \}$. From Eq.~\eqref{eq:multitime expectation}, the conditional probability for $x_+$ to occur given $x_-$ is then $\mathbb{P}_{\Upsilon|\textbf{A}_{\pm}} (x_+|x_-) = \tr[\Upsilon \textbf{A}_{x_+}^{\mathrm{T}} \otimes \textbf{A}_{x_-}^{\mathrm{T}}] / \tr[\Upsilon \, \textbf{I}_+ \otimes \textbf{A}_{x_-}^{\mathrm{T}} ]$, where $\textbf{I}_+$ is the identity on $+$ part of the process.

\begin{figure}[t]
\centering
\includegraphics[width=0.48\textwidth]{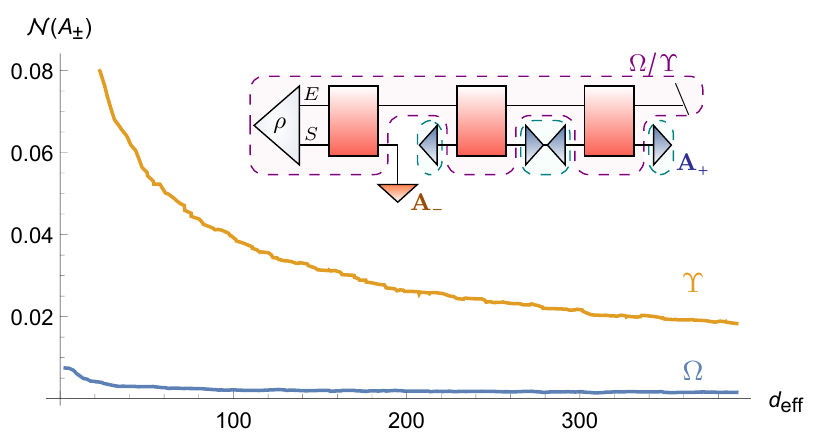}
\caption{\footnotesize{Non-Markovianity of a 3-step process $\Upsilon$ and the corresponding $\Omega$ of a single qubit coupled to a random matrix environment, with a varying effective dimension $d_{\textrm{eff}}$ (details can be found in Appendix~\ref{ap:numerics}). One can see that the non-Markovianity of a general process $\mc{N}(\Upsilon)$ trends to an equilibrium quantity ($\mc{N}(\Omega)$) with increasing $d_{\textrm{eff}}$. Note that strikingly the equilibrium non-Markovianity is also small, and decays with $d_{\textrm{eff}}$. The instruments $\textbf{A}_\pm$ used to probe these processes involved a repeated measurement and independently-prepare protocol, with a system-level causal break between $\textbf{A}_-$ and $\textbf{A}_+$, as displayed in the upper right. The red boxes in this circuit represent unitary evolution, with uniformly sampled $5 \leq \Delta t_1,\, \Delta t_2, \, \Delta t_3 \leq 50$, and dephasing, for $\Upsilon$ and $\Omega$ respectively. This was maximized over the difference in outcomes $\textbf{A}_-$, and averaged over $\textbf{A}_+$ to determine the degree of non-Markovianity of each process.}}
\label{fig:NMnumerics}
\end{figure}

The instruments are chosen to be causally separate, i.e., $\textbf{A}_-:=\textbf{A}_{j_-:1}$ ends on an output of the process and $\textbf{A}_+:=\textbf{A}_{k:j_+}$ begins on the next input. This ensures that no information is transmitted via the system from the past to the future (see the inset of Fig.~\ref{fig:NMnumerics}). Then, a process is Markovian according to this instrument if and only if the above conditional probability is independent of the outcome $x_-$: $\mathbb{P}_{\Upsilon|\textbf{A}_{\pm}}(x_+|x_-) = \mathbb{P}_{\Upsilon_+|\textbf{A}_{+}}(x_+)$. Conversely, we define the degree of non-Markovianity by $\mc{N}_\Upsilon(\textbf{A}_\pm) := \sum_{x_+} \max_{x_-,y_-} \big[ \mathbb{P}_{\Upsilon|\textbf{A}_{\pm}} (x_+| x_-) - \mathbb{P}_{\Upsilon|\textbf{A}_{\pm}}(x_+| y_-) \big]$. In other words, this quantifies the memory transmitted through the environment, which is measurable by the apparatus $\textbf{A}_\pm$. Note that there could be hidden non-Markovian effects, which would only be detectable if $\textbf{A}_\pm$ encompassed some later times and/or many instruments~\cite{Burgarth2021,Burgarth2021a}. Indeed, in such a many-time case, divisible quantum dynamics is not enough to determine quantum non-Markovianity necessarily~\cite{processtensor2,Milz2019completePos}.

This allows us to use Result~\ref{result:nCheby} to make a statement about the detectable non-Markovianity of a general process.
\\

\begin{result} \label{result:nonMark} For any quantum process $\Upsilon$, with equilibrated process $\Omega$, and causal-break instrument $\textup{\textbf{A}}_\pm$,
    \begin{equation} 
    \begin{split}\label{eq:NonMark}
        &\mathbb{P}\Bigg\{\!|\mc{N}_\Upsilon(\textup{\textbf{A}}_\pm\!)\!-\!\mc{N}_\Omega(\textup{\textbf{A}}_\pm\!)\!|\!\geq\! \frac{\eta_\mathrm{k}} {d_{\mathrm{eff}}^{{1}/{3}}[\rho]} \!\Bigg\}\! \leq\! \frac{2}{ d_{\mathrm{eff}}^{{1}/{3}}[\rho]}
    \end{split}
    \end{equation}
\end{result}
A full definition of the term $\eta_\mathrm{k}$ is given in Eq.~\eqref{eq:ckdef} in Appendix~\ref{ap:nonMarkProof}, alongside a proof of this result. The numerator of $\eta_\mathrm{k}$ goes as  $(\sqrt{2} d_S)^k$ plus $C_\mathrm{k}$, which is the likelihood ratio of outcome $w$ for instrument $\textbf{A}_-$ from which $\mc{N}$ is computed, i.e., $\underset{w,\Lambda}{\mbox{max}} \{ \mathbb{P}_{\Lambda|\textbf{A}_-}(w) / \mathbb{P}_{\bar{\Lambda}|\textbf{A}_-}(w)\}$ with either $(\Lambda,\overline{\Lambda}) =(\Upsilon,\Omega)$ or $(\Lambda,\overline{\Lambda}) =(\Omega,\Upsilon)$. The denominator of $\eta_\mathrm{k}$ is $\braket{\textbf{A}_{w}}_{\Lambda}$, which accounts for rare events, i.e., $1/\braket{\textbf{A}_{w}}_{\Lambda} \ll d_{\mathrm{eff}}^{1/3}[\rho]$.

This means that a given causal break instrument will likely show the equilibration of Markovianity, given that it has no outcome which occurs with probability $\sim \mc{O}(1)$ for one process out of $\{\Upsilon, \Omega \}$, and with probability $\sim \mc{O}(1/d_{\mathrm{eff}}^{1/3}[\rho])$ for the other. This is additionally taking the effective dimension to dominate the other terms, which occurs under similar conditions to Result~\ref{result:nCheby}. In this case the degree of non-Markovianity measurable by some $\textbf{A}_\pm$ of any $\Upsilon$ will be close to that of $\Omega$. This does not mean that $\Omega$ is necessarily Markovian. However, there is no dynamics in $\Omega$, only dephasing, and so the non-Markovianity therein must be time-independent. Therefore, on average the transient non-Markovianity of $\Upsilon$ is bounded by the effective dimension. We also expect the time-independent non-Markovianity to decay with effective system size, due to the prevalence of Markovian phenomena in macroscopic systems. Indeed, we numerically confirm this for a spin coupled to a random matrix environment; see Fig.~\ref{fig:NMnumerics}. The model, detailed in the Supplemental Material, is representative of a generic many-body evolution. Note that such a model can also be used to confirm the previous results.

These results are complementary to the recent work of Ref.~\cite{Alhambra2020-fj}, where it is shown that thermal two-time correlation functions factorize on average, assuming the system satisfies the Eigenstate Thermalization Hypothesis (ETH)~\cite{Brandao2019,Deutsch1991, Srednicki, Srednicki_1999, Rigol2007,Rigol2008, Turner2018, Deutsch_2018,Richter2019}. This directly implies that memory encoded in two point correlations washes out for a thermal state. We expect that methods from this work could be applicable to our results, in that a property such as the ETH may be necessary to arrive at entirely Markovian phenomena, in comparison to the small steady-state quantity generically observed (as seen in Fig.~\ref{fig:NMnumerics}). This would be an interesting avenue for future research.

Result~\ref{result:nonMark} can be extended to geometric measures of coherence, entanglement, etc., which confines the process in the classical domain~\cite{strasberg2019, Milz2020prx, 10.21468/SciPostPhys.10.6.141}. For related multitime equilibration bounds based on Result~\ref{result:2}, see~\cite{finitetime}. Here, we focus on Markovianization because its direct relevance to thermalization, which we discuss in our concluding remarks.

\section{Conclusions}
Our results connect equilibration with quantum stochastic processes, which encompass correlations and (quantum) memory, in addressing the problem of the quantum foundations of thermodynamics. It has been long-known that multitime non-Markovian correlations in a process vanish for a system in contact with a thermal bath in the weak system-bath coupling limit~\cite{Duemcke1983}. It was recently shown, using typicality arguments, that almost all multitime non-Markovian correlations vanish for Haar random unitary evolution~\cite{AlmostMarkov} and for unitary t-design processes~\cite{Pedro2021}. This is dubbed `process Markovianization' where a subpart of an isolated quantum system closely resembles a Markovian process even though it is non-Markovian. Here, we argued that process equilibration also implies a form of Markovianization. The strength of these results lie in their robustness against (a reasonable number of) perturbations, remaining close to an equilibrium for most times, given a sufficient coarse graining in both time (small $k$) and space (small $d_S$ and $\mc{S}_{\mc{M}}$).

This in turn leads to a better understanding of the extra ingredients needed to obtain thermalization from isolated quantum theory. Process equilibration and Markovianization are necessary conditions for a stable steady-state, and thus also thermalization. How then is the emergence of Markovianity connected to the ETH? and how does quantum chaos fit into the picture of multitime correlations? 

Regarding chaos, recently out-of-time-order correlators (OTOCs)~\cite{Kitaev} have been explored in the quantum combs framework~\cite{zonnios2021signatures,Dowling2022}. Indeed, we expect that, with careful modifications, our results may extend to time-unordered correlators; see also Ref.~\cite{Styliaris2021} for equilibration results on a certain kind of OTOC. Finally, we have extended the current results to finite-time process equilibration~\cite{finitetime}, which may provide new ways to explore equilibration-time bounds~\cite{ShortFinite, Riera2012,Malabarba2014,XXEisert}.

\begin{acknowledgments}
ND is supported by an Australian Government Research Training Program Scholarship and the Monash Graduate Excellence Scholarship. PS acknowledges financial support from a fellowship from “la Caixa” Foundation (ID 100010434,
fellowship code LCF/BQ/PR21/11840014) and from the Spanish Agencia Estatal de Investigación (project no. PID2019-107609GB-I00), the Spanish MINECO (FIS2016-80681-P, AEI/FEDER, UE), and the Generalitat de Catalunya (CIRIT 2017-SGR-1127). KM is supported through Australian Research Council Future Fellowship FT160100073, Discovery Project DP210100597, and the International Quantum U Tech Accelerator award by the US Air Force Research Laboratory.
\end{acknowledgments}
%

\newpage
\onecolumngrid
\appendix

\section{Decomposition of Correlation Functions (Eq.~\eqref{eq:GammaDef})} \label{appendix:nPoint}
We fix $t_0$ to be the time where the Schrödinger and Heisenberg pictures coincide. Then, an observable in the Heisenberg picture is defined as
\begin{equation}
    X(t_n) := U_{n,0}^\dagger X U_{n,0},
\end{equation}
where $U_{n,0} := e^{-iH(t_n-t_0)}$, and $X$ is the observable in the Schrödinger picture.
Then, for an initial state $\rho(t_0)$, 
\begin{equation}
    \begin{split}
        \braket{X_n(t_n)\dots X_1(t_1)} &= \tr[X_n(t_n)\dots X_1(t_1)\rho(t_0)] \\
        &=\tr[[U_{0,n} X_n U_{n,0}] \dots  [U_{0,2} X_2 U_{2,0}] [U_{0,1} X_1 U_{1,0}] \rho(t_0)] \\
        &=\tr[X_n U_{n,n-1} \dots  U_{3,2} X_2 U_{2,1} X_1 U_{1,0} \rho(t_0) U_{0,n}] \\
        &=\tr[X_n U_{n,n-1} \dots  U_{3,2} X_2 U_{2,1} X_1 U_{1,0} \rho(t_0) U_{0,1} U_{1,2} \dots  U_{n-1,n}] \\
        &= \tr[X_n \mc{U}_n ( \dots \mc{U}_3 (X_2 \mc{U}_2 (X_1 \mc{U}_1 (\rho(t_0))\dots)] \\
        &=\tr[\mc{X}_n \mc{U}_n \cdots \mc{U}_3 \mc{X}_2 \mc{U}_2 \mc{X}_1 \mc{U}_1 (\rho(t_0))]
    \end{split}
\end{equation}
where $\mc{U}_i(\cdot) := e^{-iH(t_i-t_{i-1})}(\cdot)e^{iH(t_i-t_{i-1})}$ is the usual unitary evolution superoperator, and $\mc{X}_i(\cdot) := X_i (\cdot) \id $ is the operator sum representation of the linear map superoperator $\mc{X}_i$ \cite{OperationalQDynamics}. Any linear map can be written as a (complex) linear combination of CP maps, so by the linearity of trace, we can write $\braket{X_n(t_n)\dots X_1(t_1)}$ as a linear combination of multitime instrument expectation values over a quantum process,
\begin{equation} \label{eq:multitime expectation1}
    \tr[\mc{A}_{k} \mc{U}_k \cdots \mc{A}_{1}\mc{U}_1\,(\rho)],
\end{equation}
where $\mc{A}_{i}$ are CP and $t_0$ is fixed. This can be measured as an expectation value of an instrument $\textbf{A}_\textbf{k}$ on a real physical process which is described by the process tensor $\Upsilon$. 

\section{The Process Tensor} \label{ap:link}
Here we give an explicit definition of quantum processes, corresponding to Fig.~\ref{fig:ProcessTensor} in the text. 
Decomposing Eq.~\eqref{eq:multitime expectation} from the text yields
\begin{gather} \label{eq:LinkProdProcessTensor}
\Upsilon_{\textbf{k}} := \tr_E [\mathtt{U}_k * \dots * \mathtt{U}_1 * \rho], \
\textbf{A}_{\textbf{k}} := \mathtt{A}_{k} \otimes \dots \otimes \mathtt{A}_{1},
\end{gather}
where $*$ is the  \textit{link product}~\cite{Chiribella2009physrevA, milz2020quantum}, corresponding to a tensor product on the $S$ space and a matrix product on $E$.
The left hand side of Eq.~\eqref{eq:multitime expectation} in the main text contains $\mathcal{A}_{i}, \ \mathcal{U}_{i}$, which are abstract maps with many representations~\cite{watrous_2018,OperationalQDynamics,milz2020quantum}; here we use the \emph{Choi} state representation for these maps, $\mathtt{U}_i$ and $\mathtt{A}_i$, resulting in the multitime Choi states $\Upsilon_\textbf{k}$ and $\textbf{A}_\textbf{k}$. With this machinery, we can define precisely what we mean by an equilibrated process, as in Eq.~\eqref{eq:LinkProdProcessTensor1}.

To see how one arrives at Eq.~\eqref{eq:LinkProdProcessTensor} from Eq.~\eqref{eq:multitime expectation} in the main body, we here present a simpler example constructing a link product for matrix multiplication - as opposed to Choi states as in Eq.~\eqref{eq:multitime expectation}. Consider the following trace equation with composite-space matrices $\mu,\nu,\pi \in \mathbb{M}_F \otimes \mathbb{M}_G $ and single-space matrices $\mathsf{x},\mathsf{y},\mathsf{z} \in \mathbb{M}_F$:
\begin{gather}
\label{eq:tensormanipulation}
\tr[\mu \, \mathsf{x} \, \nu \, \mathsf{y} \, \pi \, \mathsf{z}] = \sum_{\substack{a \dots f \\
\alpha \dots \gamma}}
\mu^{\alpha\beta}_{a b}
\mathsf{x}_{b c} \,
\nu^{\beta\gamma}_{c d}
\mathsf{y}_{d e} \,
\pi^{\gamma\alpha}_{e f}
\mathsf{z}_{f a}
\end{gather}
with the superscript (subscript) indices belonging to the spaces $G \, (F)$. Eq.~\eqref{eq:tensormanipulation} can be expressed as $\tr[\Xi \, R]$ by grouping the `Latin' and the `Greek' operators to define tensors on $\mathbb{M}_F^{\otimes 3}$: $R:= \sum_{a\dots f} \mathsf{x}_{b c} \mathsf{y}_{d e} \mathsf{z}_{f a} \ket{bdf}\!\bra{ace}$ and $\Xi := \sum_{\alpha \dots \gamma, a \dots f}\mu^{\alpha\beta}_{ab} \nu^{\beta\gamma}_{cd} \pi^{\gamma\alpha}_{ef} \ket{ace}\!\bra{bdf}$. To succinctly express the contraction of `Greek' indices we employ the \textit{link product}~\cite{Chiribella2009physrevA, milz2020quantum}, which is a matrix product and trace on the $G$ spaces and tensor product on the $F$ spaces: $\Xi = \tr_G[\mu*\nu*\pi]$. The link product employed to define the process tensor and instruments as in Eq.~\eqref{eq:LinkProdProcessTensor} can be constructed in an equivalent way, but in the Choi representation as opposed to simply matrix multiplication; for further details see~\cite{finitetime,Chiribella2009physrevA, milz2020quantum}. This has a number of advantages including that $\Upsilon$ is a positive operator, entirely analogous to a single time density operator~\cite{processtensor,processtensor2,milz2020quantum}.

\section{Proof of Bound on Average Multitime Correlation Variance (Eq.~\eqref{eq:result1})} \label{ap:MainResultProof}
Consider a partition of an isolated quantum system into a system ($S$) of interest and the rest, which we call an environment ($E$). The instruments $\mc{A}$ are then taken to act on $S$, and we suppress the notation of a tensor product with the identity on $E$: $\mc{A}\equiv \mc{A} \otimes \id_\mathrm{E}$. Consider the process $\Upsilon$, together with the unitary evolution, to encompass a pure initial state $\rho$, from which the results can be generalized to mixed states via purification \cite{ShortSystemsAndSub}. The time-independent Hamiltonian $H=\sum{E}_{n}P_{n}$ describes the evolution between instruments, where $P_{n}$ is the projector onto the energy eigenstate with energy $E_{n}$. We may in addition take these projectors to be rank one. This we can ensure, via a similar argument to \cite{ShortSystemsAndSub}, by choosing a basis for any degenerate subspace such that the $SE$ state at any particular time step may overlap only with one of the degenerate basis states $\ket{n}$ for each distinct energy. Therefore with this choice, the total $SE$ state between each instrument evolves unitarily in the subspace spanned by $\{\ket{n} \}$ as if according to a non-degenerate Hamiltonian $H^\prime  = \sum E_{n} \ket{n}\bra{n}$, where $\{\ket{n} \}$ could be different for different `steps' in the process. This argument holds for pure states at all times, and so only for pure instruments $\mc{A}_i$, i.e., the action $\mc{A}_i$ preserves the purity of the input state. Physically, non-pure instruments may only equilibrate the system more, as they increase mixing. More precisely, the object of interest, $ \overline{|\langle\textbf{A}\rangle_\Upsilon-\langle\textbf{A}\rangle_\Omega|^2}^\infty$, is convex in mixtures of pure instruments, so any bound for pure instruments directly implies a bound for mixed instruments. Therefore, without loss of generality we consider only the limiting worst case scenario of exclusively pure instruments, thus allowing degenerate energy levels. Note that we do not allow energy gap degeneracies which do not correspond to energy degeneracies, as defined above Eq.~\eqref{eq:nPointCheby}.

Recalling the definitions seen in Eqs.~\eqref{eq:multitime expectation}, \eqref{eq:LinkProdProcessTensor1} and \eqref{eq:equilExp}, the expectation values of $\textbf{A}$ over the quantum processes  $\Upsilon$ and $\Omega$ are, respectively, 
\begin{equation}
    \begin{split}
        &{\langle\textbf{A}\rangle_\Upsilon}=\tr[\mc{A}_k\mc{U}_k\mc{A}_{k-1}\mc{U}_{k-1}\cdots\mc{A}_1\mc{U}_1(\rho)],\\
        &\langle\textbf{A}\rangle_\Omega := \overline{\langle\textbf{A}\rangle_\Upsilon}^\infty=\tr[\mc{A}_k\$_k\mc{A}_{k-1}\$_{k-1}\cdots\mc{A}_1\$_1(\rho)].
    \end{split}
\end{equation}
Here $\$_i$ is the dephasing operation with respect to the non-degenerate Hamiltonian $H^\prime_i$, which describes the evolution between instruments $\mc{A}_{i-1}$ and $\mc{A}_i$. Similarly, we define the effective dimension $d_{\mathrm{eff}_i}$ as in Eq.~\eqref{eq:deff} but in terms of $\$_i$. We keep track of this as the Hamiltonian may be different between different instruments, due to the degenerate energies modifications we have made above. Indeed, the results that follow may allow for a Hamiltonian which changes arbitrarily between steps, but this has nontrivial implications for the interpretation in terms of the process tensor.

We wish to bound $\overline{|\langle\textbf{A}\rangle_\Upsilon-\langle\textbf{A}\rangle_\Omega|^2}^\infty$. Expanding the unitary $SE$ evolution in the energy eigenbases, for arbitrary $k$ instruments we have
\begin{equation}
    \begin{split} \label{eq:k=any sum}
         \langle\textbf{A}\rangle_\Upsilon-\langle\textbf{A}\rangle_\Omega&=\sum_{n_i,m_i}\tr[\mc{A}_k(P_{n_k}\dots \mc{A}_1(P_{n_1}\rho{P}_{m_1})\dots P_{m_k})]\left[\prod_{i=1}^k \ex^{-it_i(E_{n_i}-E_{m_i})}-\prod_{i=1}^k \delta_{m_in_i}\right].
    \end{split}
\end{equation}
Consider first the case with only two instruments ($k=2$), 
\begin{align} 
         \langle\textbf{A}\rangle_\Upsilon-\langle\textbf{A}\rangle_\Omega&=\sum_{n_i,m_i}\tr[\mc{A}_2(P_{n_2}\mc{A}_1(P_{n_1}\rho{P}_{m_1})P_{m_2})][\ex^{-i \Delta t_2(E_{n_2}-E_{m_2})}\ex^{-i \Delta t_1(E_{n_1}-E_{m_1})}-\delta_{m_1n_1}\delta_{m_2n_2}]\nn\\
        &=\sum_{\substack{n_1\neq{m}_1\\n_2\neq{m}_2}}\tr[\mc{A}_2(P_{n_2}\mc{A}_1(P_{n_1}\rho{P}_{m_1})P_{m_2})]\ex^{-i \Delta t_2(E_{n_2}-E_{m_2})}\ex^{-i \Delta t_1(E_{n_1}-E_{m_1})}\nn\\
        &\hspace{0.5in}+\sum_{n_1\neq{m}_1}\tr[\mc{A}_1\$_1\mc{A}_1(P_{n_1}\rho{P}_{m_1})]\ex^{-i \Delta t_1(E_{n_1}-E_{m_1})} \label{eq:k=1 sum}\\
         &\hspace{1in}+\sum_{n_2\neq{m}_2}\tr[\mc{A}_2(P_{n_2}\mc{A}_1(\omega_1)P_{m_2})]\ex^{-i \Delta t_2(E_{n_2}-E_{m_2})},\nn
\end{align}
where we have defined the multitime dephased state,
\begin{equation}
    \begin{split} \label{eq:dephasedi}
        \omega_j&= \$_j\mc{A}_{j-1}\$_{j-1}\cdots\mc{A}_1\$_1(\rho).
    \end{split}
\end{equation}
If we label the energy levels in the complex conjugate to Eq.~\eqref{eq:k=1 sum} with primed indices, $n_i^\prime$ and $m_i^\prime$, taking the modulus square will give rise to exponentials on $-i \Delta t_\ell(E_{n_\ell}-E_{n^\prime_\ell}-E_{m_\ell}+E_{m^\prime_\ell})$. Then after time-averaging over $\Delta t_i$ independently for all times, given that we have assumed that there are no energy-gap degeneracies, the integrals kill all cross terms and we arrive at a sum over modulus-squared terms without an explicit energy phase dependence. For the other terms, we have exponentials on $-i \Delta t_\ell(E_{n_\ell}-E_{m_\ell})$ and $-i \Delta t_\ell(E_{n^\prime_\ell}-E_{m^\prime_\ell})$, which go to zero with the integral over $\Delta t_i$, given the no energy-gap degeneracies condition and that $n_\ell \neq m_\ell $ and $n^\prime_\ell \neq m^\prime_\ell $. Therefore for $k=2$,
\begin{equation}
    \begin{split}\label{eq: inf avg k1}
         \overline{|\langle\textbf{A}\rangle_\Upsilon-\langle\textbf{A}\rangle_\Omega|^2}^\infty&=\sum_{{\substack{n_1\neq{m}_1\\n_2\neq{m}_2}}}\left|\tr[\mc{A}_2(P_{n_2}\mc{A}_1(P_{n_1}\rho{P}_{m_1})P_{m_2})]\right|^2+\sum_{n_2\neq{m}_2}\left|\tr[\mc{A}_2(P_{n_2}\mc{A}_1(\omega_1)P_{m_2})]\right|^2\\
    &\hspace{0.5in}+\sum_{n_1\neq{m}_1}\left|\tr[\mc{A}_2\$_2\mc{A}_1(P_{n_1}\rho{P}_{m_1})]\right|^2.
    \end{split}
\end{equation}
Every instrument $\mc{A}_i$ can be expanded in its Kraus representation, such that $\mc{A}_i(\cdot)\equiv \sum_\beta K_i^\beta (\cdot) K_i^{\beta \dg}$ and $\sum_\beta K_i^{\beta \dg} K_i^{\beta} \leq \id $ \cite{OperationalQDynamics,Wilde_2011}. Looking at the terms in Eq.~\eqref{eq: inf avg k1} with a sum over $n_i\neq m_i$ at a single $\Delta t_i$, we now prove the following identity: For $\sigma$ a general density operator and $\mc{A}$ CP, 
\begin{equation}
    \begin{split} \label{eq: single sum identity}
        \sum_{n_i\neq{m}_i}\left|\tr[\mc{A}(P_{n_i}\sigma P_{m_i})]\right|^2 &=  \sum_{n_i\neq{m}_i} \left|\sum_{\beta }\tr[ K^\beta(P_{n_i}\sigma P_{m_i})K^{\beta \dg}]\right| ^2\\
        &= \sum_{n_i\neq{m}_i} \left|\tr[\mathsf{A} P_{n_i}\sigma P_{m_i}]\right| ^2\\
        &= \sum_{n_i\neq{m}_i} \tr\left[ \mathsf{A} \sigma_{n_i m_i} \ket{n_i} \bra{m_i}\right]\left( \tr\left[ \mathsf{A} \sigma_{m_i n_i} \ket{n_i} \bra{m_i}\right]\right)^* \\
        &= \sum_{n_i\neq{m}_i} |\sigma_{n_i m_i}|^2  \bra{m_i}\mathsf{A}  \ket{n_i} \bra{n_i}  \mathsf{A}^\dg \ket{m_i} \\
        &\leq \sum_{n_i\neq{m}_i} \sigma_{n_i n_i} \sigma_{m_i m_i} \bra{m_i}\mathsf{A}  \ket{n_i} \bra{n_i}  \mathsf{A} \ket{m_i} \\
        & \leq \sum_{n_i, m_i} \tr \left[\mathsf{A} \sigma_{n_i n_i} P_{n_i} \mathsf{A} \sigma_{m_i m_i} P_{m_i} \right] \\
        &= \tr \left[\mathsf{A} \$_i(\sigma)\mathsf{A} \$_i(\sigma) \right] \\
        &\leq \sqrt{\tr \left[  \mathsf{A}\$_i(\sigma) \$_i(\sigma) \mathsf{A}  \right] \tr \left[ \$_i(\sigma) \mathsf{A} \mathsf{A}  \$_i(\sigma) \right]} \\
        &=\sqrt{\tr \left[\mathsf{A}  \mathsf{A}\left(\$_i(\sigma)\right)^2  \right] \tr \left[  \mathsf{A} \mathsf{A}  \left(\$_i(\sigma)\right)^2 \right]} \\
        &\leq \| \mathsf{A}^2 \|_\mathrm{p} \tr\left[\left(\$_i(\sigma)\right)^2 \right] =\| \mathsf{A} \|_\mathrm{p}^2 d_{\mathrm{eff}_1}^{-1}[\sigma],
        \end{split}
\end{equation}
where we have introduced the POVM element $\mathsf{A} := \sum_\beta K^{\beta \dg} {K}^{\beta} = \mathsf{A}^\dg$, the energy eigenstate decomposition $\sigma := \sum_{n_i, m_i} \sigma_{n_i m_i} \ket{n_i} \bra{m_i}$, and the effective dimension with respect to an intermediate Hamiltonian $d_{\mathrm{eff}_i}^{-1}[\cdot]:= 1/\tr[\$_i (\cdot)^2 ]$. 
In the fifth line we have used the fact that $|\sigma_{n m} | ^2 \leq \sigma_{n n} \sigma_{mm}$ for any density operator $\sigma$, which follows directly from the Cauchy-Schwarz inequality and the fact that $\sigma$ is positive hermitian (equality for pure $\sigma$). In the sixth line we have simply added the positive terms where $m_i=n_i$ to the sum. In the eighth we have again used Cauchy-Schwarz, but for the Hilbert-Schmidt inner product, $\tr[A^\dg B] \leq \|A \|_{\text{HS}} \|B \|_{\text{HS}}$ with $\|A \|_{\text{HS}} := \sqrt{\tr[A^\dg A] }$. Finally, we have used the identity $\tr(XY)\leq\|X\|_\mathrm{p} \tr(Y)$ for positive operators $X$ and $Y$, and that operator norms satisfy $\| X^\dg X\|_\mathrm{p} = \|X \|_\mathrm{p}^2 $. This identity is due to H\"older's inequality for the Hilbert Schmidt inner product $\tr(XY)$, with the 1-norm $\tr(Y)$ and infinity norm $\|X \|_\mathrm{p}$ on the right hand side~\cite{watrous_2018}. We call this operator norm the \emph{POVM norm}, so as not to mistake it with other norms used in this work, and it is explicitly defined as
\begin{equation}
\begin{split} \label{eq:norms}
    &\|\mathsf{A} \|_\mathrm{p}  := \max_{\|\psi \|_2 =1} \| \mathsf{A} \ket{\psi}\|_2 = \max_{\sqrt{\braket{\psi | \psi}} =1} \sqrt{\bra{\psi}\mathsf{A}^2 \ket{\psi}} \leq 1.
\end{split}
\end{equation}
Note that this single sum identity Eq.~\eqref{eq: single sum identity} is essentially the same as that proved in Ref.~\cite{ShortSystemsAndSub}.

We may apply this identity \eqref{eq: single sum identity} directly to the second and third terms of Eq.~\eqref{eq: inf avg k1},
\begin{equation}
    \begin{split}\label{eq:k=1 second term}
        &\sum_{n_2\neq{m}_2}\left|\tr[\mc{A}_2(P_{n_2}\mc{A}_1(\omega_1)P_{m_2})]\right|^2 +\sum_{n_1\neq{m}_1}\left|\tr[\mc{A}_2\$_2\mc{A}_1(P_{n_1}\rho{P}_{m_1})]\right|^2\\
        &\hspace{1.5in}\leq \| \mathsf{A}_1 \|_\mathrm{p}^2  d_{\mathrm{eff}_2}^{-1}[\mc{A}_1(\omega_1)]+\| \mathsf{A}_{2:1}\|_\mathrm{p}^2 d_{\mathrm{eff}_1}^{-1}[\rho]
    \end{split}
\end{equation}
with $\mc{A}_1(\omega_1) \equiv \sigma$ and $\mc{A} \equiv \mc{A}_2$ for the second term, and  $\rho \equiv \sigma$ with $\mc{A} \equiv \mc{A}_2 \$_2 \mc{A}_1$ for the third. Note that as the composition of trace-non-increasing CP maps is again trace-non-increasing CP, the map $\mc{A}_2 \$_2 \mc{A}_1$ admits a Kraus representation and thus a corresponding POVM element which we have called $\mathsf{A}_{2:1}$.

Next we need a new identity to handle the first term of \eqref{eq: inf avg k1}, 
\begin{align}
         &\sum_{{\substack{n_1\neq{m}_1\\n_2\neq{m}_2}}}\left|\tr[\mc{A}_2(P_{n_2}\mc{A}_1(P_{n_1}\rho{P}_{m_1})P_{m_2})]\right|^2 =  \sum_{{\substack{n_1\neq{m}_1\\n_2\neq{m}_2}}} \left|\sum_\alpha  \tr[\mathsf{A}_2 P_{n_2}  K_1^\alpha P_{n_1}\rho{P}_{m_1}K_1^{\alpha \dg}P_{m_2}]\right| ^2\nn\\
         &\quad= \sum_{{\substack{n_1\neq{m}_1\\n_2\neq{m}_2}}} \sum_{\alpha,\beta} |\rho_{n_1 m_1}|^2 \bra{m_2}\mathsf{A}_2 \ket{n_2}\bra{n_2}K_1^\alpha \ket{n_1} \bra{m_1}K_1^{\alpha \dg} \ket{m_2} \bra{m_2}K_1^{\beta} \ket{m_1}\bra{n_1}K_1^{\beta \dg}\ket{n_2}\bra{n_2}\mathsf{A}_2\ket{m_2} \nn\\
         &\quad\leq\sum_{{\substack{n_1,\,{m}_1\\n_2,\,{m}_2}}} \sum_{\alpha} \rho_{n_1 n_1} \rho_{m_1 m_1} |\bra{m_2}\mathsf{A}_2 \ket{n_2}|^2 \bra{n_2}K_1^{\alpha}\left(\ket{n_1} \bra{n_1}\right)K_1^{\alpha \dg} \ket{n_2} \bra{m_2}K_1^{\alpha}(\ket{m_1} \bra{m_1})K_1^{\alpha \dg} \ket{m_2} \nn\\
         &\quad\quad \quad+ \sum_{{\substack{n_1, {m}_1\\n_2, {m}_2}}} \sum_{\alpha \neq \beta} \rho_{n_1 n_1} \rho_{m_1 m_1}|\bra{m_2}\mathsf{A}_2 \ket{n_2}|^2 \bra{n_2}K_1^\alpha \ket{n_1} \bra{m_1}K_1^{\alpha \dg} \ket{m_2} \bra{m_2}K_1^{\beta} \ket{m_1}\bra{n_1}K_1^{\beta \dg}\ket{n_2}  \nn\\
         &\quad\leq \sum_\alpha \sum_{n_2,\, m_2} |\bra{m_2}\mathsf{A}_2 \ket{n_2}|^2 \bra{n_2}K_1^{\alpha}(\omega_1)K_1^{\alpha \dg} \ket{n_2} \bra{m_2}K_1^{\alpha}(\omega_1)K_1^{\alpha \dg} \ket{m_2} \nn\\
         &\quad\quad \quad+ \underset{i,j}{\max}|(\mathsf{A}_2)_{ij}|^2 \sum_{\alpha \neq \beta} \sum_{{\substack{n_1, {m}_1\\n_2, {m}_2}}}  \bra{m_2}K_1^{\beta} (\ket{m_1}\bra{m_1})K_1^{\alpha \dg} \ket{m_2}  \bra{n_2}K_1^\alpha (\ket{n_1}\bra{n_1})K_1^{\beta \dg}\ket{n_2} \label{eq:k=1 first term}\\
         &\quad= \sum_\alpha \sum_{n_2,\, m_2} (K_1^{\alpha}(\omega_1)K_1^{\alpha \dg} )_{n_2 n_2} (K_1^{\alpha}(\omega_1)K_1^{\alpha \dg} )_{m_2 m_2} \tr\left[\mathsf{A}_2 \ket{n_2}\bra{n_2} \mathsf{A}_2 \ket{m_2}\bra{m_2}\right] \nn\\
         &\quad \quad \quad+\underset{i,j}{\max}|(\mathsf{A}_2)_{ij}|^2 \sum_{\alpha \neq \beta} \tr[K_1^{\beta}K_1^{\alpha \dg}] \tr[K_1^{\alpha}K_1^{\beta \dg}] \nn\\
         &\quad= \sum_\alpha \tr\left[\mathsf{A}_2 \$_2(K_1^{\alpha}(\omega_1)K_1^{\alpha \dg} ) \mathsf{A}_2 \$_2 (K_1^{\alpha}(\omega_1)K_1^{\alpha \dg} )\right] \nn\\
         &\quad\leq \sum_\alpha \| \mathsf{A}_2\|_\mathrm{p}^2 \tr [ \left(\$_2(K_1^{\alpha}(\omega_1)K_1^{\alpha \dg} )\right)^2] \nn\\
         &\quad\leq  \| \mathsf{A}_2\|_\mathrm{p}^2 \tr [ \left(\sum_\alpha \$_2(K_1^{\alpha}(\omega_1)K_1^{\alpha \dg} )\right)^2] = \| \mathsf{A}_2\|_\mathrm{p}^2 d_{\mathrm{eff}_2}^{-1}[\mc{A}_1(\omega_1)]\nn
\end{align}
where in the third line we have used that $|\rho_{n_1 m_1}|^2 = \rho_{n_1 n_1} \rho_{m_1 m_1}$ for pure states, split the sums $\sum_{\alpha, \, \beta} = \sum_{\alpha \neq \beta} +\sum_\alpha \delta_{\alpha \beta}$, and added the (positive) extra terms $n_1=m_1$ and $n_2=m_2$ in the summations. In the third last line we have chosen an orthogonal (\emph{canonical}) Kraus representation for $\mc{A}_1$, a minimal representation such that $\tr[K_1^{\alpha \dg}K_1^{\beta}]\propto \delta^{\alpha \beta}$ \cite{Wilde_2011}. At that point we have a term of the form of the seventh line of Eq.~\eqref{eq: single sum identity} and so can use that result to arrive at the penultimate line. In the final line we bring the sum inside by the linearity of the trace, and as $\sum|x_i|^2\leq|\sum{x}_i|^2$ for positive $x_i$. Therefore for $k=2$ we have in the end,
\begin{equation}
    \begin{split} \label{eq:k=1 first bound final}
        \overline{|\langle\textbf{A}\rangle_\Upsilon-\langle\textbf{A}\rangle_\Omega|^2}^\infty&\leq\|\mathsf{A}_{2 :1} \|_\mathrm{p}^2  d_{\mathrm{eff}_1}^{-1}[\rho] + 2\| \mathsf{A}_2 \|_\mathrm{p}^2  d_{\mathrm{eff}_2}^{-1}[\mc{A}_1(\omega_1)].
    \end{split}
\end{equation}

We now generalize the result \eqref{eq:k=1 first bound final} to arbitrary $k$ instruments. Starting from Eq.~\eqref{eq:k=any sum}, just as in the $k=2$ case we multiply it by its complex conjugate and take the infinite time average over each $\{\Delta t_i \}_{i=1}^k$. We arrive at $\overline{|\langle\textbf{A}\rangle_\Upsilon-\langle\textbf{A}\rangle_\Omega|^2}^\infty$ being bounded above by $2^{k}-1$ terms of the form 
\begin{equation}
    \begin{split} \label{eq:GeneralTerm}
        \sum_{\substack{n_i\neq{m}_i\\\cdots\\ n_j \neq m_j}}\big|\tr[\mc{A}_k \$_k \mc{A}_{k-1}  \cdots \$_{j+1} \mc{A}_j(P_{n_j}C_{\{n,m\}} P_{m_j})] \big|^2 \leq \|\mathsf{A}_{k:\dots :j} \|_\mathrm{p}^2  d_{\mathrm{eff}_j}^{-1}[\mc{A}_{j-1}(\omega_{j-1})],
    \end{split}
\end{equation}
with $1\leq i < j \leq k$, such that $\|\mathsf{A}_{k:\dots :j} \|_\mathrm{p}^2$ is POVM norm of the composition of CP maps $\mc{A}_k \$_k \mc{A}_{k-1} \cdots \$_{j+1} \mc{A}_j$, and $C_{\{n,m\}}$ is a composition of CP maps, dephasing maps, and projections all acting on $\rho$, where its exact form does not matter. This inequality \eqref{eq:GeneralTerm} is found using Eq.~\eqref{eq: single sum identity} for a single sum over $n_i\neq m_i$, and the generalization of Eq.~\eqref{eq:k=1 first term} for multiple sums. In other words, the final (leftmost) projector is what determines the form of the inequality for each term. We here give the proof of Eq.~\eqref{eq:GeneralTerm} for a triple sum over $n_i \neq m_i$, which one can see will generalize to higher sums,
\begin{equation}
    \begin{split} \label{eq:triple-sum term}
         &\sum_{{\substack{n_1\neq{m}_1\\n_2\neq{m}_2 \\n_3 \neq m_3}}}\left|\tr[\mc{A}_3(P_{n_3}\mc{A}_2(P_{n_2}\mc{A}_1 (P_{n_1} \rho P_{m_1}) {P}_{m_2})P_{m_3})]\right|^2 =  \sum_{{\substack{n_1\neq{m}_1\\n_2\neq{m}_2 \\n_3 \neq m_3}}} \left|\sum_{\alpha , \alpha^\prime}   \tr[\mathsf{A}_3 P_{n_3}  K_2^\alpha P_{n_2}K_1^{\alpha^\prime} P_{n_1} \rho P_{m_1} K_1^{\alpha^\prime \dg} {P}_{m_2}K_2^{\alpha \dg}P_{m_3}]\right| ^2\\
         &\quad= \sum_{{\substack{n_1,{m}_1\\n_2,{m}_2 \\n_3 , m_3}}} \sum_{{\substack{\alpha,\beta \\ \alpha^\prime, \beta^\prime}}} |\rho_{n_1 m_1}|^2 |\bra{m_3}\mathsf{A}_3 \ket{n_3}|^2 \bra{n_3}K_2^{\alpha^\prime} \ket{n_2} \bra{n_2}K_1^\alpha \ket{n_1}  \bra{m_1}K_1^{\alpha \dg} \ket{m_2} \bra{m_2}K_2^{\alpha^\prime \dg} \ket{m_3} \\
         &\hspace{2.3in}  \times \bra{m_3}K_2^{\beta^\prime} \ket{m_2} \bra{m_2}K_1^{\beta} \ket{m_1}\bra{n_1}K_1^{\beta \dg}\ket{n_2} \bra{n_2}K_2^{\beta^\prime \dg}\ket{n_3}  \\
         &\quad\leq \sum_{{\substack{n_2,{m}_2 \\n_3, m_3}}} \sum_{\alpha,\alpha^\prime} |\bra{m_3}\mathsf{A}_3 \ket{n_3}|^2 \bra{n_3}K_2^{\alpha^\prime} \ket{n_2} \bra{n_2}K_1^{\alpha}(\omega_1)K_1^{\alpha \dg} \ket{n_2} \bra{n_2}K_2^{\alpha^\prime \dg} \ket{n_3}   \\
         &\hspace{1.7in}  \times   \bra{m_3}K_2^{\alpha^\prime} \ket{m_2} \bra{m_2}K_1^{\alpha}(\omega_1)K_1^{\alpha \dg} \ket{m_2} \bra{m_2}K_2^{\alpha^\prime \dg} \ket{m_3} \\
         &\quad\quad \quad+ \sum_{{\substack{n_2, {m}_2\\n_3, {m}_3}}} \sum_{{\substack{\alpha \neq \beta \\ \alpha^\prime \neq \beta^\prime}}} \bra{n_3}K_2^{\alpha^\prime} \ket{n_2} \bra{n_2}K_1^{\alpha}K_1^{\beta \dg} \ket{n_2} \bra{n_2}K_2^{\beta^\prime \dg} \ket{n_3}      \bra{m_3}K_2^{\beta^\prime} \ket{m_2} \bra{m_2}K_1^{\beta}K_1^{\alpha \dg} \ket{m_2} \bra{m_2}K_2^{\alpha^\prime \dg} \ket{m_3} \\
         &\quad\leq \sum_{n_3 ,m_3} \sum_{\alpha,\alpha^\prime}  \left(K_2^{\alpha^\prime} \$_2(K_1^{\alpha}(\omega_1)K_1^{\alpha \dg})K_2^{\alpha^\prime \dg} \right)_{n_3 n_3} \left(K_2^{\alpha^\prime} \$_2(K_1^{\alpha}(\omega_1)K_1^{\alpha \dg})K_2^{\alpha^\prime \dg} \right)_{m_3 m_3} \tr\left[\mathsf{A}_3 \ket{n_3}\bra{n_3} \mathsf{A}_3 \ket{m_3}\bra{m_3}\right]
         \\
         &\quad\quad \quad+\sum_{n_2,m_2}\sum_{{\substack{\alpha \neq \beta \\ \alpha^\prime \neq \beta^\prime}}} \tr[R^{\alpha^\prime n_2 \alpha} R^{\beta^\prime n_2 \beta \dg}]     \tr[R^{\beta^\prime m_2 \beta}R^{\alpha^\prime m_2 \alpha \dg} ]\\
         & \quad\leq\sum_{\alpha,\alpha^\prime}    \tr\left[\mathsf{A}_3 \$_3\left(K_2^{\alpha^\prime} \$_2(K_1^{\alpha}(\omega_1)K_1^{\alpha \dg})K_2^{\alpha^\prime \dg} \right) \mathsf{A}_3 \$_3 \left(K_2^{\alpha^\prime} \$_2(K_1^{\alpha}(\omega_1)K_1^{\alpha \dg})K_2^{\alpha^\prime \dg} \right)\right] \\ 
         &\quad\quad \quad+\sum_{{\substack{n_2,n_2^\prime \\ m_2,m_2^\prime }}}\sum_{{\substack{\alpha \neq \beta \\ \alpha^\prime \neq \beta^\prime}}} |\tr[R^{\alpha^\prime n_2 \alpha} R^{\beta^\prime n^\prime_2 \beta \dg}]     \tr[R^{\beta^\prime m_2 \beta}R^{\alpha^\prime m^\prime_2 \alpha \dg} ]| \\
         &\quad\leq  \| \mathsf{A}_3\|_\mathrm{p}^2 \tr [ \left(\sum_{\alpha, \alpha^\prime} \$_3\left(K_2^{\alpha^\prime} \$_2(K_1^{\alpha}(\omega_1)K_1^{\alpha \dg})K_2^{\alpha^\prime \dg} \right)\right)^2] \\
         &\quad\quad \quad+\sum_{{\substack{n_2,n_2^\prime \\ m_2,m_2^\prime }}}\sum_{{\substack{\alpha \neq \beta \\ \alpha^\prime \neq \beta^\prime}}} |(\delta_{\alpha^\prime \beta^\prime} \delta_{\alpha \beta } \delta_{n_2 n_2^\prime})     (\delta_{\alpha^\prime \beta^\prime} \delta_{\alpha \beta } \delta_{m_2 m_2^\prime} )| = \| \mathsf{A}_3\|_\mathrm{p}^2 d_{\mathrm{eff}_2}^{-1}[\mc{A}_2(\omega_2)].
     \end{split}
\end{equation}
Here, in the second line we have included the positive diagonal terms $n_i=m_i$, and from the third line onward we have applied the reasoning behind the first steps of Eq.~\eqref{eq:k=1 first term}, to eliminate the $\alpha \neq \beta$ cross terms. This we have done by choosing canonical Kraus representations for the combined Kraus operator $R^{\alpha^\prime n_2 \alpha} := K_2^{\alpha^\prime} \ket{n_2}\bra{n_2} K_1^{\alpha}$. Note that this is indeed a Kraus operator as $\sum_{n_2,\alpha,\alpha^\prime} R^{\alpha^\prime n_2 \alpha} R^{\alpha^\prime n_2 \alpha \dg} = \sum_{n_2, \alpha,\alpha^\prime}K_2^{\alpha^\prime \dg} \ket{n_2}\bra{n_2} K_1^{\alpha \dg} K_1^{\alpha} \ket{n_2}\bra{n_2} K_2^{\alpha^\prime}  = \id$. In the second last inequality we have used that $\sum_i x_i \leq \sum_i |x_i|$, and included extra positive terms when $n_2 \,(m_2) \neq n^\prime_2 \,(m^\prime_2) $. The rest of the proof follows essentially the same steps as Eq.~\eqref{eq:k=1 first term}. Note that this argumentation will generalize to larger sums.

Now, we can count all the terms in the expansion of Eq.~\eqref{eq:k=any sum}, for arbitrary $k$ instruments by noting that the leftmost projector is what determines the bound on the term. There are exactly $2^k-1$ terms, with for instance exactly one term in the expansion that has no projectors (only dephasing); two terms with the leftmost projector at the first position (acting directly on $\mc{A}_1$) with either dephasing or another projector at the zero position (acting on $\rho)$; $ \binom{2}{0} + \binom{2}{1} + \binom{2}{2} = 4$ terms with the leftmost projector being at the second position (with a choice of dephasing or projector for positions zero and one); and so on. We arrive at 
\begin{equation}
    \begin{split} \label{eq:finalbound}
         \overline{|\langle\textbf{A}\rangle_\Upsilon-\langle\textbf{A}\rangle_\Omega|^2}^\infty&\leq  \|\mathsf{A}_{k:\dots :1} \|_\mathrm{p}^2 d_{\mathrm{eff}_1}^{-1}[\rho]+ 
         (1+1)\|\mathsf{A}_{k:\dots :2} \|_\mathrm{p}^2  d_{\mathrm{eff}_2}^{-1}[\mc{A}_1(\omega_1)] +
         \dots \\
         &\quad+(1+(k-2)+\frac{(k-2)(k-3)}{2}+\dots) \|\mathsf{A}_{k:(k-1)} \|_\mathrm{p}^2 d_{\mathrm{eff}_{k-1}}^{-1}[\mc{A}_{k-2}(\omega_{k-2})]\\
         &\quad +(1+(k-1)+\frac{(k-1)(k-2)}{2}+\dots)\| \mathsf{A}_k \|_\mathrm{p}^2 d_{\mathrm{eff}_k}^{-1}[\mc{A}_{k-1}(\omega_{k-1})] \\
         &=\sum_{j=0}^{k-1} 2^j \|\mathsf{A}_{k: \dots  :(j+1)} \|_\mathrm{p}^2 d_{\mathrm{eff}_{(j+1)}}^{-1}[\mc{A}_{j}(\omega_{j})] \\
         &\leq  \sum_{j=0}^{k-1} 2^j \|\mathsf{A}_{k: \dots  :(j+1)} \|_\mathrm{p}^2 \| \mc{A}_{j}\|_\mathrm{s}^2 \dots  \| \mc{A}_{1}\|_\mathrm{s}^2 \Big) d_{\mathrm{eff}_1}^{-1}[\rho] \leq \sum_{j=0}^{k-1} 2^j \|\mathsf{A}_{k: \dots  :(j+1)} \|_\mathrm{p}^2 d_S^{2j}  d_{\mathrm{eff}_1}^{-1}[\rho]  \\
         &\leq (2^k-1) d_S^{2k}  d_{\mathrm{eff}}^{-1}[\rho] 
    \end{split}
\end{equation}
where for convenience we have defined $\mc{A}_{0}(\omega_{0}):=\rho$ and $d_{\mathrm{eff}_1} = d_{\mathrm{eff}}$, and summed the partial geometric series of coefficients. In the final line we have used that an instrument $\mc{A}_j$ may change the inverse effective dimension by at most the system dimension $d_S$, 
\begin{equation} \label{eq:deff_inequality k=any}
    d_{\mathrm{eff}_j}^{-1}[\mc{A}_{j-1}(\omega_{j-1})] \leq \|\mc{A}_{j-1} \|_\mathrm{s}^2 \dots \|\mc{A}_{1} \|_\mathrm{s}^2 d_{\mathrm{eff}}^{-1}[\rho] \leq d_S^{2j} d_{\mathrm{eff}}^{-1}[\rho].
\end{equation}
Here we have used that dephasing may not increase the purity of a state, and defined the operator norm induced by the Schatten 2-norm on density matrices~\cite{Watrous2004,Wilde_2011},
\begin{equation}
    \|\mc{A} \|_\mathrm{s} := \underset{\|\sigma \|_2 =1}{\max} \|\mc{A}(\sigma) \|_2 = \max_{\sqrt{\tr[\sigma^2] } =1}\sqrt{\tr[\mc{A}(\sigma)^2]} \leq d_S.
\end{equation}
This is equal to the largest singular value of $\sum_\alpha K_\alpha  \otimes K_\alpha^*$, where $K_\alpha$ are Kraus operators. We have also used that the instruments act only on the system, $\|\mc{A} \|_\mathrm{s} \equiv \|\mc{A}_S \otimes \mc{I}_E \|_\mathrm{s} = \|\mc{A}_S \|_\mathrm{s}$, and that the largest singular value of a Choi matrix of a CP map $\mc{A} \leq 1$ is $d_S$.

Using the third last line of Eq.~\eqref{eq:finalbound}, we arrive at an alternative (tighter) bound in comparison to Eq.~\eqref{eq:nPointCheby},
\begin{equation}
    \label{eq:result1_alt} \overline{|\langle\textbf{A}_{\textbf{k}}\rangle_\Upsilon-\langle \textbf{A}_{\textbf{k}} \rangle_\Omega|^2}^\infty  \leq  (2^{k}-1) \underset{j\in[0,k-1]}{\mathrm{max}} \frac{\|\mathsf{A}_{k: \dots  :(j+1)} \|_\mathrm{p}^2 } {d_{\mathrm{eff}_{(j+1)}}[\mc{A}_{j}(\omega_{j})]}\, .
\end{equation}
Substituting this bound into the results allows one to relax the restriction of coarse instruments: $d_S \ll d_E$. Instead, the requirement for equilibration is that given an instrument $\textbf{A}_{\textbf{k}}$, \emph{no effective dimension at any stage of the equilibriated process $\Omega$ is small.} In other words, this means that if the process $\Upsilon$ begins in a quantum state equilibration circumstance (large $d_{\mathrm{eff}}[\rho]$), then the \emph{process} $\Upsilon$ will also equilibrate as long as no instrument $\mc{A}_i$ significantly reduces the effective dimension. An example of an instrument that does this is:
\begin{equation}
    \mc{A}_0(\cdot) = \sum_\alpha \ket{0}\bra{\alpha} \cdot \ket{\alpha}\bra{0},
\end{equation}
which maps all input states (with arbitrary $d_{\mathrm{eff}}[\rho]$) to the pure $\ket{0}\bra{0}$ energy eigenstate (with a minimum $d_{\mathrm{eff}}[\ket{0}\bra{0}] =1$).

Note that for a sequence of observables, the expectation value over the equilibrium process reduces to $\braket{\mathsf{O}_{\textbf{k}}}_{\Omega} = \sum_n (X_k )_{nn} \dots (X_1 )_{nn} \rho_{nn}$, where $M_{nn}$ denotes the $n^{\mathrm{th}}$ diagonal matrix element of $M$ in the energy eigenbasis. For this particular choice of left-operators $\mc{A}(\cdot ) \equiv X (\cdot) \id$, and in addition considering only $2-$point functions, assuming a form of the Eigenstate Thermalization Hypothesis, and taking $\rho$ to be thermal, our result Eq.~\eqref{eq:finalbound} reduces to a similar one as found in Ref.~\cite{Alhambra2020-fj}. Such a situation does not however correspond to a single sequence of physical measurements, and does not incorporate all the structure of multitime temporal correlation functions like the minimal setup using the process tensor framework does.

Finally, we may also extend these results and consider measurements using correlated instruments $\{ \mc{A}_i\}$, which are represented by appending an ancilla space $W$ to $SE$ \cite{Taranto2019FiniteMarkov}. Instruments are then taken to act on the experimentally accessible space of $SW$, in which case they are often called testers \cite{Chiribella2009physrevA}. Projective measurements are then performed on the ancilla space at some stage of the process, in order to determine the correlations. Considering our previous results, this means for an ancilla initial state $\gamma_{_W}$, we take $\rho \to \rho_{_{SE}} \otimes \gamma_{_W}$, and $\mc{A}(\cdot) \to \mc{A}_{SW}\otimes \mc{I}_E (\cdot)$, where we take the ancilla to be a perfectly controlled space such that only $SE$ evolves unitarily: $P_n \to (P_{n})_{SE} \otimes \id_W$. It is straightforward to see that all previous results hold in this case.

\section{Proof of Result~\ref{result:nCheby}} \label{ap:chebyProof}
Chebyshev's inequality states that for any random variable $x$ with mean $\mu$ and variance $\sigma^2$, and for any $\delta>0$,
\begin{equation}
    \mathbb{P}[|x-\mu|\geq\delta\sigma]\leq\delta^{-2}.
\end{equation} 
We use that Eq.~\eqref{eq:nPointCheby} is a bound on the variance, and choose $\delta=\sigma^{-1} \sqrt{2^{k}-1}d_S^k  \left(d_{\mathrm{eff}_1}^{-1}[\rho]\right)^{-1/3} \geq \left(d_{\mathrm{eff}_1}^{-1}[\rho]\right)^{1/6}$ to arrive at
\begin{equation} 
    \begin{split}\label{eq:Cheby}
        &\mathbb{P}\left\{|\braket{\textbf{A}_\textbf{k}}_{\Upsilon-\Omega}|\geq \frac{\sqrt{2^{k}-1} d_S^k}{  d_{\mathrm{eff}}^{1/3}[\rho]} \right\} \leq \left(\sigma^{-1} \sqrt{2^{k}-1}d_S^k  \left(d_{\mathrm{eff}_1}^{-1}[\rho]\right)^{-1/3} \right)^{-2} \leq  \frac{1}{d_{\mathrm{eff}}^{1/3}[\rho]}.
    \end{split}
\end{equation}
Recalling the decomposition Eq.~\eqref{eq:GammaDef}, one can see that each $\textbf{A}_\textbf{k}^{(i)}$ in this expansion satisfies the equilibration result of Eq.~\eqref{eq:Cheby}. Applying H\"older's inequality for $|\vec{\alpha}\cdot \vec{\textbf{A}}_\textbf{k}  | \leq \| \vec{\alpha}\|_1 \|\vec{\textbf{A}}_\textbf{k} \|_{\infty}$, where $\vec{\textbf{A}}_\textbf{k}:= \big( \langle \textbf{A}_\textbf{k}^{(1)} \rangle_{\Upsilon-\Omega},\langle \textbf{A}_\textbf{k}^{(2)} \rangle_{\Upsilon-\Omega},\dots \big)$ and $\vec{\alpha}:= (\alpha_1,\alpha_2,\dots)$, we arrive at the equilibration of correlation functions Result~\ref{result:nCheby},
\begin{equation}
    \begin{split}
        \mathbb{P}\left\{|\braket{\mathsf{O}_{\textbf{k}}}_{\Upsilon-\Omega}|\geq \frac{{\sum_i | \alpha_i| } \sqrt{2^{k}-1}d_S^k}{  d_{\mathrm{eff}}^{1/3}[\rho]} \right\} &=\mathbb{P}\left\{|\vec{\alpha}\cdot \vec{\textbf{A}}_\textbf{k}|\geq \frac{{\sum_i | \alpha_i| } \sqrt{2^{k}-1}d_S^k}{  d_{\mathrm{eff}}^{1/3}[\rho]} \right\} \\
        &\leq \mathbb{P}\left\{\| \vec{\alpha}\|_1 \|\vec{\textbf{A}}_\textbf{k} \|_{\infty} \geq \frac{{\sum_i | \alpha_i| } \sqrt{2^{k}-1}d_S^k}{  d_{\mathrm{eff}}^{1/3}[\rho]} \right\}\\
        &=\mathbb{P}\left\{ \|\vec{\textbf{A}}_\textbf{k} \|_{\infty} \geq \frac{\sqrt{2^{k}-1}d_S^k}{  d_{\mathrm{eff}_1}^{1/3}[\rho]} \right\} \leq d_{\mathrm{eff}}^{-1/3}[\rho].
    \end{split}
\end{equation}

\section{Proof of Result~\ref{result:2}} \label{ap:distinProof}
We first provide a proof for Eq.~\eqref{eq:Distinguishability}, from which Result~\ref{result:2} follows directly. Note that this proof follows closely to one given in Ref.~\cite{ShortSystemsAndSub}. 

Consider that each $\textbf{A}_\textbf{w} \in \mc{M}_k$ has outcomes $\{ \vec{x} \} = \{(x_1,x_2,\dots,x_w)\}$ corresponding to the instrument $\textbf{A}_{\vec{x}}$, where $w\leq k$. We can then bound the time averaged diamond norm, as defined in Eq.~\eqref{eq:diamond},
\begin{equation}
    \begin{split} \label{eq:DistinguishabilityProof}
         \overline{D_{\mc{M}_k}(\Upsilon,\Omega)}^\infty &:=
         \overline{\max_{\textbf{A}_\textbf{w} \in \mc{M}_k} (1/2) \sum_{\vec{x}} \left|\tr[\textbf{A}_{\vec{x}} \left( \Upsilon_\textbf{w} - \Omega_\textbf{w} \right)] \right| }^\infty\\
         & \leq \frac{1}{2} \sum_{\textbf{A}_\textbf{w} \in \mc{M}_k}  \sum_{x_i}  \overline{ | \langle\textbf{A}_{\vec{x}} \rangle_\Upsilon -\langle\textbf{A}_{\vec{x}} \rangle_\Omega | }^\infty \\ 
         &\leq \frac{1}{2} \sum_{\textbf{A}_\textbf{w} \in \mc{M}_k} \sum_{\vec{x} } \sqrt{ \overline{|\langle\textbf{A}_{\vec{x}}\rangle_\Upsilon -\langle\textbf{A}_{\vec{x}} \rangle_\Omega |^2}^\infty}\\
          &\leq \frac{1}{2} \sum_{\textbf{A}_\textbf{w}\in \mc{M}_k} \sum_{\vec{x} }\sqrt{(2^{w}-1)d_S^{2w}  d_{\mathrm{eff}}^{-1}[\rho]} \\ 
         &\leq \frac{1}{2} \sum_{\textbf{A}_\textbf{w}\in \mc{M}_k} \sum_{\vec{x} } \sqrt{(2^{k}-1)d_S^{2k}  d_{\mathrm{eff}}^{-1}[\rho]}\\
         &= \frac{\mc{S}_{\mc{M}_{k}}d_S^k\sqrt{(2^{k}-1)}}{2\sqrt{d_{\mathrm{eff}}[\rho]}},
    \end{split}
\end{equation}
where in the second line we have used that $\underset{i}{\mathrm{max}} | a_i| \leq \sum_i |a_i| $, in the fourth we have used Eq.~\eqref{eq:result1}, and in the penultimate line that $w\leq k$. We have defined the total number of outcomes for all instruments in the set of (up to) $k-$time measurements $\mc{M}_{k}$
\begin{equation} \label{eq:S_def}
    \mc{S}_{\mc{M}_{k}} = \sum_{\textbf{A}_\textbf{w}\in \mc{M}_k} \mathrm{card}(\textbf{A}_\textbf{w})
\end{equation}
where $ \mathrm{card}( M)$ is the cardinality of the instrument $M$ (also called the Kraus rank), i.e. the number of Kraus operators in the minimal (canonical) Kraus representation. 

Markov's inequality states that for any non-negative random variable $X$ with mean $\mu$, and for any $\delta > 0 $, $\mathbb{P}(X\geq \delta ) \leq  \mu / \delta $. Using the bound on the mean Eq.~\eqref{eq:DistinguishabilityProof}, we arrive at Result~\ref{result:2} through the choice of $\delta = \mu \, d_{\mathrm{eff}}^{1/4}[\rho]$.

\section{Proof of Result~\ref{result:nonMark}} \label{ap:nonMarkProof}
Consider the instruments $\textbf{A}_\pm$ and the definition of non-Markovianity as described above Result~\ref{result:nonMark}. Define $\braket{\textbf{A}_{x_-}}_{\Upsilon} := \braket{\textbf{I}_+ \otimes \textbf{A}_{x_-} }_{\Upsilon}$. Consistent with the definition of $\mc{N}$, take the non-Markovianity of $\Upsilon$ and $\Omega$ to be, respectively, $\mc{N}_\Upsilon(\textbf{A}_\pm) = \braket{ \textbf{A}_+ \otimes \textbf{A}_{x_-}}_\Upsilon/\braket{\textbf{A}_{x_-}}_{\Upsilon} - \braket{\textbf{A}_+ \otimes \textbf{A}_{y_-} }_\Upsilon/\braket{\textbf{A}_{y_-}}_{\Upsilon}$ and $\mc{N}_\Omega(\textbf{A}_\pm) = \braket{\textbf{A}_+ \otimes \textbf{A}_{x^\prime_-}}_\Omega/\braket{\textbf{A}_{x^\prime_-}}_{\Omega} - \braket{\textbf{A}_+ \otimes \textbf{A}_{y^\prime_-}}_\Omega/\braket{\textbf{A}_{y^\prime_-}}_{\Omega}$. Then, 
\begin{align}
        |\mc{N}_\Upsilon(\textbf{A}_\pm)-&\mc{N}_\Omega(\textbf{A}_\pm)| = \mc{N}_\Upsilon(\textbf{A}_\pm)-\mc{N}_\Omega(\textbf{A}_\pm)\nn\\
        &= \frac{\braket{ \textbf{A}_+ \otimes \textbf{A}_{x_-} }_\Upsilon}{\braket{\textbf{A}_{x_-}}_{\Upsilon}} - \frac{\braket{\textbf{A}_+ \otimes \textbf{A}_{y_-} }_\Upsilon}{\braket{\textbf{A}_{y_-}}_{\Upsilon}} -\frac{\braket{\textbf{A}_+\otimes\textbf{A}_{x^\prime_-}}_\Omega}{\braket{\textbf{A}_{x^\prime_-}}_{\Omega}} + \frac{\braket{\textbf{A}_+ \otimes\textbf{A}_{y^\prime_-} }_\Omega}{\braket{\textbf{A}_{y^\prime_-}}_{\Omega}}\nn\\
        &\leq \frac{\braket{ \textbf{A}_+ \otimes \textbf{A}_{x_-} }_\Upsilon}{\braket{\textbf{A}_{x_-}}_{\Upsilon}} - \frac{\braket{\textbf{A}_+ \otimes \textbf{A}_{y_-} }_\Upsilon}{\braket{\textbf{A}_{y_-}}_{\Upsilon}} -\frac{\braket{\textbf{A}_+\otimes\textbf{A}_{x_-}}_\Omega}{\braket{\textbf{A}_{x_-}}_{\Omega}} + \frac{\braket{\textbf{A}_+ \otimes\textbf{A}_{y_-} }_\Omega}{\braket{\textbf{A}_{y_-}}_{\Omega}}\nn\\
        &= \frac{\braket{ \textbf{A}_+ \otimes \textbf{A}_{x_-} }_\Upsilon}{\braket{\textbf{A}_{x_-}}_{\Omega} +\braket{\textbf{A}_{x_-}}_{\Upsilon-\Omega}} - \frac{\braket{\textbf{A}_+ \otimes \textbf{A}_{y_-} }_\Upsilon}{\braket{\textbf{A}_{y_-}}_{\Omega} + \braket{\textbf{A}_{y_-}}_{\Upsilon-\Omega}} -\frac{\braket{\textbf{A}_+\otimes\textbf{A}_{x_-}}_\Omega}{\braket{\textbf{A}_{x_-}}_{\Omega}} + \frac{\braket{\textbf{A}_+ \otimes\textbf{A}_{y_-} }_\Omega}{\braket{\textbf{A}_{y_-}}_{\Omega}}\nn\\ 
        &= \left| \frac{\braket{ \textbf{A}_+ \otimes \textbf{A}_{x_-} }_\Upsilon}{\left(\frac{\braket{\textbf{A}_{x_-}}_{\Upsilon-\Omega}}{\braket{\textbf{A}_{x_-}}_\Omega} + 1 \right)\braket{\textbf{A}_{x_-}}_\Omega} - \frac{\braket{\textbf{A}_+ \otimes \textbf{A}_{y_-} }_\Upsilon}{\left(\frac{\braket{\textbf{A}_{y_-}}_{\Upsilon-\Omega}}{\braket{\textbf{A}_{y_-}}_\Omega} + 1 \right)\braket{\textbf{A}_{y_-}}_\Omega} -\frac{\braket{\textbf{A}_+\otimes\textbf{A}_{x_-}}_\Omega}{\braket{\textbf{A}_{x_-}}_{\Omega}} + \frac{\braket{\textbf{A}_+ \otimes\textbf{A}_{y_-} }_\Omega}{\braket{\textbf{A}_{y_-}}_{\Omega}} \right| \nn\\ 
        &= \left| \frac{\braket{\textbf{A}_+ \otimes \textbf{A}_{x_-} }_{\Upsilon-\Omega}}{\braket{\textbf{A}_{x_-}}_{\Omega}}  - \frac{\braket{\textbf{A}_+ \otimes \textbf{A}_{y_-} }_{\Upsilon-\Omega}}{\braket{\textbf{A}_{y_-}}_{\Omega}}  \right.\nn\\
        &\quad\left. +\sum_{i=1}^\infty \left(-\frac{\braket{\textbf{A}_{x_-}}_{\Upsilon-\Omega}}{\braket{\textbf{A}_{x_-}}_\Omega} \right)^i \frac{\braket{\textbf{A}_+ \otimes \textbf{A}_{x_-} }_{\Upsilon}}{\braket{\textbf{A}_{x_-}}_{\Omega}}-\sum_{j=1}^\infty \left(-\frac{\braket{\textbf{A}_{y_-}}_{\Upsilon-\Omega}}{\braket{\textbf{A}_{y_-}}_\Omega} \right)^j \frac{\braket{\textbf{A}_+ \otimes \textbf{A}_{y_-} }_{\Upsilon}}{\braket{\textbf{A}_{y_-}}_{\Omega}} \right|\nn\\
        &\leq 2\, \mathrm{max} \left(\left |\frac{\braket{\textbf{A}_+ \otimes \textbf{A}_{x_-} }_{\Upsilon-\Omega}}{\braket{\textbf{A}_{x_-}}_{\Omega}}\right|,\left|\frac{\braket{\textbf{A}_+ \otimes \textbf{A}_{y_-} }_{\Upsilon-\Omega}}{\braket{\textbf{A}_{y_-}}_{\Omega}}\right| \right) \label{eq:NonMarkProofPrelim}\\
        &\quad+ 2\, \mathrm{max} \left(\left|\frac{\braket{\textbf{A}_+ \otimes \textbf{A}_{x_-} }_{\Upsilon}}{\braket{\textbf{A}_{x_-}}_{\Omega}} \sum_{i=1}^\infty \left(\frac{\braket{\textbf{A}_{x_-}}_{\Upsilon-\Omega}}{\braket{\textbf{A}_{x_-}}_\Omega} \right)^i\right|,\left|\frac{\braket{\textbf{A}_+ \otimes \textbf{A}_{y_-} }_{\Upsilon}}{\braket{\textbf{A}_{y_-}}_{\Omega}}\sum_{j=1}^\infty \left(\frac{\braket{\textbf{A}_{y_-}}_{\Upsilon-\Omega}}{\braket{\textbf{A}_{y_-}}_\Omega} \right)^j \right| \right)\nn\\
        &\leq 2 \underset{w\in \{x^{(\prime)}_-,y^{(\prime)}_-\}}{\mathrm{max}} \left( \left|\frac{\braket{\textbf{A}_+ \otimes \textbf{A}_{w} }_{\Upsilon-\Omega}}{\braket{\textbf{A}_{w}}_{\Omega}}\right| \right)+ 2 \underset{w\in \{x^{(\prime)}_-,y^{(\prime)}_-\}}{\mathrm{max}} \left( \left|\frac{\braket{\textbf{A}_+ \otimes \textbf{A}_{w} }_{\Upsilon}}{\braket{\textbf{A}_{w}}_{\Omega}} \sum_{i=1}^\infty \left(\frac{\braket{\textbf{A}_{w}}_{\Upsilon-\Omega}}{\braket{\textbf{A}_{w}}_\Omega} \right)^i \right| \right)\nn\\
        &\leq 2 \underset{\substack{w\in \{x^{(\prime)}_-,y^{(\prime)}_-\}\\\Lambda \in \{\Upsilon,\Omega \}}}{\mathrm{max}} \left( \left| \frac{\braket{\textbf{A}_+ \otimes \textbf{A}_{w} }_{\Upsilon-\Omega}}{\braket{\textbf{A}_{w}}_{\Lambda}}\right| \right) + 2  \underset{\substack{w\in \{x^{(\prime)}_-,y^{(\prime)}_-\}\\\Lambda \in \{\Upsilon,\Omega \}}}{\mathrm{max}} \left( \left|\frac{\braket{\textbf{A}_+ \otimes \textbf{A}_{w} }_{\bar{\Lambda}}}{\braket{\textbf{A}_{w}}_{\Lambda}} \sum_{i=1}^\infty \left(\frac{\braket{\textbf{A}_{w}}_{\Upsilon-\Omega}}{\braket{\textbf{A}_{w}}_\Lambda} \right)^i\right| \right),\nn 
\end{align}
where in the first line without loss of generality we have assumed that $\mc{N}_\Upsilon(\textbf{A}_\pm) \geq \mc{N}_\Omega(\textbf{A}_\pm)$ -- otherwise all $x/y_-$ are replaced with $x^\prime/y^\prime_-$, which is accounted for by the maximum in the penultimate line. In the third line we used that $\mc{N}_\Omega(\textbf{A}_\pm)$ is defined as a maximum over $x^\prime$ and $y^\prime$, and so any other choice $x$ and $y$ will give a smaller number. In the fifth line we have reintroduced the absolute value (as in the first line we have assumed the entire right hand side is positive), and factored the denominators of the first two terms. This allows us in the following line to expand the geometric series with respect to the term $r=\braket{\textbf{A}_{x/y_-}}_{\Upsilon-\Omega}/\braket{\textbf{A}_{x/y_-}}_\Omega $. This requires the assumption that, without loss of generality, $|r|=|\braket{\textbf{A}_{x/y_-}}_{\Upsilon-\Omega}/\braket{\textbf{A}_{x/y_-}}_\Omega | = |\braket{\textbf{A}_{x/y_-}}_\Upsilon/\braket{\textbf{A}_{x/y_-}}_\Omega - 1| < 1$. In the alternative case ($r>1$ for this choice of $r$) there will be a geometric expansion for the third and fourth terms instead (where $r^\prime=\braket{\textbf{A}_{x/y_-}}_{\Omega-\Upsilon}/\braket{\textbf{A}_{x/y_-}}_\Upsilon $ and so $|r^\prime| <1 $ is then ensured), and this is accounted for by the maximum in the last line (if $\Lambda = \Upsilon \, (\Omega)$, then its complement is $\bar{\Lambda} = \Omega \, (\Upsilon)$). Finally, we have used the triangle inequality, and that $|x-y|\leq 2\, \mathrm{max}\{|x|,|y|\}$. 

We now explicitly define $\eta_\mathrm{k}$ and $C_\mathrm{k}$ from Result~\ref{result:nonMark},
\begin{equation}
\begin{split} \label{eq:ckdef}
    &\eta_\mathrm{k}:= 2 \underset{\substack{w\in \{x^{(\prime)}_-,y^{(\prime)}_-\}\\\Lambda \in \{\Upsilon,\Omega \}}}{\mathrm{max}} \left( \frac{ \sqrt{2^{k}-1}d_S^k +  C_\mathrm{k} }{  \braket{\textup{\textbf{A}}_{w}}_{\Lambda} } \right)\\
    &C_\mathrm{k} :=\left( \left|\braket{\textbf{A}_+ \otimes \textbf{A}_{w} }_{\bar{\Lambda}}  \right|\right) \sum_{i=1}^\infty  \frac{\left(\sqrt{2^{k_-}-1} d_S^{k_-}\right)^i}{  d_{\mathrm{eff}}^{{(i-1)}/3}[\rho]\braket{\textbf{A}_{w}}_{\Lambda}^i},
\end{split}
\end{equation}
where $k_-$ is the number of times that the instrument $\textbf{A}_-$ acts at. From now on we will drop the notation for the maximums, noting that wherever $w$ and $\Lambda$ appear, a maximum over $\{x,y,x^{\prime}_-,y^{\prime}_-\}$ and $\{\Upsilon, \Omega \}$ is implied respectively.

Now we are in a position to prove Result~\ref{result:nonMark}. Subbing in the definitions of $\eta_\mathrm{k}$ and $C_\mathrm{k}$ \eqref{eq:ckdef} into the left hand side of Result~\ref{result:nonMark} we get,
\begin{align}
        &\mathbb{P}\left\{|\mc{N}_\Upsilon(\textbf{A}_\pm)-\mc{N}_\Omega(\textbf{A}_\pm)|\geq \frac{2 \sqrt{2^{k}-1}d_S^k}{ \braket{\textbf{A}_{w}}_{\Lambda} d_{\mathrm{eff}}^{1/3}[\rho]} + \, \left|\frac{\braket{\textbf{A}_+ \otimes \textbf{A}_{w} }_{\bar{\Lambda}}}{\braket{\textbf{A}_{w}}_{\Lambda}}\right| \sum_{i=1}^\infty \left( \frac{\sqrt{2^{k_-}-1} d_S^{k_-} }{  \braket{\textbf{A}_{w}}_{\Lambda} d_{\mathrm{eff}}^{1/3}[\rho]}\right)^i \right\} \nn\\
         &\leq \mathbb{P}\left\{   \left|\frac{\braket{\textbf{A}_+ \otimes \textbf{A}_{w} }_{\Upsilon-\Omega}}{\braket{\textbf{A}_{w}}_{\Lambda}}\right|  +  \left|\frac{\braket{\textbf{A}_+ \otimes \textbf{A}_{w} }_{\bar{\Lambda}}}{\braket{\textbf{A}_{w}}_{\Lambda}} \sum_{i=1}^\infty \left(\frac{\braket{\textbf{A}_{w}}_{\Upsilon-\Omega}}{\braket{\textbf{A}_{w}}_\Lambda} \right)^i\right| \geq \frac{ \sqrt{2^{k}-1} d_S^{k} }{  \braket{\textbf{A}_{w}}_{\Lambda}d_{\mathrm{eff}}^{1/3}[\rho]} \right. \nn\\
         &\hspace{3in}+ \left|\frac{\braket{\textbf{A}_+ \otimes \textbf{A}_{w} }_{\bar{\Lambda}}}{\braket{\textbf{A}_{x_-}}_{\Lambda}} \right| \sum_{i=1}^\infty \left( \frac{\sqrt{2^{k_-}-1} d_S^{k_-} \braket{\textbf{A}_{w}}_{\Lambda}}{  \braket{\textbf{A}_{w}}_{\Lambda} d_{\mathrm{eff}}^{1/3}[\rho]}\right)^i \Bigg\} \nn\\
         &\leq \mathbb{P}\Bigg\{ |\braket{\textbf{A}_+ \otimes \textbf{A}_{w} }_{\Upsilon-\Omega}| \geq \frac{ \sqrt{2^{k}-1} d_S^{k} }{  d_{\mathrm{eff}}^{1/3}[\rho]} \Bigg\} +\mathbb{P}\Bigg\{  |\frac{\braket{\textbf{A}_+ \otimes \textbf{A}_{w} }_{\bar{\Lambda}}}{\braket{\textbf{A}_{w}}_{\Lambda}} \sum_{i=1}^\infty \left(\frac{\braket{\textbf{A}_{w}}_{\Upsilon-\Omega}}{\braket{\textbf{A}_{w}}_\Lambda} \right)^i|  \nn\\
         &\hspace{4in} \geq |\frac{\braket{\textbf{A}_+ \otimes \textbf{A}_{w} }_{\bar{\Lambda}}}{\braket{\textbf{A}_{x_-}}_{\Lambda}} | \sum_{i=1}^\infty \left( \frac{\sqrt{2^{k_-}-1} d_S^{k_-} }{  \braket{\textbf{A}_{w}}_{\Lambda} d_{\mathrm{eff}}^{1/3}[\rho]}\right)^i \Bigg\}\nn\\
         &\leq d_{\mathrm{eff}}^{-1/3}[\rho] +\mathbb{P}\Bigg\{  | \sum_{i=1}^\infty \left(\frac{\braket{\textbf{A}_{w}}_{\Upsilon-\Omega}}{\braket{\textbf{A}_{w}}_\Lambda} \right)^i| \geq  \sum_{i=1}^\infty \left( \frac{\sqrt{2^{k_-}-1} d_S^{k_-} }{  \braket{\textbf{A}_{w}}_{\Lambda} d_{\mathrm{eff}}^{1/3}[\rho]}\right)^i \Bigg\} \nn\\
         &\leq d_{\mathrm{eff}}^{-1/3}[\rho] + \mathbb{P}\Bigg\{ \underset{n \to \infty}{\mathrm{lim}} \sum_{i=1}^n  \left( |\frac{\braket{\textbf{A}_{w}}_{\Upsilon-\Omega}}{\braket{\textbf{A}_{w}}_\Lambda}| \right)^i \geq \underset{m \to \infty}{\mathrm{lim}} \sum_{i=1}^m\left( \frac{\sqrt{2^{k_-}-1} d_S^{k_-} }{  \braket{\textbf{A}_{w}}_{\Lambda} d_{\mathrm{eff}}^{1/3}[\rho]}\right)^i \Bigg\}\label{eq:NonMarkProof} \\
         &\leq d_{\mathrm{eff}}^{-1/3}[\rho] + \mathbb{P}\Bigg\{ \bigcup_{i=1}^\infty \left(| \frac{\braket{\textbf{A}_{w}}_{\Upsilon-\Omega}}{\braket{\textbf{A}_{w}}_\Lambda}|^i \geq  \big( \frac{\sqrt{2^{k_-}-1} d_S^{k_-} }{  \braket{\textbf{A}_{w}}_{\Lambda}d_{\mathrm{eff}}^{1/3}[\rho]}\big)^i\right) \Bigg\} \nn\\
         &= d_{\mathrm{eff}}^{-1/3}[\rho] + \mathbb{P}\Bigg\{ \bigcup_{i=1}^\infty \left(  |\frac{\braket{\textbf{A}_{w}}_{\Upsilon-\Omega}}{\braket{\textbf{A}_{w}}_\Lambda} | \geq   \frac{\sqrt{2^{k_-}-1} d_S^{k_-} }{  \braket{\textbf{A}_{w}}_{\Lambda} d_{\mathrm{eff}}^{1/3}[\rho]}\right) \Bigg\} \nn\\
         &=d_{\mathrm{eff}}^{-1/3}[\rho] + \mathbb{P}\Bigg\{  \left(  |\braket{\textbf{A}_{w}}_{\Upsilon-\Omega}| \geq   \frac{\sqrt{2^{k_-}-1} d_S^{k_-} }{  d_{\mathrm{eff}}^{1/3}[\rho]}\right) \Bigg\} \leq 2\, d_{\mathrm{eff}}^{-1/3}[\rho]\nn
\end{align}
For the second inequality we have used that, for positive functions $a$ and $b$ of a random variable $t$, and for $A,B >0$,   
\begin{equation}
    \mathbb{P}\{a(t) + b(t) > A + B \} \leq \mathbb{P}\left\{(a(t) > A) \cup (b(t) >B) \right\} \leq \mathbb{P}\{a(t)> A \}+\mathbb{P}\{b(t) > B \}, 
\end{equation}
where the second inequality here is due to Boole's inequality. For the third inequality we have applied the Chebyshev bound Result~\ref{result:nCheby} for the first term, and for the second term divided by the common factor (note that $\mathrm{max}(f g) \leq \mathrm{max}(f) \times \mathrm{max}(g) $). In the third last line we use again that $\mathbb{P}\{a(t) + b(t) > A + B \} \leq \mathbb{P}\{(a(t) > A) \cup (b(t) >B) \}$ (\emph{ad infinitum}) and in the penultimate line that $|x|^i \geq |y|^i \iff |x| \geq |y| $ for $i \in \mathbb{Z}_{\geq 1}$. Finally, we use that the union of many copies of the same logical statement is equivalent to a single copy of that statement, and again apply Result~\ref{result:nCheby}.

\section{Numerical Analysis of Equilibrium Process Non-Markovianity (Fig.~\ref{fig:NMnumerics})} \label{ap:numerics}
\begin{figure}[t]
\centering
\includegraphics[width=0.4\textwidth]{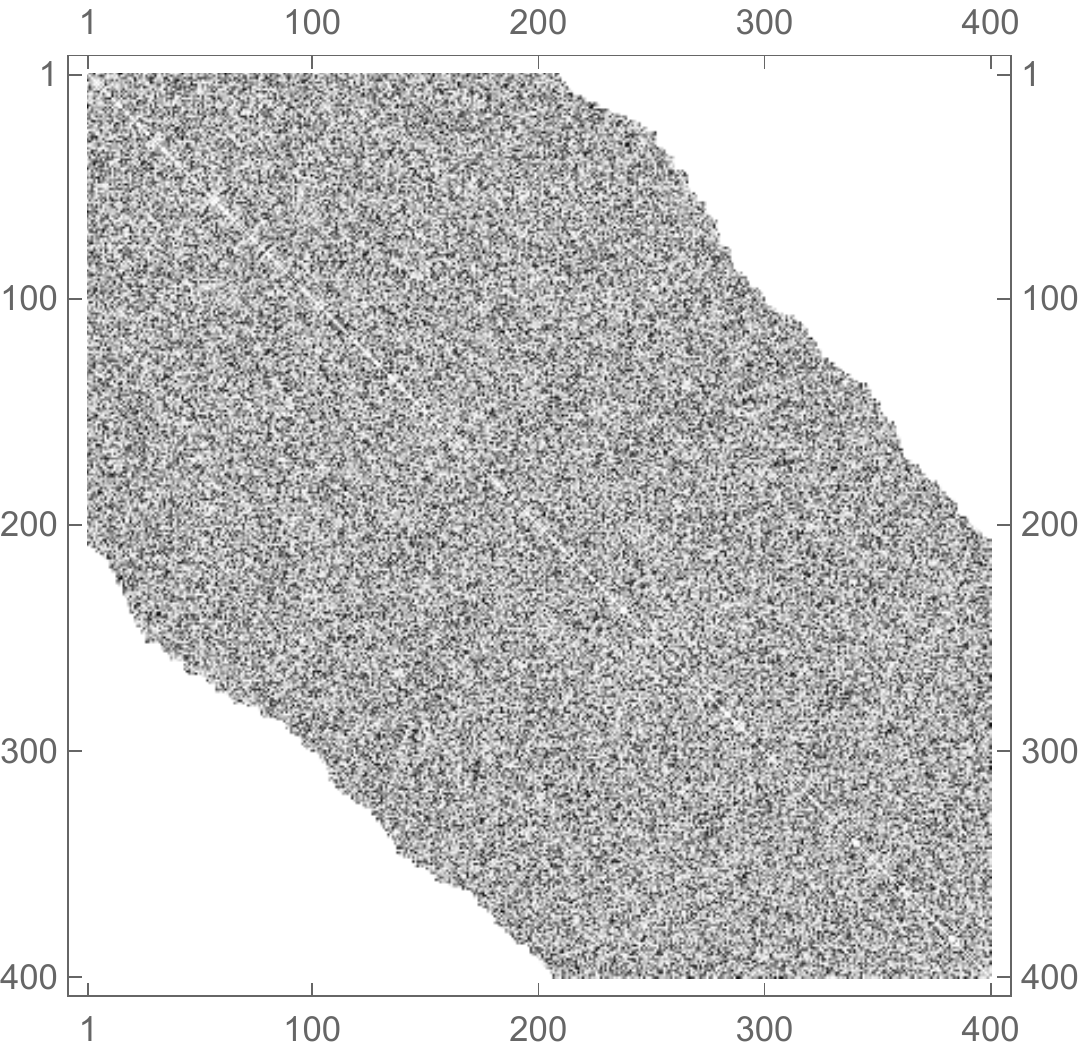}
\caption{\footnotesize{An example of the randomly generated bath coupling matrix $B$ for $d_B=400$, where a matrix element is non-zero only if the difference in energies corresponding to the element is less than the level splitting $\omega$ of the system spin, in addition to the diagonal being set to zero.}}
\label{fig:B_matrix}
\end{figure}

Our model for the numerical simulation is a single spin coupled to a random matrix environment, with the system-environment Hamiltonian specified as follows:
\begin{equation}
    H_{SE} = H_S +H_E + \lambda \sigma_x \otimes B, \quad H_S =\frac{\omega}{2} \sigma_z + \frac{\Delta}{2}\sigma_x, \quad H_E = \sum_{k=1}^{d_E} \epsilon_k \ket{k} \bra{k}.
\end{equation}
To mimic a large bath as well as possible with finite numerical resources, random matrix theory is used to sample the bath coupling operator $B$ and the free bath Hamiltonian $H_E$. Specifically, the energies $k$ of the bath are randomly sampled from the interval $[0, 1]$. Furthermore, the bath coupling operator is then constructed as $B = (R+R^T)/2 - \text{diag}[(R + R^T)/2]$, where $R$ is a random matrix with entries sampled from the interval $[-1, 1]$, and also the system splitting $\omega$ is used as a sharp cutoff for deciding which bath energies may couple in the matrix $R$ (see Fig.~\ref{fig:B_matrix}). Consequently, $B$ is a Hermitian operator with diagonal elements set equal to zero (to make the bath ‘passive’), with the far-from-diagonal elements also set to zero (to simulate a more physical interaction between energy eigenstates, compared to if $B$ was fully populated). The other numerical parameters are the level splitting of the qubit, which is set to $\omega = 0.5$ (note that $\omega$ should be smaller than one, otherwise the system energy dominates the bath energy), the tunneling of the qubit, set to $\Delta = 0.2$, and the system-bath coupling strength, set to $\lambda = 0.1$ (note that $\lambda$ should not be too small to let the system `talk’ to many levels of the finite bath). Hence, the only free parameter of the Hamiltonian in the following is the dimension $d_E$ of the bath Hilbert space. Note that there is no averaging over the random matrices in the Hamiltonian: only a single instance is used for them. 

A random pure initial $SE$ state $\rho$ is generated via a random complex vector in the computational basis, followed by normalization. This is to simulate a generic situation, with a relatively large $d_{\textrm{eff}}$ on average.

To simulate the instrument $\textbf{A}_\pm$, a random rank-1 measurement on the (single qubit) system state (with the identity acting on the variable sized environment) is generated for steps 2 and 3, with a new system state $\rho_{S,i}$  randomly generated independently after each step, such that e.g. the input to step 3 is $\rho_{3} = \rho_{S,3} \otimes \tr_S [ \mc{A}_2 ( \$ (\rho_2 ))]$. By a random measurement, we mean a POVM element constructed via choosing a (normalized) Kraus operator with uniformly random complex components. Across the three steps, together this is the instrument $\mc{A}_+$. 20 iterations of a step 1 measurement $\mc{A}_-$ are also randomly generated, and for a given $\mc{A}_+$, $\rho$, and $H_{SE}$, the largest difference in the expectation value $\braket{\textbf{A}_\pm}_{\Upsilon/\Omega}$ for varying $\mc{A}_-$ is computed. This simulates the different possible outcomes one could possibly obtain for a step 1 measurement, and what effect this may have on the final measurement. For a given $H_{SE}$ and $\rho$, this is repeated and averaged over for separately 25 sampled $\mc{A}_+$. 

\begin{figure}[t]
\centering
\includegraphics[width=0.5\textwidth]{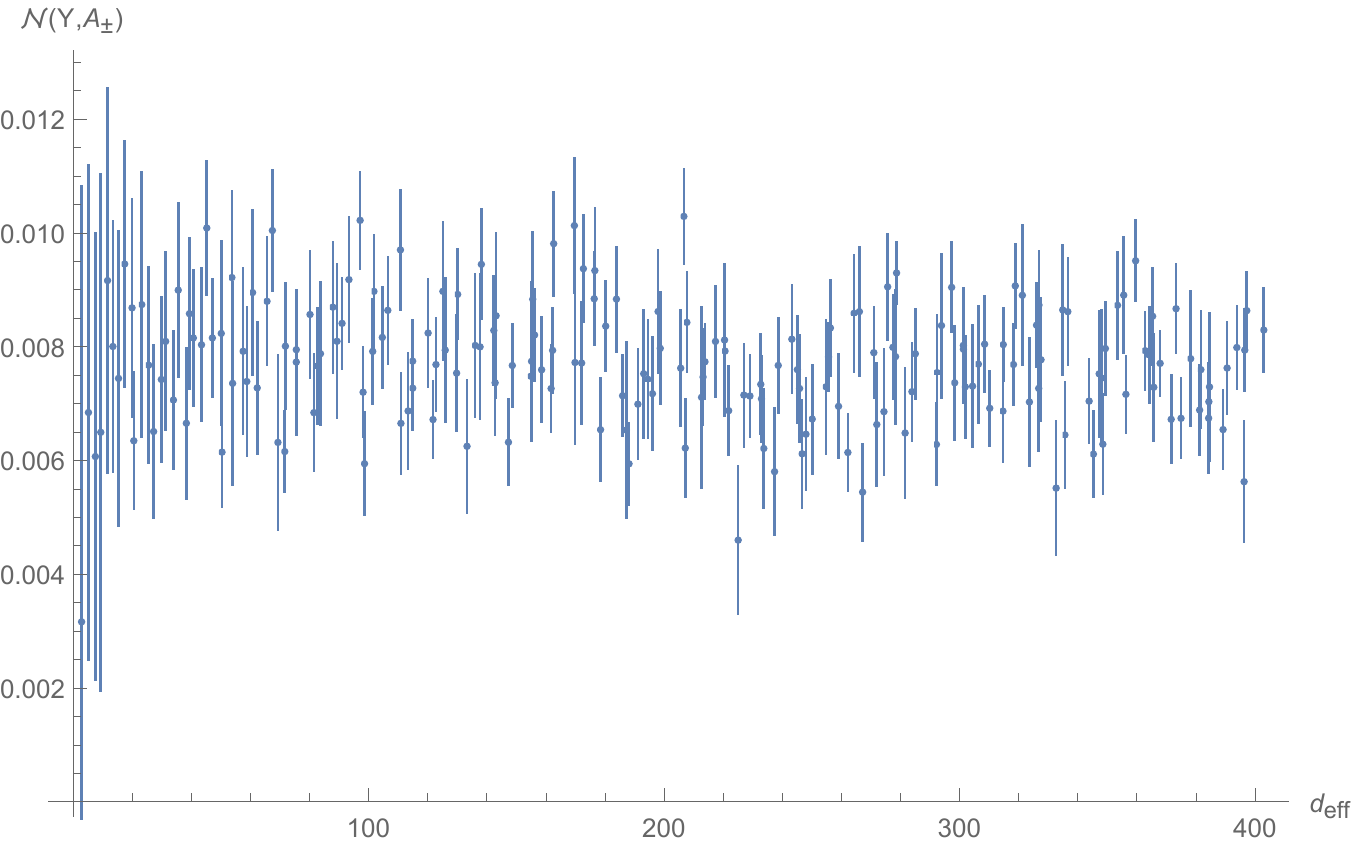}
\caption{\footnotesize{Non-Markovianity of a 3-step process $\Upsilon_s$ of a single qubit coupled to a random matrix environment, with a varying $d_{\textrm{eff}}$. The instruments $\textbf{A}_\pm$ used to probe these processes involved a repeated measurement and independently-prepare protocol, with a system-level causal break between $\textbf{A}_-$ and $\textbf{A}_+$, as displayed in the upper right of Fig.~\ref{fig:NMnumerics}. The unitary evolution of this process was generated with uniformly sampled $0.01 \leq \Delta t_1,\, \Delta t_2, \, \Delta t_3 \leq 0.5$. This was maximized over the difference in outcomes $\textbf{A}_-$, and averaged over $\textbf{A}_+$ to determine the non-Markovianity. This was produced using \cite{Mathematica,QIpackage}.}}
\label{fig:NMnumericsEXTRA}

\end{figure}

Two unitary processes $\Upsilon_\ell$ and $\Upsilon_s$ were used in the analysis, in addition to the dephased process $\Omega$. Random long time intervals were used for $\Upsilon_\ell$, picked uniformly such that $5 \leq \Delta t_1,\, \Delta t_2, \, \Delta t_3 \leq 50$, and short time intervals for $\Upsilon_s$, such that $0.01 \leq \Delta t_1,\, \Delta t_2, \, \Delta t_3 \leq 0.5$. $\Upsilon_s$ produced relatively random results for the non-Markovianity, which is expected as the process does not have enough time to relax from its initial conditions. The results for $\Upsilon_s$ are shown in Fig.~\ref{fig:NMnumericsEXTRA}, while the results for $\Upsilon_\ell$ are a part of Fig.~\ref{fig:NMnumerics}. The standard error for the data depicted in Fig.~\ref{fig:NMnumerics} was too small to include as error bars, and the average for $\Upsilon_\ell$ was $\sigma_{\rm av}=1.0\times 10^{-3}$, and for $\Omega$ it was $\sigma_{\rm av}=1.8 \times 10^{-4}$.

For each $d_E$ increasing by two, up to $d_E=400$, the entire computation is repeated for 40 iterations of generating a random $H_{SE}$ and random initial state $\rho$, with the final non-Markovianity being the average of these runs. Fig.~\ref{fig:NMnumerics} was plotted using a moving average in bins of $5$ points, so the trend could be more easily discernible. 


\begin{thebibliography}{87}%
\makeatletter
\providecommand \@ifxundefined [1]{%
 \@ifx{#1\undefined}
}%
\providecommand \@ifnum [1]{%
 \ifnum #1\expandafter \@firstoftwo
 \else \expandafter \@secondoftwo
 \fi
}%
\providecommand \@ifx [1]{%
 \ifx #1\expandafter \@firstoftwo
 \else \expandafter \@secondoftwo
 \fi
}%
\providecommand \natexlab [1]{#1}%
\providecommand \enquote  [1]{``#1''}%
\providecommand \bibnamefont  [1]{#1}%
\providecommand \bibfnamefont [1]{#1}%
\providecommand \citenamefont [1]{#1}%
\providecommand \href@noop [0]{\@secondoftwo}%
\providecommand \href [0]{\begingroup \@sanitize@url \@href}%
\providecommand \@href[1]{\@@startlink{#1}\@@href}%
\providecommand \@@href[1]{\endgroup#1\@@endlink}%
\providecommand \@sanitize@url [0]{\catcode `\\12\catcode `\$12\catcode
  `\&12\catcode `\#12\catcode `\^12\catcode `\_12\catcode `\%12\relax}%
\providecommand \@@startlink[1]{}%
\providecommand \@@endlink[0]{}%
\providecommand \url  [0]{\begingroup\@sanitize@url \@url }%
\providecommand \@url [1]{\endgroup\@href {#1}{\urlprefix }}%
\providecommand \urlprefix  [0]{URL }%
\providecommand \Eprint [0]{\href }%
\providecommand \doibase [0]{http://dx.doi.org/}%
\providecommand \selectlanguage [0]{\@gobble}%
\providecommand \bibinfo  [0]{\@secondoftwo}%
\providecommand \bibfield  [0]{\@secondoftwo}%
\providecommand \translation [1]{[#1]}%
\providecommand \BibitemOpen [0]{}%
\providecommand \bibitemStop [0]{}%
\providecommand \bibitemNoStop [0]{.\EOS\space}%
\providecommand \EOS [0]{\spacefactor3000\relax}%
\providecommand \BibitemShut  [1]{\csname bibitem#1\endcsname}%
\let\auto@bib@innerbib\@empty
\bibitem [{\citenamefont {Rivas}\ and\ \citenamefont {van
  Huelga}(2012)}]{Rivas2012}%
  \BibitemOpen
  \bibfield  {author} {\bibinfo {author} {\bibfnamefont {A.}~\bibnamefont
  {Rivas}}\ and\ \bibinfo {author} {\bibfnamefont {S.~F.}\ \bibnamefont {van
  Huelga}},\ }\href {\doibase 10.1007/978-3-642-23354-8} {\emph {\bibinfo
  {title} {Open Quantum Systems}}}\ (\bibinfo  {publisher} {Springer-Verlag},\
  \bibinfo {year} {2012})\BibitemShut {NoStop}%
\bibitem [{\citenamefont {Rotter}\ and\ \citenamefont
  {Bird}(2015)}]{Rotter2015}%
  \BibitemOpen
  \bibfield  {author} {\bibinfo {author} {\bibfnamefont {I.}~\bibnamefont
  {Rotter}}\ and\ \bibinfo {author} {\bibfnamefont {J.~P.}\ \bibnamefont
  {Bird}},\ }\href {\doibase 10.1088/0034-4885/78/11/114001} {\bibfield
  {journal} {\bibinfo  {journal} {Rep. Prog. Phys.}\ }\textbf {\bibinfo
  {volume} {78}},\ \bibinfo {pages} {114001} (\bibinfo {year}
  {2015})}\BibitemShut {NoStop}%
\bibitem [{\citenamefont {Pottier}(2010)}]{pottier2010nonequilibrium}%
  \BibitemOpen
  \bibfield  {author} {\bibinfo {author} {\bibfnamefont {N.}~\bibnamefont
  {Pottier}},\ }\href@noop {} {\emph {\bibinfo {title} {Nonequilibrium
  Statistical Physics: Linear Irreversible Processes}}},\ Oxford Graduate
  Texts\ (\bibinfo  {publisher} {Oxford University Press},\ \bibinfo {year}
  {2010})\BibitemShut {NoStop}%
\bibitem [{\citenamefont {Kubo}(1966)}]{Kubo1966}%
  \BibitemOpen
  \bibfield  {author} {\bibinfo {author} {\bibfnamefont {R.}~\bibnamefont
  {Kubo}},\ }\href {\doibase 10.1088/0034-4885/29/1/306} {\bibfield  {journal}
  {\bibinfo  {journal} {Rep. Prog. Phys.}\ }\textbf {\bibinfo {volume} {29}},\
  \bibinfo {pages} {255} (\bibinfo {year} {1966})}\BibitemShut {NoStop}%
\bibitem [{\citenamefont {Weiss}(2012)}]{Weiss2012}%
  \BibitemOpen
  \bibfield  {author} {\bibinfo {author} {\bibfnamefont {U.}~\bibnamefont
  {Weiss}},\ }\href {\doibase 10.1142/8334} {\emph {\bibinfo {title} {Quantum
  Dissipative Systems}}},\ \bibinfo {edition} {4th}\ ed.\ (\bibinfo
  {publisher} {World Scientific},\ \bibinfo {year} {2012})\BibitemShut
  {NoStop}%
\bibitem [{\citenamefont {Stefanucci}\ and\ \citenamefont {van
  Leeuwen}(2013)}]{stefanuccivanleeuwen2013}%
  \BibitemOpen
  \bibfield  {author} {\bibinfo {author} {\bibfnamefont {G.}~\bibnamefont
  {Stefanucci}}\ and\ \bibinfo {author} {\bibfnamefont {R.}~\bibnamefont {van
  Leeuwen}},\ }\href {\doibase 10.1017/CBO9781139023979} {\emph {\bibinfo
  {title} {Nonequilibrium Many-Body Theory of Quantum Systems: A Modern
  Introduction}}}\ (\bibinfo  {publisher} {Cambridge University Press},\
  \bibinfo {year} {2013})\BibitemShut {NoStop}%
\bibitem [{\citenamefont {Lax}(1967)}]{Lax1967}%
  \BibitemOpen
  \bibfield  {author} {\bibinfo {author} {\bibfnamefont {M.}~\bibnamefont
  {Lax}},\ }\href {\doibase 10.1103/PhysRev.157.213} {\bibfield  {journal}
  {\bibinfo  {journal} {Phys. Rev.}\ }\textbf {\bibinfo {volume} {157}},\
  \bibinfo {pages} {213} (\bibinfo {year} {1967})}\BibitemShut {NoStop}%
\bibitem [{\citenamefont {Pollock}\ \emph
  {et~al.}(2018{\natexlab{a}})\citenamefont {Pollock}, \citenamefont
  {Rodr\'{\i}guez-Rosario}, \citenamefont {Frauenheim}, \citenamefont
  {Paternostro},\ and\ \citenamefont {Modi}}]{processtensor}%
  \BibitemOpen
  \bibfield  {author} {\bibinfo {author} {\bibfnamefont {F.~A.}\ \bibnamefont
  {Pollock}}, \bibinfo {author} {\bibfnamefont {C.}~\bibnamefont
  {Rodr\'{\i}guez-Rosario}}, \bibinfo {author} {\bibfnamefont {T.}~\bibnamefont
  {Frauenheim}}, \bibinfo {author} {\bibfnamefont {M.}~\bibnamefont
  {Paternostro}}, \ and\ \bibinfo {author} {\bibfnamefont {K.}~\bibnamefont
  {Modi}},\ }\href {\doibase 10.1103/PhysRevA.97.012127} {\bibfield  {journal}
  {\bibinfo  {journal} {Phys. Rev. A}\ }\textbf {\bibinfo {volume} {97}},\
  \bibinfo {pages} {012127} (\bibinfo {year} {2018}{\natexlab{a}})}\BibitemShut
  {NoStop}%
\bibitem [{\citenamefont {Pollock}\ \emph
  {et~al.}(2018{\natexlab{b}})\citenamefont {Pollock}, \citenamefont
  {Rodr\'{\i}guez-Rosario}, \citenamefont {Frauenheim}, \citenamefont
  {Paternostro},\ and\ \citenamefont {Modi}}]{processtensor2}%
  \BibitemOpen
  \bibfield  {author} {\bibinfo {author} {\bibfnamefont {F.~A.}\ \bibnamefont
  {Pollock}}, \bibinfo {author} {\bibfnamefont {C.}~\bibnamefont
  {Rodr\'{\i}guez-Rosario}}, \bibinfo {author} {\bibfnamefont {T.}~\bibnamefont
  {Frauenheim}}, \bibinfo {author} {\bibfnamefont {M.}~\bibnamefont
  {Paternostro}}, \ and\ \bibinfo {author} {\bibfnamefont {K.}~\bibnamefont
  {Modi}},\ }\href {\doibase 10.1103/PhysRevLett.120.040405} {\bibfield
  {journal} {\bibinfo  {journal} {Phys. Rev. Lett.}\ }\textbf {\bibinfo
  {volume} {120}},\ \bibinfo {pages} {040405} (\bibinfo {year}
  {2018}{\natexlab{b}})}\BibitemShut {NoStop}%
\bibitem [{\citenamefont {Li}\ \emph {et~al.}(2018)\citenamefont {Li},
  \citenamefont {Hall},\ and\ \citenamefont {Wiseman}}]{Wiseman2018}%
  \BibitemOpen
  \bibfield  {author} {\bibinfo {author} {\bibfnamefont {L.}~\bibnamefont
  {Li}}, \bibinfo {author} {\bibfnamefont {M.~J.}\ \bibnamefont {Hall}}, \ and\
  \bibinfo {author} {\bibfnamefont {H.~M.}\ \bibnamefont {Wiseman}},\ }\href
  {\doibase https://doi.org/10.1016/j.physrep.2018.07.001} {\bibfield
  {journal} {\bibinfo  {journal} {Phys. Rep.}\ }\textbf {\bibinfo {volume}
  {759}},\ \bibinfo {pages} {1} (\bibinfo {year} {2018})},\ \bibinfo {note}
  {concepts of quantum non-Markovianity: A hierarchy}\BibitemShut {NoStop}%
\bibitem [{\citenamefont {Milz}\ \emph
  {et~al.}(2020{\natexlab{a}})\citenamefont {Milz}, \citenamefont {Sakuldee},
  \citenamefont {Pollock},\ and\ \citenamefont
  {Modi}}]{Milz2020kolmogorovextension}%
  \BibitemOpen
  \bibfield  {author} {\bibinfo {author} {\bibfnamefont {S.}~\bibnamefont
  {Milz}}, \bibinfo {author} {\bibfnamefont {F.}~\bibnamefont {Sakuldee}},
  \bibinfo {author} {\bibfnamefont {F.~A.}\ \bibnamefont {Pollock}}, \ and\
  \bibinfo {author} {\bibfnamefont {K.}~\bibnamefont {Modi}},\ }\href {\doibase
  10.22331/q-2020-04-20-255} {\bibfield  {journal} {\bibinfo  {journal}
  {{Quantum}}\ }\textbf {\bibinfo {volume} {4}},\ \bibinfo {pages} {255}
  (\bibinfo {year} {2020}{\natexlab{a}})}\BibitemShut {NoStop}%
\bibitem [{\citenamefont {Milz}\ and\ \citenamefont
  {Modi}(2021)}]{milz2020quantum}%
  \BibitemOpen
  \bibfield  {author} {\bibinfo {author} {\bibfnamefont {S.}~\bibnamefont
  {Milz}}\ and\ \bibinfo {author} {\bibfnamefont {K.}~\bibnamefont {Modi}},\
  }\href {\doibase 10.1103/PRXQuantum.2.030201} {\bibfield  {journal} {\bibinfo
   {journal} {PRX Quantum}\ }\textbf {\bibinfo {volume} {2}},\ \bibinfo {pages}
  {030201} (\bibinfo {year} {2021})}\BibitemShut {NoStop}%
\bibitem [{\citenamefont {Dowling}\ \emph {et~al.}(2021)\citenamefont
  {Dowling}, \citenamefont {Figueroa-Romero}, \citenamefont {Pollock},
  \citenamefont {Strasberg},\ and\ \citenamefont {Modi}}]{finitetime}%
  \BibitemOpen
  \bibfield  {author} {\bibinfo {author} {\bibfnamefont {N.}~\bibnamefont
  {Dowling}}, \bibinfo {author} {\bibfnamefont {P.}~\bibnamefont
  {Figueroa-Romero}}, \bibinfo {author} {\bibfnamefont {F.}~\bibnamefont
  {Pollock}}, \bibinfo {author} {\bibfnamefont {P.}~\bibnamefont {Strasberg}},
  \ and\ \bibinfo {author} {\bibfnamefont {K.}~\bibnamefont {Modi}},\ }\href
  {\doibase 10.48550/arXiv.2112.01099} {\enquote {\bibinfo {title}
  {Equilibration of non-markovian quantum processes in finite time
  intervals},}\ } (\bibinfo {year} {2021}),\ \Eprint
  {http://arxiv.org/abs/2112.01099} {arXiv:2112.01099 [quant-ph]} \BibitemShut
  {NoStop}%
\bibitem [{\citenamefont {Linden}\ \emph {et~al.}(2009)\citenamefont {Linden},
  \citenamefont {Popescu}, \citenamefont {Short},\ and\ \citenamefont
  {Winter}}]{Linden2008}%
  \BibitemOpen
  \bibfield  {author} {\bibinfo {author} {\bibfnamefont {N.}~\bibnamefont
  {Linden}}, \bibinfo {author} {\bibfnamefont {S.}~\bibnamefont {Popescu}},
  \bibinfo {author} {\bibfnamefont {A.~J.}\ \bibnamefont {Short}}, \ and\
  \bibinfo {author} {\bibfnamefont {A.}~\bibnamefont {Winter}},\ }\href
  {\doibase 10.1103/PhysRevE.79.061103} {\bibfield  {journal} {\bibinfo
  {journal} {Phys. Rev. E}\ }\textbf {\bibinfo {volume} {79}},\ \bibinfo
  {pages} {061103} (\bibinfo {year} {2009})}\BibitemShut {NoStop}%
\bibitem [{\citenamefont {Neuenhahn}\ and\ \citenamefont
  {Marquardt}(2012)}]{Neuenhahn2012}%
  \BibitemOpen
  \bibfield  {author} {\bibinfo {author} {\bibfnamefont {C.}~\bibnamefont
  {Neuenhahn}}\ and\ \bibinfo {author} {\bibfnamefont {F.}~\bibnamefont
  {Marquardt}},\ }\href {\doibase 10.1103/PhysRevE.85.060101} {\bibfield
  {journal} {\bibinfo  {journal} {Phys. Rev. E}\ }\textbf {\bibinfo {volume}
  {85}},\ \bibinfo {pages} {060101(R)} (\bibinfo {year} {2012})}\BibitemShut
  {NoStop}%
\bibitem [{\citenamefont {Campos~Venuti}\ and\ \citenamefont
  {Zanardi}(2010)}]{Zanardi2010}%
  \BibitemOpen
  \bibfield  {author} {\bibinfo {author} {\bibfnamefont {L.}~\bibnamefont
  {Campos~Venuti}}\ and\ \bibinfo {author} {\bibfnamefont {P.}~\bibnamefont
  {Zanardi}},\ }\href {\doibase 10.1103/PhysRevA.81.022113} {\bibfield
  {journal} {\bibinfo  {journal} {Phys. Rev. A}\ }\textbf {\bibinfo {volume}
  {81}},\ \bibinfo {pages} {022113} (\bibinfo {year} {2010})}\BibitemShut
  {NoStop}%
\bibitem [{\citenamefont {Bocchieri}\ and\ \citenamefont
  {Loinger}(1957)}]{Bochieri1957}%
  \BibitemOpen
  \bibfield  {author} {\bibinfo {author} {\bibfnamefont {P.}~\bibnamefont
  {Bocchieri}}\ and\ \bibinfo {author} {\bibfnamefont {A.}~\bibnamefont
  {Loinger}},\ }\href {\doibase 10.1103/PhysRev.107.337} {\bibfield  {journal}
  {\bibinfo  {journal} {Phys. Rev.}\ }\textbf {\bibinfo {volume} {107}},\
  \bibinfo {pages} {337} (\bibinfo {year} {1957})}\BibitemShut {NoStop}%
\bibitem [{\citenamefont {Gogolin}\ and\ \citenamefont
  {Eisert}(2016)}]{Gogolin}%
  \BibitemOpen
  \bibfield  {author} {\bibinfo {author} {\bibfnamefont {C.}~\bibnamefont
  {Gogolin}}\ and\ \bibinfo {author} {\bibfnamefont {J.}~\bibnamefont
  {Eisert}},\ }\href {\doibase 10.1088/0034-4885/79/5/056001} {\bibfield
  {journal} {\bibinfo  {journal} {Rep. Prog. Phys.}\ }\textbf {\bibinfo
  {volume} {79}},\ \bibinfo {pages} {056001} (\bibinfo {year}
  {2016})}\BibitemShut {NoStop}%
\bibitem [{\citenamefont {Venuti}(2015)}]{venuti2015recurrence}%
  \BibitemOpen
  \bibfield  {author} {\bibinfo {author} {\bibfnamefont {L.~C.}\ \bibnamefont
  {Venuti}},\ }\href {\doibase 10.48550/arXiv.1509.04352} {\enquote {\bibinfo
  {title} {The recurrence time in quantum mechanics},}\ } (\bibinfo {year}
  {2015}),\ \Eprint {http://arxiv.org/abs/1509.04352} {arXiv:1509.04352
  [quant-ph]} \BibitemShut {NoStop}%
\bibitem [{\citenamefont {Reimann}(2008)}]{Reimann_2008}%
  \BibitemOpen
  \bibfield  {author} {\bibinfo {author} {\bibfnamefont {P.}~\bibnamefont
  {Reimann}},\ }\href {\doibase 10.1103/PhysRevLett.101.190403} {\bibfield
  {journal} {\bibinfo  {journal} {Phys. Rev. Lett.}\ }\textbf {\bibinfo
  {volume} {101}},\ \bibinfo {pages} {190403} (\bibinfo {year}
  {2008})}\BibitemShut {NoStop}%
\bibitem [{\citenamefont {Alhambra}\ \emph {et~al.}(2020)\citenamefont
  {Alhambra}, \citenamefont {Riddell},\ and\ \citenamefont
  {Garc{\'\i}a-Pintos}}]{Alhambra2020-fj}%
  \BibitemOpen
  \bibfield  {author} {\bibinfo {author} {\bibfnamefont {{\'A}.~M.}\
  \bibnamefont {Alhambra}}, \bibinfo {author} {\bibfnamefont {J.}~\bibnamefont
  {Riddell}}, \ and\ \bibinfo {author} {\bibfnamefont {L.~P.}\ \bibnamefont
  {Garc{\'\i}a-Pintos}},\ }\href {\doibase 10.1103/PhysRevLett.124.110605}
  {\bibfield  {journal} {\bibinfo  {journal} {Phys. Rev. Lett.}\ }\textbf
  {\bibinfo {volume} {124}},\ \bibinfo {pages} {110605} (\bibinfo {year}
  {2020})}\BibitemShut {NoStop}%
\bibitem [{\citenamefont {Figueroa-Romero}\ \emph {et~al.}(2021)\citenamefont
  {Figueroa-Romero}, \citenamefont {Pollock},\ and\ \citenamefont
  {Modi}}]{Pedro2021}%
  \BibitemOpen
  \bibfield  {author} {\bibinfo {author} {\bibfnamefont {P.}~\bibnamefont
  {Figueroa-Romero}}, \bibinfo {author} {\bibfnamefont {F.~A.}\ \bibnamefont
  {Pollock}}, \ and\ \bibinfo {author} {\bibfnamefont {K.}~\bibnamefont
  {Modi}},\ }\href {\doibase 10.1038/s42005-021-00629-w} {\bibfield  {journal}
  {\bibinfo  {journal} {Commun. Phys.}\ }\textbf {\bibinfo {volume} {4}},\
  \bibinfo {pages} {127} (\bibinfo {year} {2021})}\BibitemShut {NoStop}%
\bibitem [{\citenamefont {Gemmer}\ \emph {et~al.}(2009)\citenamefont {Gemmer},
  \citenamefont {Michel},\ and\ \citenamefont {Mahler}}]{gemmer2009}%
  \BibitemOpen
  \bibfield  {author} {\bibinfo {author} {\bibfnamefont {J.}~\bibnamefont
  {Gemmer}}, \bibinfo {author} {\bibfnamefont {M.}~\bibnamefont {Michel}}, \
  and\ \bibinfo {author} {\bibfnamefont {G.}~\bibnamefont {Mahler}},\ }\href
  {\doibase 10.1007/b98082} {\emph {\bibinfo {title} {Quantum Thermodynamics:
  Emergence of Thermodynamic Behavior Within Composite Quantum Systems}}},\
  Lecture Notes in Physics\ (\bibinfo  {publisher} {Springer Berlin
  Heidelberg},\ \bibinfo {year} {2009})\BibitemShut {NoStop}%
\bibitem [{\citenamefont {D'Alessio}\ \emph {et~al.}(2016)\citenamefont
  {D'Alessio}, \citenamefont {Kafri}, \citenamefont {Polkovnikov},\ and\
  \citenamefont {Rigol}}]{Rigol2016}%
  \BibitemOpen
  \bibfield  {author} {\bibinfo {author} {\bibfnamefont {L.}~\bibnamefont
  {D'Alessio}}, \bibinfo {author} {\bibfnamefont {Y.}~\bibnamefont {Kafri}},
  \bibinfo {author} {\bibfnamefont {A.}~\bibnamefont {Polkovnikov}}, \ and\
  \bibinfo {author} {\bibfnamefont {M.}~\bibnamefont {Rigol}},\ }\href
  {\doibase 10.1080/00018732.2016.1198134} {\bibfield  {journal} {\bibinfo
  {journal} {Adv. Phys.}\ }\textbf {\bibinfo {volume} {65}},\ \bibinfo {pages}
  {239} (\bibinfo {year} {2016})}\BibitemShut {NoStop}%
\bibitem [{\citenamefont {Mori}\ \emph {et~al.}(2018)\citenamefont {Mori},
  \citenamefont {Ikeda}, \citenamefont {Kaminishi},\ and\ \citenamefont
  {Ueda}}]{Mori_2018}%
  \BibitemOpen
  \bibfield  {author} {\bibinfo {author} {\bibfnamefont {T.}~\bibnamefont
  {Mori}}, \bibinfo {author} {\bibfnamefont {T.~N.}\ \bibnamefont {Ikeda}},
  \bibinfo {author} {\bibfnamefont {E.}~\bibnamefont {Kaminishi}}, \ and\
  \bibinfo {author} {\bibfnamefont {M.}~\bibnamefont {Ueda}},\ }\href {\doibase
  10.1088/1361-6455/aabcdf} {\bibfield  {journal} {\bibinfo  {journal} {J.
  Phys. B: At. Mol. Opt.}\ }\textbf {\bibinfo {volume} {51}},\ \bibinfo {pages}
  {112001} (\bibinfo {year} {2018})}\BibitemShut {NoStop}%
\bibitem [{\citenamefont {Costa}\ and\ \citenamefont
  {Shrapnel}(2016)}]{Costa2016}%
  \BibitemOpen
  \bibfield  {author} {\bibinfo {author} {\bibfnamefont {F.}~\bibnamefont
  {Costa}}\ and\ \bibinfo {author} {\bibfnamefont {S.}~\bibnamefont
  {Shrapnel}},\ }\href {\doibase 10.1088/1367-2630/18/6/063032} {\bibfield
  {journal} {\bibinfo  {journal} {New J. Phys.}\ }\textbf {\bibinfo {volume}
  {18}},\ \bibinfo {pages} {063032} (\bibinfo {year} {2016})}\BibitemShut
  {NoStop}%
\bibitem [{\citenamefont {Chiribella}\ \emph {et~al.}(2009)\citenamefont
  {Chiribella}, \citenamefont {D'Ariano},\ and\ \citenamefont
  {Perinotti}}]{Chiribella2009physrevA}%
  \BibitemOpen
  \bibfield  {author} {\bibinfo {author} {\bibfnamefont {G.}~\bibnamefont
  {Chiribella}}, \bibinfo {author} {\bibfnamefont {G.~M.}\ \bibnamefont
  {D'Ariano}}, \ and\ \bibinfo {author} {\bibfnamefont {P.}~\bibnamefont
  {Perinotti}},\ }\href {\doibase 10.1103/PhysRevA.80.022339} {\bibfield
  {journal} {\bibinfo  {journal} {Phys. Rev. A}\ }\textbf {\bibinfo {volume}
  {80}},\ \bibinfo {pages} {022339} (\bibinfo {year} {2009})}\BibitemShut
  {NoStop}%
\bibitem [{\citenamefont {Tasaki}(1998)}]{Tasaki_1998}%
  \BibitemOpen
  \bibfield  {author} {\bibinfo {author} {\bibfnamefont {H.}~\bibnamefont
  {Tasaki}},\ }\href {\doibase 10.1103/PhysRevLett.80.1373} {\bibfield
  {journal} {\bibinfo  {journal} {Phys. Rev. Lett.}\ }\textbf {\bibinfo
  {volume} {80}},\ \bibinfo {pages} {1373} (\bibinfo {year}
  {1998})}\BibitemShut {NoStop}%
\bibitem [{\citenamefont {Short}(2011)}]{ShortSystemsAndSub}%
  \BibitemOpen
  \bibfield  {author} {\bibinfo {author} {\bibfnamefont {A.~J.}\ \bibnamefont
  {Short}},\ }\href {\doibase 10.1088/1367-2630/13/5/053009} {\bibfield
  {journal} {\bibinfo  {journal} {New J. Phys.}\ }\textbf {\bibinfo {volume}
  {13}},\ \bibinfo {pages} {053009} (\bibinfo {year} {2011})}\BibitemShut
  {NoStop}%
\bibitem [{\citenamefont {Ueda}(2020)}]{Ueda2020}%
  \BibitemOpen
  \bibfield  {author} {\bibinfo {author} {\bibfnamefont {M.}~\bibnamefont
  {Ueda}},\ }\href {\doibase 10.1038/s42254-020-0237-x} {\bibfield  {journal}
  {\bibinfo  {journal} {Nat. Rev. Phys.}\ }\textbf {\bibinfo {volume} {2}},\
  \bibinfo {pages} {669} (\bibinfo {year} {2020})}\BibitemShut {NoStop}%
\bibitem [{\citenamefont {Davies}\ and\ \citenamefont
  {Lewis}(1970)}]{Davies1970}%
  \BibitemOpen
  \bibfield  {author} {\bibinfo {author} {\bibfnamefont {E.~B.}\ \bibnamefont
  {Davies}}\ and\ \bibinfo {author} {\bibfnamefont {J.~T.}\ \bibnamefont
  {Lewis}},\ }\href {\doibase 10.1007/BF01647093} {\bibfield  {journal}
  {\bibinfo  {journal} {Commun. Math. Phys.}\ }\textbf {\bibinfo {volume}
  {17}},\ \bibinfo {pages} {239} (\bibinfo {year} {1970})}\BibitemShut
  {NoStop}%
\bibitem [{\citenamefont {Chiribella}\ \emph {et~al.}(2008)\citenamefont
  {Chiribella}, \citenamefont {D`Ariano},\ and\ \citenamefont
  {Perinotti}}]{Chiribella_2008}%
  \BibitemOpen
  \bibfield  {author} {\bibinfo {author} {\bibfnamefont {G.}~\bibnamefont
  {Chiribella}}, \bibinfo {author} {\bibfnamefont {G.~M.}\ \bibnamefont
  {D`Ariano}}, \ and\ \bibinfo {author} {\bibfnamefont {P.}~\bibnamefont
  {Perinotti}},\ }\href {\doibase 10.1209/0295-5075/83/30004} {\bibfield
  {journal} {\bibinfo  {journal} {{EPL} (Europhysics Letters)}\ }\textbf
  {\bibinfo {volume} {83}},\ \bibinfo {pages} {30004} (\bibinfo {year}
  {2008})}\BibitemShut {NoStop}%
\bibitem [{\citenamefont {Hardy}(2007)}]{Hardy_2007}%
  \BibitemOpen
  \bibfield  {author} {\bibinfo {author} {\bibfnamefont {L.}~\bibnamefont
  {Hardy}},\ }\href {\doibase 10.1088/1751-8113/40/12/s12} {\bibfield
  {journal} {\bibinfo  {journal} {J. Phys. A-Math. Theor.}\ }\textbf {\bibinfo
  {volume} {40}},\ \bibinfo {pages} {3081} (\bibinfo {year}
  {2007})}\BibitemShut {NoStop}%
\bibitem [{\citenamefont {Hardy}(2012)}]{hardy2012}%
  \BibitemOpen
  \bibfield  {author} {\bibinfo {author} {\bibfnamefont {L.}~\bibnamefont
  {Hardy}},\ }\href {\doibase 10.1098/rsta.2011.0326} {\bibfield  {journal}
  {\bibinfo  {journal} {Philos. T. R. Soc. A}\ }\textbf {\bibinfo {volume}
  {370}},\ \bibinfo {pages} {3385} (\bibinfo {year} {2012})}\BibitemShut
  {NoStop}%
\bibitem [{\citenamefont {Hardy}(2016)}]{hardy2016operational}%
  \BibitemOpen
  \bibfield  {author} {\bibinfo {author} {\bibfnamefont {L.}~\bibnamefont
  {Hardy}},\ }\href {\doibase 10.48550/arXiv.1608.06940} {\enquote {\bibinfo
  {title} {Operational general relativity: Possibilistic, probabilistic, and
  quantum},}\ } (\bibinfo {year} {2016}),\ \Eprint
  {http://arxiv.org/abs/1608.06940} {arXiv:1608.06940 [gr-qc]} \BibitemShut
  {NoStop}%
\bibitem [{\citenamefont {Cotler}\ \emph {et~al.}(2018)\citenamefont {Cotler},
  \citenamefont {Jian}, \citenamefont {Qi},\ and\ \citenamefont
  {Wilczek}}]{Cotler2018}%
  \BibitemOpen
  \bibfield  {author} {\bibinfo {author} {\bibfnamefont {J.}~\bibnamefont
  {Cotler}}, \bibinfo {author} {\bibfnamefont {C.-M.}\ \bibnamefont {Jian}},
  \bibinfo {author} {\bibfnamefont {X.-L.}\ \bibnamefont {Qi}}, \ and\ \bibinfo
  {author} {\bibfnamefont {F.}~\bibnamefont {Wilczek}},\ }\href {\doibase
  10.1007/JHEP09(2018)093} {\bibfield  {journal} {\bibinfo  {journal} {J. High
  Energy Phys.}\ }\textbf {\bibinfo {volume} {2018}},\ \bibinfo {pages} {93}
  (\bibinfo {year} {2018})}\BibitemShut {NoStop}%
\bibitem [{\citenamefont {Kretschmann}\ and\ \citenamefont
  {Werner}(2005)}]{Werner2005}%
  \BibitemOpen
  \bibfield  {author} {\bibinfo {author} {\bibfnamefont {D.}~\bibnamefont
  {Kretschmann}}\ and\ \bibinfo {author} {\bibfnamefont {R.~F.}\ \bibnamefont
  {Werner}},\ }\href {\doibase 10.1103/PhysRevA.72.062323} {\bibfield
  {journal} {\bibinfo  {journal} {Phys. Rev. A}\ }\textbf {\bibinfo {volume}
  {72}},\ \bibinfo {pages} {062323} (\bibinfo {year} {2005})}\BibitemShut
  {NoStop}%
\bibitem [{\citenamefont {Caruso}\ \emph {et~al.}(2014)\citenamefont {Caruso},
  \citenamefont {Giovannetti}, \citenamefont {Lupo},\ and\ \citenamefont
  {Mancini}}]{Caruso2014}%
  \BibitemOpen
  \bibfield  {author} {\bibinfo {author} {\bibfnamefont {F.}~\bibnamefont
  {Caruso}}, \bibinfo {author} {\bibfnamefont {V.}~\bibnamefont {Giovannetti}},
  \bibinfo {author} {\bibfnamefont {C.}~\bibnamefont {Lupo}}, \ and\ \bibinfo
  {author} {\bibfnamefont {S.}~\bibnamefont {Mancini}},\ }\href {\doibase
  10.1103/RevModPhys.86.1203} {\bibfield  {journal} {\bibinfo  {journal} {Rev.
  Mod. Phys.}\ }\textbf {\bibinfo {volume} {86}},\ \bibinfo {pages} {1203}
  (\bibinfo {year} {2014})}\BibitemShut {NoStop}%
\bibitem [{\citenamefont {Portmann}\ \emph {et~al.}(2017)\citenamefont
  {Portmann}, \citenamefont {Matt}, \citenamefont {Maurer}, \citenamefont
  {Renner},\ and\ \citenamefont {Tackmann}}]{Portmann2017}%
  \BibitemOpen
  \bibfield  {author} {\bibinfo {author} {\bibfnamefont {C.}~\bibnamefont
  {Portmann}}, \bibinfo {author} {\bibfnamefont {C.}~\bibnamefont {Matt}},
  \bibinfo {author} {\bibfnamefont {U.}~\bibnamefont {Maurer}}, \bibinfo
  {author} {\bibfnamefont {R.}~\bibnamefont {Renner}}, \ and\ \bibinfo {author}
  {\bibfnamefont {B.}~\bibnamefont {Tackmann}},\ }\href {\doibase
  10.1109/TIT.2017.2676805} {\bibfield  {journal} {\bibinfo  {journal} {IEEE
  Transactions on Information Theory}\ }\textbf {\bibinfo {volume} {63}},\
  \bibinfo {pages} {3277} (\bibinfo {year} {2017})}\BibitemShut {NoStop}%
\bibitem [{\citenamefont {Shrapnel}\ \emph {et~al.}(2018)\citenamefont
  {Shrapnel}, \citenamefont {Costa},\ and\ \citenamefont
  {Milburn}}]{CostaBornRule}%
  \BibitemOpen
  \bibfield  {author} {\bibinfo {author} {\bibfnamefont {S.}~\bibnamefont
  {Shrapnel}}, \bibinfo {author} {\bibfnamefont {F.}~\bibnamefont {Costa}}, \
  and\ \bibinfo {author} {\bibfnamefont {G.}~\bibnamefont {Milburn}},\ }\href
  {\doibase 10.1088/1367-2630/aabe12} {\bibfield  {journal} {\bibinfo
  {journal} {New J. Phys.}\ }\textbf {\bibinfo {volume} {20}},\ \bibinfo
  {pages} {053010} (\bibinfo {year} {2018})}\BibitemShut {NoStop}%
\bibitem [{\citenamefont {Oreshkov}\ \emph {et~al.}(2012)\citenamefont
  {Oreshkov}, \citenamefont {Costa},\ and\ \citenamefont
  {Brukner}}]{CostaOreshkov2012ER}%
  \BibitemOpen
  \bibfield  {author} {\bibinfo {author} {\bibfnamefont {O.}~\bibnamefont
  {Oreshkov}}, \bibinfo {author} {\bibfnamefont {F.}~\bibnamefont {Costa}}, \
  and\ \bibinfo {author} {\bibfnamefont {{\v{C}}.}~\bibnamefont {Brukner}},\
  }\href {\doibase 10.1038/ncomms2076} {\bibfield  {journal} {\bibinfo
  {journal} {Nat. Commun.}\ }\textbf {\bibinfo {volume} {3}},\ \bibinfo {pages}
  {1092} (\bibinfo {year} {2012})}\BibitemShut {NoStop}%
\bibitem [{\citenamefont
  {Strasberg}(2019{\natexlab{a}})}]{PhysRevE.100.022127}%
  \BibitemOpen
  \bibfield  {author} {\bibinfo {author} {\bibfnamefont {P.}~\bibnamefont
  {Strasberg}},\ }\href {\doibase 10.1103/PhysRevE.100.022127} {\bibfield
  {journal} {\bibinfo  {journal} {Phys. Rev. E}\ }\textbf {\bibinfo {volume}
  {100}},\ \bibinfo {pages} {022127} (\bibinfo {year}
  {2019}{\natexlab{a}})}\BibitemShut {NoStop}%
\bibitem [{\citenamefont {Giarmatzi}\ and\ \citenamefont
  {Costa}(2021)}]{arXiv:1811.03722}%
  \BibitemOpen
  \bibfield  {author} {\bibinfo {author} {\bibfnamefont {C.}~\bibnamefont
  {Giarmatzi}}\ and\ \bibinfo {author} {\bibfnamefont {F.}~\bibnamefont
  {Costa}},\ }\href {\doibase 10.22331/q-2021-04-26-440} {\bibfield  {journal}
  {\bibinfo  {journal} {{Quantum}}\ }\textbf {\bibinfo {volume} {5}},\ \bibinfo
  {pages} {440} (\bibinfo {year} {2021})}\BibitemShut {NoStop}%
\bibitem [{\citenamefont {Strasberg}\ and\ \citenamefont
  {Winter}(2019)}]{PhysRevE.100.022135}%
  \BibitemOpen
  \bibfield  {author} {\bibinfo {author} {\bibfnamefont {P.}~\bibnamefont
  {Strasberg}}\ and\ \bibinfo {author} {\bibfnamefont {A.}~\bibnamefont
  {Winter}},\ }\href {\doibase 10.1103/PhysRevE.100.022135} {\bibfield
  {journal} {\bibinfo  {journal} {Phys. Rev. E}\ }\textbf {\bibinfo {volume}
  {100}},\ \bibinfo {pages} {022135} (\bibinfo {year} {2019})}\BibitemShut
  {NoStop}%
\bibitem [{\citenamefont
  {Strasberg}(2019{\natexlab{b}})}]{PhysRevLett.123.180604}%
  \BibitemOpen
  \bibfield  {author} {\bibinfo {author} {\bibfnamefont {P.}~\bibnamefont
  {Strasberg}},\ }\href {\doibase 10.1103/PhysRevLett.123.180604} {\bibfield
  {journal} {\bibinfo  {journal} {Phys. Rev. Lett.}\ }\textbf {\bibinfo
  {volume} {123}},\ \bibinfo {pages} {180604} (\bibinfo {year}
  {2019}{\natexlab{b}})}\BibitemShut {NoStop}%
\bibitem [{\citenamefont {Strasberg}\ and\ \citenamefont
  {D\'{\i}az}(2019)}]{strasberg2019}%
  \BibitemOpen
  \bibfield  {author} {\bibinfo {author} {\bibfnamefont {P.}~\bibnamefont
  {Strasberg}}\ and\ \bibinfo {author} {\bibfnamefont {M.~G.}\ \bibnamefont
  {D\'{\i}az}},\ }\href {\doibase 10.1103/PhysRevA.100.022120} {\bibfield
  {journal} {\bibinfo  {journal} {Phys. Rev. A}\ }\textbf {\bibinfo {volume}
  {100}},\ \bibinfo {pages} {022120} (\bibinfo {year} {2019})}\BibitemShut
  {NoStop}%
\bibitem [{\citenamefont {Milz}\ \emph
  {et~al.}(2020{\natexlab{b}})\citenamefont {Milz}, \citenamefont {Egloff},
  \citenamefont {Taranto}, \citenamefont {Theurer}, \citenamefont {Plenio},
  \citenamefont {Smirne},\ and\ \citenamefont {Huelga}}]{Milz2020prx}%
  \BibitemOpen
  \bibfield  {author} {\bibinfo {author} {\bibfnamefont {S.}~\bibnamefont
  {Milz}}, \bibinfo {author} {\bibfnamefont {D.}~\bibnamefont {Egloff}},
  \bibinfo {author} {\bibfnamefont {P.}~\bibnamefont {Taranto}}, \bibinfo
  {author} {\bibfnamefont {T.}~\bibnamefont {Theurer}}, \bibinfo {author}
  {\bibfnamefont {M.~B.}\ \bibnamefont {Plenio}}, \bibinfo {author}
  {\bibfnamefont {A.}~\bibnamefont {Smirne}}, \ and\ \bibinfo {author}
  {\bibfnamefont {S.~F.}\ \bibnamefont {Huelga}},\ }\href {\doibase
  10.1103/PhysRevX.10.041049} {\bibfield  {journal} {\bibinfo  {journal} {Phys.
  Rev. X}\ }\textbf {\bibinfo {volume} {10}},\ \bibinfo {pages} {041049}
  (\bibinfo {year} {2020}{\natexlab{b}})}\BibitemShut {NoStop}%
\bibitem [{\citenamefont {Chernyak}\ \emph {et~al.}(2006)\citenamefont
  {Chernyak}, \citenamefont {\ifmmode~\check{S}\else \v{S}\fi{}anda},\ and\
  \citenamefont {Mukamel}}]{Chernyak2006}%
  \BibitemOpen
  \bibfield  {author} {\bibinfo {author} {\bibfnamefont {V.}~\bibnamefont
  {Chernyak}}, \bibinfo {author} {\bibfnamefont {F.~c.~v.}\ \bibnamefont
  {\ifmmode~\check{S}\else \v{S}\fi{}anda}}, \ and\ \bibinfo {author}
  {\bibfnamefont {S.}~\bibnamefont {Mukamel}},\ }\href {\doibase
  10.1103/PhysRevE.73.036119} {\bibfield  {journal} {\bibinfo  {journal} {Phys.
  Rev. E}\ }\textbf {\bibinfo {volume} {73}},\ \bibinfo {pages} {036119}
  (\bibinfo {year} {2006})}\BibitemShut {NoStop}%
\bibitem [{\citenamefont {Engel}\ \emph {et~al.}(2007)\citenamefont {Engel},
  \citenamefont {Calhoun}, \citenamefont {Read}, \citenamefont {Ahn},
  \citenamefont {Man{\v{c}}al}, \citenamefont {Cheng}, \citenamefont
  {Blankenship},\ and\ \citenamefont {Fleming}}]{Engel2007}%
  \BibitemOpen
  \bibfield  {author} {\bibinfo {author} {\bibfnamefont {G.~S.}\ \bibnamefont
  {Engel}}, \bibinfo {author} {\bibfnamefont {T.~R.}\ \bibnamefont {Calhoun}},
  \bibinfo {author} {\bibfnamefont {E.~L.}\ \bibnamefont {Read}}, \bibinfo
  {author} {\bibfnamefont {T.-K.}\ \bibnamefont {Ahn}}, \bibinfo {author}
  {\bibfnamefont {T.}~\bibnamefont {Man{\v{c}}al}}, \bibinfo {author}
  {\bibfnamefont {Y.-C.}\ \bibnamefont {Cheng}}, \bibinfo {author}
  {\bibfnamefont {R.~E.}\ \bibnamefont {Blankenship}}, \ and\ \bibinfo {author}
  {\bibfnamefont {G.~R.}\ \bibnamefont {Fleming}},\ }\href {\doibase
  10.1038/nature05678} {\bibfield  {journal} {\bibinfo  {journal} {Nature}\
  }\textbf {\bibinfo {volume} {446}},\ \bibinfo {pages} {782} (\bibinfo {year}
  {2007})}\BibitemShut {NoStop}%
\bibitem [{\citenamefont {Krumm}\ \emph {et~al.}(2016)\citenamefont {Krumm},
  \citenamefont {Sperling},\ and\ \citenamefont {Vogel}}]{Krumm2016}%
  \BibitemOpen
  \bibfield  {author} {\bibinfo {author} {\bibfnamefont {F.}~\bibnamefont
  {Krumm}}, \bibinfo {author} {\bibfnamefont {J.}~\bibnamefont {Sperling}}, \
  and\ \bibinfo {author} {\bibfnamefont {W.}~\bibnamefont {Vogel}},\ }\href
  {\doibase 10.1103/PhysRevA.93.063843} {\bibfield  {journal} {\bibinfo
  {journal} {Phys. Rev. A}\ }\textbf {\bibinfo {volume} {93}},\ \bibinfo
  {pages} {063843} (\bibinfo {year} {2016})}\BibitemShut {NoStop}%
\bibitem [{\citenamefont {Moreva}\ \emph {et~al.}(2017)\citenamefont {Moreva},
  \citenamefont {Gramegna}, \citenamefont {Brida}, \citenamefont {Maccone},\
  and\ \citenamefont {Genovese}}]{Moreva2017}%
  \BibitemOpen
  \bibfield  {author} {\bibinfo {author} {\bibfnamefont {E.}~\bibnamefont
  {Moreva}}, \bibinfo {author} {\bibfnamefont {M.}~\bibnamefont {Gramegna}},
  \bibinfo {author} {\bibfnamefont {G.}~\bibnamefont {Brida}}, \bibinfo
  {author} {\bibfnamefont {L.}~\bibnamefont {Maccone}}, \ and\ \bibinfo
  {author} {\bibfnamefont {M.}~\bibnamefont {Genovese}},\ }\href {\doibase
  10.1103/PhysRevD.96.102005} {\bibfield  {journal} {\bibinfo  {journal} {Phys.
  Rev. D}\ }\textbf {\bibinfo {volume} {96}},\ \bibinfo {pages} {102005}
  (\bibinfo {year} {2017})}\BibitemShut {NoStop}%
\bibitem [{\citenamefont {Duan}\ \emph {et~al.}(2017)\citenamefont {Duan},
  \citenamefont {Prokhorenko}, \citenamefont {Cogdell}, \citenamefont {Ashraf},
  \citenamefont {Stevens}, \citenamefont {Thorwart},\ and\ \citenamefont
  {Miller}}]{Miller2017}%
  \BibitemOpen
  \bibfield  {author} {\bibinfo {author} {\bibfnamefont {H.~G.}\ \bibnamefont
  {Duan}}, \bibinfo {author} {\bibfnamefont {V.~I.}\ \bibnamefont
  {Prokhorenko}}, \bibinfo {author} {\bibfnamefont {R.~J.}\ \bibnamefont
  {Cogdell}}, \bibinfo {author} {\bibfnamefont {K.}~\bibnamefont {Ashraf}},
  \bibinfo {author} {\bibfnamefont {A.~L.}\ \bibnamefont {Stevens}}, \bibinfo
  {author} {\bibfnamefont {M.}~\bibnamefont {Thorwart}}, \ and\ \bibinfo
  {author} {\bibfnamefont {R.~J.~D.}\ \bibnamefont {Miller}},\ }\href {\doibase
  10.1073/pnas.1702261114} {\bibfield  {journal} {\bibinfo  {journal} {Proc
  Natl Acad Sci U S A}\ }\textbf {\bibinfo {volume} {114}},\ \bibinfo {pages}
  {8493} (\bibinfo {year} {2017})}\BibitemShut {NoStop}%
\bibitem [{\citenamefont {Ringbauer}\ \emph {et~al.}(2018)\citenamefont
  {Ringbauer}, \citenamefont {Costa}, \citenamefont {Goggin}, \citenamefont
  {White},\ and\ \citenamefont {Fedrizzi}}]{Ringbauer2018}%
  \BibitemOpen
  \bibfield  {author} {\bibinfo {author} {\bibfnamefont {M.}~\bibnamefont
  {Ringbauer}}, \bibinfo {author} {\bibfnamefont {F.}~\bibnamefont {Costa}},
  \bibinfo {author} {\bibfnamefont {M.~E.}\ \bibnamefont {Goggin}}, \bibinfo
  {author} {\bibfnamefont {A.~G.}\ \bibnamefont {White}}, \ and\ \bibinfo
  {author} {\bibfnamefont {A.}~\bibnamefont {Fedrizzi}},\ }\href {\doibase
  10.1038/s41534-018-0086-y} {\bibfield  {journal} {\bibinfo  {journal} {npj
  Quantum Information}\ }\textbf {\bibinfo {volume} {4}},\ \bibinfo {pages}
  {37} (\bibinfo {year} {2018})}\BibitemShut {NoStop}%
\bibitem [{\citenamefont {White}\ \emph {et~al.}(2020)\citenamefont {White},
  \citenamefont {Hill}, \citenamefont {Pollock}, \citenamefont {Hollenberg},\
  and\ \citenamefont {Modi}}]{White2020}%
  \BibitemOpen
  \bibfield  {author} {\bibinfo {author} {\bibfnamefont {G.~A.~L.}\
  \bibnamefont {White}}, \bibinfo {author} {\bibfnamefont {C.~D.}\ \bibnamefont
  {Hill}}, \bibinfo {author} {\bibfnamefont {F.~A.}\ \bibnamefont {Pollock}},
  \bibinfo {author} {\bibfnamefont {L.~C.~L.}\ \bibnamefont {Hollenberg}}, \
  and\ \bibinfo {author} {\bibfnamefont {K.}~\bibnamefont {Modi}},\ }\href
  {\doibase 10.1038/s41467-020-20113-3} {\bibfield  {journal} {\bibinfo
  {journal} {Nature Communications}\ }\textbf {\bibinfo {volume} {11}},\
  \bibinfo {pages} {6301} (\bibinfo {year} {2020})}\BibitemShut {NoStop}%
\bibitem [{\citenamefont {White}\ \emph {et~al.}(2022)\citenamefont {White},
  \citenamefont {Pollock}, \citenamefont {Hollenberg}, \citenamefont {Hill},\
  and\ \citenamefont {Modi}}]{white2022manybody}%
  \BibitemOpen
  \bibfield  {author} {\bibinfo {author} {\bibfnamefont {G.~A.~L.}\
  \bibnamefont {White}}, \bibinfo {author} {\bibfnamefont {F.~A.}\ \bibnamefont
  {Pollock}}, \bibinfo {author} {\bibfnamefont {L.~C.~L.}\ \bibnamefont
  {Hollenberg}}, \bibinfo {author} {\bibfnamefont {C.~D.}\ \bibnamefont
  {Hill}}, \ and\ \bibinfo {author} {\bibfnamefont {K.}~\bibnamefont {Modi}},\
  }\href {\doibase 10.48550/arXiv.2107.13934} {\enquote {\bibinfo {title} {From
  many-body to many-time physics},}\ } (\bibinfo {year} {2022}),\ \Eprint
  {http://arxiv.org/abs/2107.13934} {arXiv:2107.13934 [quant-ph]} \BibitemShut
  {NoStop}%
\bibitem [{\citenamefont {Knipschild}\ and\ \citenamefont
  {Gemmer}(2020)}]{Gemmer2020}%
  \BibitemOpen
  \bibfield  {author} {\bibinfo {author} {\bibfnamefont {L.}~\bibnamefont
  {Knipschild}}\ and\ \bibinfo {author} {\bibfnamefont {J.}~\bibnamefont
  {Gemmer}},\ }\href {\doibase 10.1103/PhysRevE.101.062205} {\bibfield
  {journal} {\bibinfo  {journal} {Phys. Rev. E}\ }\textbf {\bibinfo {volume}
  {101}},\ \bibinfo {pages} {062205} (\bibinfo {year} {2020})}\BibitemShut
  {NoStop}%
\bibitem [{\citenamefont {Taranto}\ \emph {et~al.}(2021)\citenamefont
  {Taranto}, \citenamefont {Pollock},\ and\ \citenamefont
  {Modi}}]{taranto2019memory}%
  \BibitemOpen
  \bibfield  {author} {\bibinfo {author} {\bibfnamefont {P.}~\bibnamefont
  {Taranto}}, \bibinfo {author} {\bibfnamefont {F.~A.}\ \bibnamefont
  {Pollock}}, \ and\ \bibinfo {author} {\bibfnamefont {K.}~\bibnamefont
  {Modi}},\ }\href {\doibase 10.1038/s41534-021-00481-4} {\bibfield  {journal}
  {\bibinfo  {journal} {npj Quantum Information}\ }\textbf {\bibinfo {volume}
  {7}},\ \bibinfo {pages} {149} (\bibinfo {year} {2021})}\BibitemShut {NoStop}%
\bibitem [{\citenamefont {Milz}\ \emph {et~al.}(2019)\citenamefont {Milz},
  \citenamefont {Kim}, \citenamefont {Pollock},\ and\ \citenamefont
  {Modi}}]{Milz2019completePos}%
  \BibitemOpen
  \bibfield  {author} {\bibinfo {author} {\bibfnamefont {S.}~\bibnamefont
  {Milz}}, \bibinfo {author} {\bibfnamefont {M.~S.}\ \bibnamefont {Kim}},
  \bibinfo {author} {\bibfnamefont {F.~A.}\ \bibnamefont {Pollock}}, \ and\
  \bibinfo {author} {\bibfnamefont {K.}~\bibnamefont {Modi}},\ }\href {\doibase
  10.1103/PhysRevLett.123.040401} {\bibfield  {journal} {\bibinfo  {journal}
  {Phys. Rev. Lett.}\ }\textbf {\bibinfo {volume} {123}},\ \bibinfo {pages}
  {040401} (\bibinfo {year} {2019})}\BibitemShut {NoStop}%
\bibitem [{\citenamefont {Burgarth}\ \emph
  {et~al.}(2021{\natexlab{a}})\citenamefont {Burgarth}, \citenamefont {Facchi},
  \citenamefont {Ligab\`o},\ and\ \citenamefont {Lonigro}}]{Burgarth2021}%
  \BibitemOpen
  \bibfield  {author} {\bibinfo {author} {\bibfnamefont {D.}~\bibnamefont
  {Burgarth}}, \bibinfo {author} {\bibfnamefont {P.}~\bibnamefont {Facchi}},
  \bibinfo {author} {\bibfnamefont {M.}~\bibnamefont {Ligab\`o}}, \ and\
  \bibinfo {author} {\bibfnamefont {D.}~\bibnamefont {Lonigro}},\ }\href
  {\doibase 10.1103/PhysRevA.103.012203} {\bibfield  {journal} {\bibinfo
  {journal} {Phys. Rev. A}\ }\textbf {\bibinfo {volume} {103}},\ \bibinfo
  {pages} {012203} (\bibinfo {year} {2021}{\natexlab{a}})}\BibitemShut
  {NoStop}%
\bibitem [{\citenamefont {Burgarth}\ \emph
  {et~al.}(2021{\natexlab{b}})\citenamefont {Burgarth}, \citenamefont {Facchi},
  \citenamefont {Lonigro},\ and\ \citenamefont {Modi}}]{Burgarth2021a}%
  \BibitemOpen
  \bibfield  {author} {\bibinfo {author} {\bibfnamefont {D.}~\bibnamefont
  {Burgarth}}, \bibinfo {author} {\bibfnamefont {P.}~\bibnamefont {Facchi}},
  \bibinfo {author} {\bibfnamefont {D.}~\bibnamefont {Lonigro}}, \ and\
  \bibinfo {author} {\bibfnamefont {K.}~\bibnamefont {Modi}},\ }\href {\doibase
  10.1103/PhysRevA.104.L050404} {\bibfield  {journal} {\bibinfo  {journal}
  {Phys. Rev. A}\ }\textbf {\bibinfo {volume} {104}},\ \bibinfo {pages}
  {L050404} (\bibinfo {year} {2021}{\natexlab{b}})}\BibitemShut {NoStop}%
\bibitem [{\citenamefont {Brand\~ao}\ \emph {et~al.}(2019)\citenamefont
  {Brand\~ao}, \citenamefont {Crosson}, \citenamefont {\ifmmode
  \mbox{\c{S}}\else \c{S}\fi{}ahino\ifmmode~\breve{g}\else \u{g}\fi{}lu},\ and\
  \citenamefont {Bowen}}]{Brandao2019}%
  \BibitemOpen
  \bibfield  {author} {\bibinfo {author} {\bibfnamefont {F.~G. S.~L.}\
  \bibnamefont {Brand\~ao}}, \bibinfo {author} {\bibfnamefont {E.}~\bibnamefont
  {Crosson}}, \bibinfo {author} {\bibfnamefont {M.~B.}\ \bibnamefont {\ifmmode
  \mbox{\c{S}}\else \c{S}\fi{}ahino\ifmmode~\breve{g}\else \u{g}\fi{}lu}}, \
  and\ \bibinfo {author} {\bibfnamefont {J.}~\bibnamefont {Bowen}},\ }\href
  {\doibase 10.1103/PhysRevLett.123.110502} {\bibfield  {journal} {\bibinfo
  {journal} {Phys. Rev. Lett.}\ }\textbf {\bibinfo {volume} {123}},\ \bibinfo
  {pages} {110502} (\bibinfo {year} {2019})}\BibitemShut {NoStop}%
\bibitem [{\citenamefont {Deutsch}(1991)}]{Deutsch1991}%
  \BibitemOpen
  \bibfield  {author} {\bibinfo {author} {\bibfnamefont {J.~M.}\ \bibnamefont
  {Deutsch}},\ }\href {\doibase 10.1103/PhysRevA.43.2046} {\bibfield  {journal}
  {\bibinfo  {journal} {Phys. Rev. A}\ }\textbf {\bibinfo {volume} {43}},\
  \bibinfo {pages} {2046} (\bibinfo {year} {1991})}\BibitemShut {NoStop}%
\bibitem [{\citenamefont {Srednicki}(1994)}]{Srednicki}%
  \BibitemOpen
  \bibfield  {author} {\bibinfo {author} {\bibfnamefont {M.}~\bibnamefont
  {Srednicki}},\ }\href {\doibase 10.1103/PhysRevE.50.888} {\bibfield
  {journal} {\bibinfo  {journal} {Phys. Rev. E}\ }\textbf {\bibinfo {volume}
  {50}},\ \bibinfo {pages} {888} (\bibinfo {year} {1994})}\BibitemShut
  {NoStop}%
\bibitem [{\citenamefont {Srednicki}(1999)}]{Srednicki_1999}%
  \BibitemOpen
  \bibfield  {author} {\bibinfo {author} {\bibfnamefont {M.}~\bibnamefont
  {Srednicki}},\ }\href {\doibase 10.1088/0305-4470/32/7/007} {\bibfield
  {journal} {\bibinfo  {journal} {J. Phys. A-Math. Gen.}\ }\textbf {\bibinfo
  {volume} {32}},\ \bibinfo {pages} {1163} (\bibinfo {year}
  {1999})}\BibitemShut {NoStop}%
\bibitem [{\citenamefont {Rigol}\ \emph {et~al.}(2007)\citenamefont {Rigol},
  \citenamefont {Dunjko}, \citenamefont {Yurovsky},\ and\ \citenamefont
  {Olshanii}}]{Rigol2007}%
  \BibitemOpen
  \bibfield  {author} {\bibinfo {author} {\bibfnamefont {M.}~\bibnamefont
  {Rigol}}, \bibinfo {author} {\bibfnamefont {V.}~\bibnamefont {Dunjko}},
  \bibinfo {author} {\bibfnamefont {V.}~\bibnamefont {Yurovsky}}, \ and\
  \bibinfo {author} {\bibfnamefont {M.}~\bibnamefont {Olshanii}},\ }\href
  {\doibase 10.1103/PhysRevLett.98.050405} {\bibfield  {journal} {\bibinfo
  {journal} {Phys. Rev. Lett.}\ }\textbf {\bibinfo {volume} {98}},\ \bibinfo
  {pages} {050405} (\bibinfo {year} {2007})}\BibitemShut {NoStop}%
\bibitem [{\citenamefont {Rigol}\ \emph {et~al.}(2008)\citenamefont {Rigol},
  \citenamefont {Dunjko},\ and\ \citenamefont {Olshanii}}]{Rigol2008}%
  \BibitemOpen
  \bibfield  {author} {\bibinfo {author} {\bibfnamefont {M.}~\bibnamefont
  {Rigol}}, \bibinfo {author} {\bibfnamefont {V.}~\bibnamefont {Dunjko}}, \
  and\ \bibinfo {author} {\bibfnamefont {M.}~\bibnamefont {Olshanii}},\ }\href
  {\doibase 10.1038/nature06838} {\bibfield  {journal} {\bibinfo  {journal}
  {Nature}\ }\textbf {\bibinfo {volume} {452}},\ \bibinfo {pages} {854 EP }
  (\bibinfo {year} {2008})}\BibitemShut {NoStop}%
\bibitem [{\citenamefont {Turner}\ \emph {et~al.}(2018)\citenamefont {Turner},
  \citenamefont {Michailidis}, \citenamefont {Abanin}, \citenamefont {Serbyn},\
  and\ \citenamefont {Papi\'{c}}}]{Turner2018}%
  \BibitemOpen
  \bibfield  {author} {\bibinfo {author} {\bibfnamefont {C.~J.}\ \bibnamefont
  {Turner}}, \bibinfo {author} {\bibfnamefont {A.~A.}\ \bibnamefont
  {Michailidis}}, \bibinfo {author} {\bibfnamefont {D.~A.}\ \bibnamefont
  {Abanin}}, \bibinfo {author} {\bibfnamefont {M.}~\bibnamefont {Serbyn}}, \
  and\ \bibinfo {author} {\bibfnamefont {Z.}~\bibnamefont {Papi\'{c}}},\ }\href
  {\doibase 10.1038/s41567-018-0137-5} {\bibfield  {journal} {\bibinfo
  {journal} {Nat. Phys.}\ }\textbf {\bibinfo {volume} {14}},\ \bibinfo {pages}
  {745} (\bibinfo {year} {2018})}\BibitemShut {NoStop}%
\bibitem [{\citenamefont {Deutsch}(2018)}]{Deutsch_2018}%
  \BibitemOpen
  \bibfield  {author} {\bibinfo {author} {\bibfnamefont {J.~M.}\ \bibnamefont
  {Deutsch}},\ }\href {\doibase 10.1088/1361-6633/aac9f1} {\bibfield  {journal}
  {\bibinfo  {journal} {Rep. Prog. Phys.}\ }\textbf {\bibinfo {volume} {81}},\
  \bibinfo {pages} {082001} (\bibinfo {year} {2018})}\BibitemShut {NoStop}%
\bibitem [{\citenamefont {Richter}\ \emph {et~al.}(2019)\citenamefont
  {Richter}, \citenamefont {Gemmer},\ and\ \citenamefont
  {Steinigeweg}}]{Richter2019}%
  \BibitemOpen
  \bibfield  {author} {\bibinfo {author} {\bibfnamefont {J.}~\bibnamefont
  {Richter}}, \bibinfo {author} {\bibfnamefont {J.}~\bibnamefont {Gemmer}}, \
  and\ \bibinfo {author} {\bibfnamefont {R.}~\bibnamefont {Steinigeweg}},\
  }\href {\doibase 10.1103/PhysRevE.99.050104} {\bibfield  {journal} {\bibinfo
  {journal} {Phys. Rev. E}\ }\textbf {\bibinfo {volume} {99}},\ \bibinfo
  {pages} {050104(R)} (\bibinfo {year} {2019})}\BibitemShut {NoStop}%
\bibitem [{\citenamefont {Milz}\ \emph {et~al.}(2021)\citenamefont {Milz},
  \citenamefont {Spee}, \citenamefont {Xu}, \citenamefont {Pollock},
  \citenamefont {Modi},\ and\ \citenamefont
  {Gühne}}]{10.21468/SciPostPhys.10.6.141}%
  \BibitemOpen
  \bibfield  {author} {\bibinfo {author} {\bibfnamefont {S.}~\bibnamefont
  {Milz}}, \bibinfo {author} {\bibfnamefont {C.}~\bibnamefont {Spee}}, \bibinfo
  {author} {\bibfnamefont {Z.-P.}\ \bibnamefont {Xu}}, \bibinfo {author}
  {\bibfnamefont {F.~A.}\ \bibnamefont {Pollock}}, \bibinfo {author}
  {\bibfnamefont {K.}~\bibnamefont {Modi}}, \ and\ \bibinfo {author}
  {\bibfnamefont {O.}~\bibnamefont {Gühne}},\ }\href {\doibase
  10.21468/SciPostPhys.10.6.141} {\bibfield  {journal} {\bibinfo  {journal}
  {SciPost Phys.}\ }\textbf {\bibinfo {volume} {10}},\ \bibinfo {pages} {141}
  (\bibinfo {year} {2021})}\BibitemShut {NoStop}%
\bibitem [{\citenamefont {Dümcke}(1983)}]{Duemcke1983}%
  \BibitemOpen
  \bibfield  {author} {\bibinfo {author} {\bibfnamefont {R.}~\bibnamefont
  {Dümcke}},\ }\href {\doibase 10.1063/1.525681} {\bibfield  {journal}
  {\bibinfo  {journal} {J. Math. Phys.}\ }\textbf {\bibinfo {volume} {24}},\
  \bibinfo {pages} {311} (\bibinfo {year} {1983})}\BibitemShut {NoStop}%
\bibitem [{\citenamefont {Figueroa-Romero}\ \emph {et~al.}(2019)\citenamefont
  {Figueroa-Romero}, \citenamefont {Modi},\ and\ \citenamefont
  {Pollock}}]{AlmostMarkov}%
  \BibitemOpen
  \bibfield  {author} {\bibinfo {author} {\bibfnamefont {P.}~\bibnamefont
  {Figueroa-Romero}}, \bibinfo {author} {\bibfnamefont {K.}~\bibnamefont
  {Modi}}, \ and\ \bibinfo {author} {\bibfnamefont {F.~A.}\ \bibnamefont
  {Pollock}},\ }\href {\doibase 10.22331/q-2019-04-30-136} {\bibfield
  {journal} {\bibinfo  {journal} {{Quantum}}\ }\textbf {\bibinfo {volume}
  {3}},\ \bibinfo {pages} {136} (\bibinfo {year} {2019})}\BibitemShut {NoStop}%
\bibitem [{\citenamefont {{Alexei Kitaev}}(2014)}]{Kitaev}%
  \BibitemOpen
  \bibfield  {author} {\bibinfo {author} {\bibnamefont {{Alexei Kitaev}}},\
  }\href@noop {} {\enquote {\bibinfo {title} {2015 breakthrough prize
  fundamental physics symposium},}\ }\bibinfo {howpublished}
  {\textsc{url:}~\url{https://breakthroughprize.org/Laureates/1/L3}} (\bibinfo
  {year} {2014})\BibitemShut {NoStop}%
\bibitem [{\citenamefont {Zonnios}\ \emph {et~al.}(2022)\citenamefont
  {Zonnios}, \citenamefont {Levinsen}, \citenamefont {Parish}, \citenamefont
  {Pollock},\ and\ \citenamefont {Modi}}]{zonnios2021signatures}%
  \BibitemOpen
  \bibfield  {author} {\bibinfo {author} {\bibfnamefont {M.}~\bibnamefont
  {Zonnios}}, \bibinfo {author} {\bibfnamefont {J.}~\bibnamefont {Levinsen}},
  \bibinfo {author} {\bibfnamefont {M.~M.}\ \bibnamefont {Parish}}, \bibinfo
  {author} {\bibfnamefont {F.~A.}\ \bibnamefont {Pollock}}, \ and\ \bibinfo
  {author} {\bibfnamefont {K.}~\bibnamefont {Modi}},\ }\href {\doibase
  10.1103/PhysRevLett.128.150601} {\bibfield  {journal} {\bibinfo  {journal}
  {Phys. Rev. Lett.}\ }\textbf {\bibinfo {volume} {128}},\ \bibinfo {pages}
  {150601} (\bibinfo {year} {2022})}\BibitemShut {NoStop}%
\bibitem [{\citenamefont {Dowling}\ and\ \citenamefont
  {Modi}(2022)}]{Dowling2022}%
  \BibitemOpen
  \bibfield  {author} {\bibinfo {author} {\bibfnamefont {N.}~\bibnamefont
  {Dowling}}\ and\ \bibinfo {author} {\bibfnamefont {K.}~\bibnamefont {Modi}},\
  }\href {\doibase 10.48550/ARXIV.2210.14926} {\enquote {\bibinfo {title}
  {Quantum chaos = volume-law spatiotemporal entanglement},}\ } (\bibinfo
  {year} {2022}),\ \Eprint {http://arxiv.org/abs/2210.14926} {arXiv:2210.14926
  [quant-ph]} \BibitemShut {NoStop}%
\bibitem [{\citenamefont {Styliaris}\ \emph {et~al.}(2021)\citenamefont
  {Styliaris}, \citenamefont {Anand},\ and\ \citenamefont
  {Zanardi}}]{Styliaris2021}%
  \BibitemOpen
  \bibfield  {author} {\bibinfo {author} {\bibfnamefont {G.}~\bibnamefont
  {Styliaris}}, \bibinfo {author} {\bibfnamefont {N.}~\bibnamefont {Anand}}, \
  and\ \bibinfo {author} {\bibfnamefont {P.}~\bibnamefont {Zanardi}},\ }\href
  {\doibase 10.1103/PhysRevLett.126.030601} {\bibfield  {journal} {\bibinfo
  {journal} {Phys. Rev. Lett.}\ }\textbf {\bibinfo {volume} {126}},\ \bibinfo
  {pages} {030601} (\bibinfo {year} {2021})}\BibitemShut {NoStop}%
\bibitem [{\citenamefont {Short}\ and\ \citenamefont
  {Farrelly}(2012)}]{ShortFinite}%
  \BibitemOpen
  \bibfield  {author} {\bibinfo {author} {\bibfnamefont {A.~J.}\ \bibnamefont
  {Short}}\ and\ \bibinfo {author} {\bibfnamefont {T.~C.}\ \bibnamefont
  {Farrelly}},\ }\href {\doibase 10.1088/1367-2630/14/1/013063} {\bibfield
  {journal} {\bibinfo  {journal} {New J. Phys.}\ }\textbf {\bibinfo {volume}
  {14}},\ \bibinfo {pages} {013063} (\bibinfo {year} {2012})}\BibitemShut
  {NoStop}%
\bibitem [{\citenamefont {Riera}\ \emph {et~al.}(2012)\citenamefont {Riera},
  \citenamefont {Gogolin},\ and\ \citenamefont {Eisert}}]{Riera2012}%
  \BibitemOpen
  \bibfield  {author} {\bibinfo {author} {\bibfnamefont {A.}~\bibnamefont
  {Riera}}, \bibinfo {author} {\bibfnamefont {C.}~\bibnamefont {Gogolin}}, \
  and\ \bibinfo {author} {\bibfnamefont {J.}~\bibnamefont {Eisert}},\ }\href
  {\doibase 10.1103/PhysRevLett.108.080402} {\bibfield  {journal} {\bibinfo
  {journal} {Phys. Rev. Lett.}\ }\textbf {\bibinfo {volume} {108}},\ \bibinfo
  {pages} {080402} (\bibinfo {year} {2012})}\BibitemShut {NoStop}%
\bibitem [{\citenamefont {Malabarba}\ \emph {et~al.}(2014)\citenamefont
  {Malabarba}, \citenamefont {Garc\'{\i}a-Pintos}, \citenamefont {Linden},
  \citenamefont {Farrelly},\ and\ \citenamefont {Short}}]{Malabarba2014}%
  \BibitemOpen
  \bibfield  {author} {\bibinfo {author} {\bibfnamefont {A.~S.~L.}\
  \bibnamefont {Malabarba}}, \bibinfo {author} {\bibfnamefont {L.~P.}\
  \bibnamefont {Garc\'{\i}a-Pintos}}, \bibinfo {author} {\bibfnamefont
  {N.}~\bibnamefont {Linden}}, \bibinfo {author} {\bibfnamefont {T.~C.}\
  \bibnamefont {Farrelly}}, \ and\ \bibinfo {author} {\bibfnamefont {A.~J.}\
  \bibnamefont {Short}},\ }\href {\doibase 10.1103/PhysRevE.90.012121}
  {\bibfield  {journal} {\bibinfo  {journal} {Phys. Rev. E}\ }\textbf {\bibinfo
  {volume} {90}},\ \bibinfo {pages} {012121} (\bibinfo {year}
  {2014})}\BibitemShut {NoStop}%
\bibitem [{\citenamefont {Wilming}\ \emph {et~al.}(2018)\citenamefont
  {Wilming}, \citenamefont {de~Oliveira}, \citenamefont {Short},\ and\
  \citenamefont {Eisert}}]{XXEisert}%
  \BibitemOpen
  \bibfield  {author} {\bibinfo {author} {\bibfnamefont {H.}~\bibnamefont
  {Wilming}}, \bibinfo {author} {\bibfnamefont {T.~R.}\ \bibnamefont
  {de~Oliveira}}, \bibinfo {author} {\bibfnamefont {A.~J.}\ \bibnamefont
  {Short}}, \ and\ \bibinfo {author} {\bibfnamefont {J.}~\bibnamefont
  {Eisert}},\ }\enquote {\bibinfo {title} {Equilibration times in closed
  quantum many-body systems},}\ in\ \href {\doibase
  10.1007/978-3-319-99046-0_18} {\emph {\bibinfo {booktitle} {Thermodynamics in
  the Quantum Regime: Fundamental Aspects and New Directions}}},\ \bibinfo
  {editor} {edited by\ \bibinfo {editor} {\bibfnamefont {F.}~\bibnamefont
  {Binder}}, \bibinfo {editor} {\bibfnamefont {L.~A.}\ \bibnamefont {Correa}},
  \bibinfo {editor} {\bibfnamefont {C.}~\bibnamefont {Gogolin}}, \bibinfo
  {editor} {\bibfnamefont {J.}~\bibnamefont {Anders}}, \ and\ \bibinfo {editor}
  {\bibfnamefont {G.}~\bibnamefont {Adesso}}}\ (\bibinfo  {publisher} {Springer
  International Publishing},\ \bibinfo {address} {Cham},\ \bibinfo {year}
  {2018})\ pp.\ \bibinfo {pages} {435--455}\BibitemShut {NoStop}%
\bibitem [{\citenamefont {Milz}\ \emph {et~al.}(2017)\citenamefont {Milz},
  \citenamefont {Pollock},\ and\ \citenamefont {Modi}}]{OperationalQDynamics}%
  \BibitemOpen
  \bibfield  {author} {\bibinfo {author} {\bibfnamefont {S.}~\bibnamefont
  {Milz}}, \bibinfo {author} {\bibfnamefont {F.~A.}\ \bibnamefont {Pollock}}, \
  and\ \bibinfo {author} {\bibfnamefont {K.}~\bibnamefont {Modi}},\ }\href
  {\doibase 10.1142/S1230161217400169} {\bibfield  {journal} {\bibinfo
  {journal} {Open Syst. Inf. Dyn.}\ }\textbf {\bibinfo {volume} {24}},\
  \bibinfo {pages} {1740016} (\bibinfo {year} {2017})}\BibitemShut {NoStop}%
\bibitem [{\citenamefont {Watrous}(2018)}]{watrous_2018}%
  \BibitemOpen
  \bibfield  {author} {\bibinfo {author} {\bibfnamefont {J.}~\bibnamefont
  {Watrous}},\ }\href {\doibase 10.1017/9781316848142} {\emph {\bibinfo {title}
  {The Theory of Quantum Information}}}\ (\bibinfo  {publisher} {Cambridge
  University Press},\ \bibinfo {year} {2018})\BibitemShut {NoStop}%
\bibitem [{\citenamefont {Wilde}(2011)}]{Wilde_2011}%
  \BibitemOpen
  \bibfield  {author} {\bibinfo {author} {\bibfnamefont {M.~M.}\ \bibnamefont
  {Wilde}},\ }\href {\doibase 10.1017/9781316809976.001} {\enquote {\bibinfo
  {title} {{From Classical to Quantum Shannon Theory}},}\ } (\bibinfo {year}
  {2011}),\ \Eprint {http://arxiv.org/abs/1106.1445} {arXiv:1106.1445
  [quant-ph]} \BibitemShut {NoStop}%
\bibitem [{\citenamefont {Watrous}(2004)}]{Watrous2004}%
  \BibitemOpen
  \bibfield  {author} {\bibinfo {author} {\bibfnamefont {J.}~\bibnamefont
  {Watrous}},\ }\href {\doibase 10.26421/QIC5.1-6} {\bibfield  {journal}
  {\bibinfo  {journal} {Quantum Inf. Comput.}\ }\textbf {\bibinfo {volume} {5}}
  (\bibinfo {year} {2004}),\ 10.26421/QIC5.1-6}\BibitemShut {NoStop}%
\bibitem [{\citenamefont {Taranto}\ \emph {et~al.}(2019)\citenamefont
  {Taranto}, \citenamefont {Milz}, \citenamefont {Pollock},\ and\ \citenamefont
  {Modi}}]{Taranto2019FiniteMarkov}%
  \BibitemOpen
  \bibfield  {author} {\bibinfo {author} {\bibfnamefont {P.}~\bibnamefont
  {Taranto}}, \bibinfo {author} {\bibfnamefont {S.}~\bibnamefont {Milz}},
  \bibinfo {author} {\bibfnamefont {F.~A.}\ \bibnamefont {Pollock}}, \ and\
  \bibinfo {author} {\bibfnamefont {K.}~\bibnamefont {Modi}},\ }\href {\doibase
  10.1103/PhysRevA.99.042108} {\bibfield  {journal} {\bibinfo  {journal} {Phys.
  Rev. A}\ }\textbf {\bibinfo {volume} {99}},\ \bibinfo {pages} {042108}
  (\bibinfo {year} {2019})}\BibitemShut {NoStop}%
\bibitem [{\citenamefont {Inc.}()}]{Mathematica}%
  \BibitemOpen
  \bibfield  {author} {\bibinfo {author} {\bibfnamefont {W.~R.}\ \bibnamefont
  {Inc.}},\ }\href@noop {} {\enquote {\bibinfo {title} {Mathematica, {V}ersion
  12.3.1},}\ }\bibinfo {note} {Champaign, IL, 2021}\BibitemShut {NoStop}%
\bibitem [{\citenamefont {Miszczak}\ \emph {et~al.}(011 )\citenamefont
  {Miszczak}, \citenamefont {Puchała},\ and\ \citenamefont
  {Gawron}}]{QIpackage}%
  \BibitemOpen
  \bibfield  {author} {\bibinfo {author} {\bibfnamefont {J.}~\bibnamefont
  {Miszczak}}, \bibinfo {author} {\bibfnamefont {Z.}~\bibnamefont {Puchała}},
  \ and\ \bibinfo {author} {\bibfnamefont {P.}~\bibnamefont {Gawron}},\ }\href
  {https://github.com/iitis/qi} {\enquote {\bibinfo {title} {Qi package for
  anaylsis of quantum systems},}\ } (\bibinfo {year} {2011-})\BibitemShut
  {NoStop}%
\end{thebibliography}
\end{document}